\documentclass[%
 aip,
 amsmath,amssymb,
 reprint,%
]{revtex4-1}

\usepackage{graphicx}
\usepackage{dcolumn}
\usepackage{bm}

\usepackage[utf8]{inputenc}
\usepackage[T1]{fontenc}
\usepackage{mathptmx}
\usepackage{etoolbox}

\usepackage{xcolor}

\makeatletter
\def\@email#1#2{%
 \endgroup
 \patchcmd{\titleblock@produce}
  {\frontmatter@RRAPformat}
  {\frontmatter@RRAPformat{\produce@RRAP{*#1\href{mailto:#2}{#2}}}\frontmatter@RRAPformat}
  {}{}
}%
\makeatother
\begin{document}

\preprint{AIP/123-QED}

\title{Quantification of Reynolds-averaged-Navier-Stokes model form uncertainty in transitional boundary layer and airfoil flows}

\author{Minghan Chu}
 \altaffiliation[Also at ]{Mechanical and Materials Engineering Department, Queen's University, Kingston, ON K7L 2V9, Canada.}
 
\author{Xiaohua Wu}%
\homepage{https://www.rmc-cmr.ca/en/mechanical-and-aerospace-engineering/xiaohua-wu}
 \email{17mc93@queensu.ca.}
\affiliation{ 
Mechanical and Aerospace Engineering, Royal Military College of Canada, Kingston, ON K7K 7B4, Canada.
}%

\author{David E. Rival}
\affiliation{%
Mechanical and Materials Engineering Department, Queen's University, Kingston, ON K7L 2V9, Canada.
}%

\date{\today}

\begin{abstract}
It is well known that Boussinesq turbulent-viscosity hypothesis can introduce uncertainty in predictions for complex flow features such as separation, reattachment, and laminar-turbulent transition. This study adopts a recent physics-based uncertainty quantification (UQ) approach to address such model form uncertainty in Reynolds-averaged Naiver-Stokes (RANS) simulations. Thus far, almost all UQ studies have focused on quantifying the model form uncertainty in turbulent flow scenarios. The focus of the study is to advance our understanding of the performance of the UQ approach on two different transitional flow scenarios: a flat plate and a SD7003 airfoil, to close this gap. For the T3A (flat-plate flow) flow, most of the model form uncertainty is concentrated in the laminar-turbulent transition region. For the SD7003 airfoil flow, the eigenvalue perturbations reveal a decrease as well as an increase in the length of the separation bubble. As a consequence, the uncertainty bounds successfully encompass the reattachment point. Likewise, the region of reverse flow that appear in the separation bubble is either suppressed or bolstered by the eigenvalue perturbations. In this context, the UQ methodology is applied to transition and show great results. This is the first successful RANS UQ study for transitional flows. 
\end{abstract}

\maketitle


\section{\label{sec:level1}INTRODUCTION}
Transitional flow regime is very frequently encountered in turbomachines and especially in aircraft engines at relatively low Reynolds numbers. As a consequence, a significant part of the flow on the blade surfaces is under the laminar-turbulent transition process. The boundary development, losses, efficiency, and momentum transfer are greatly affected by the laminar-turbulent transition. Therefore, accurate prediction for the transition process is crucial for the design of efficient as well as reliable machines \cite{pecnik2007application}.   

Reynolds-averaged Navier-Stokes (RANS) simulations remain the most commonly used computational technique for analysis of turbulent flows. There has been considerable effort spent in the past two decades to develop RANS-based transition models for engineering applications to predict various kinds of transitional flows \cite{menter2002transition,menter2004correlation,menter2006transition,langtry2009correlation,menter2015one,wei2017modeling,tousi2021active}. Each model has its strengths and weaknesses, and by far the correlation-based transition models by Langtry and Menter \cite{langtry2009correlation,menter2015one} have been widely used in engineering industries, in particular, aerospace industry. Most RANS models have adopted the Boussinesq turbulent viscosity hypothesis, i.e., anisotropy Reynolds stresses are proportional to the mean rate of strain, therefore also referred to as linear eddy viscosity models. It is well known that linear eddy viscosity models are limited due to the restrictions of the Boussinesq turbulent viscosity hypothesis, in particular, on yielding accurate predictions for complex flow features such as separation-induced transition. Large eddy simulations (LES) or Direct numerical simulations (DNS) provide high-fidelity solution for such problems, but the calculations are often too expensive in computational time and cost, especially for high-Reynolds number flows. Therefore, uncertainty quantification (UQ) for the model form uncertainty is an valuable alternative route for improving the RANS predictive capability in engineering applications. More expensive LES or DNS would only be considered necessary if the model form uncertainty is too large. 

The current study considers a physics-based approach that has been recently introduced by Emory \textit{et al.} \cite{emory2013modeling}, namely eigenspace perturbation method. This framework quantifies the model form uncertainty associated with the linear eddy viscosity model via sequential perturbations in the predicted amplitude (turbulence kinetic energy), shape (eigenvalues), and orientation (eigenvectors) of the anisotropy tensor. This perturbation method for RANS model uncertainty quantification has been applied to analyze and estimate the RANS uncertainty in flow through scramjets \cite{emory2011characterizing}, aircraft nozzle jets, turbomachinery, over stream-lined bodies \cite{gorle2019epistemic}, supersonic axisymmetric submerged jet \cite{mishra2017rans}, and canonical cases of turbulent flows over a backward-facing step \cite{iaccarino2017eigenspace,cremades2019reynolds}.  This method has been used for robust design of Organic Rankine Cycle (ORC) turbine cascades \cite{razaaly2019optimization}. In aerospace applications, this method has been used for design optimization under uncertainty\cite{cook2019optimization,mishra2020design,matha2022extending,matha2022assessment}. In civil engineering applications, this method is being used to design urban canopies \cite{garcia2014quantifying}, ensuring the ventilation of enclosed spaces, and used in the wind engineering practice for turbulent bluff body flows \cite{gorle2015quantifying}. This perturbation method for RANS model uncertainty quantification has been used in conjunction with Machine Learning algorithms to provide precise estimates of RANS model uncertainty in the presence of data \cite{xiao2016quantifying,wu2016bayesian,parish2016paradigm,xiao2017random,wang2017physics,wang2017comprehensive,heyse2021estimating}. The method is also being used for the creation of probabilistic aerodynamic databases, enabling the certification of virtual aircraft designs \cite{mukhopadhaya2020multi,nigam2021toolset}.

In contrast to the data-driven methods, the current approach does not require high-fidelity data as input, hence more generally applicable. A key feature is that most of the application of this perturbation method for RANS model uncertainty have been on fully developed turbulent flows. However, in many aerospace applications flows undergoing transition are important and we require reliable UQ for the RANS predictions for such transitional flows, while few studies of the performance of this UQ methodology in the prediction for RANS transition modeling thus far have been performed. In addition, most studies focused on the performance of the eigenspace perturbation approach on mean velocity profile, skin and/or pressure coefficient, and did not consider the turbulence quantities. The essence for eigenspace perturbation method to perturb the shape of the Reynolds stress tensor is to move a linear-eddy-viscosity predicted Reynolds stress anisotropy state to a new location on a barycentric map \cite{banerjee2007presentation}. The concept of Reynolds stress anisotropy can be better understood by analyzing the anisotropy trajectories presented on this map. However, very few studies have been performed to analyze anisotropy states for transitional flows.

The goal of this paper is therefore to advance the understanding of the performance of the eigenspace perturbation approach for quantifying the model form uncertainty in RANS simulations for two different transitional flow scenarios: flow over a flat plate (zero-pressure gradient) and flow over an SD7003 airfoil. Specifically, the objectives are (1) to quantify the model form uncertainty in three different linear eddy viscosity models (two turbulence models and one transition model) to scrutinize the differences between them; (2) to analyze the anisotropy states for the transitional boundary layer over a SD7003 airfoil based on a widely used RANS transition model and the in-house DNS data \cite{zhang2021turbulent} using both the barycentric and Lumley's invariant maps \cite{lumley1979computational,pope2001turbulent,durbin2011statistical} ; (3) to explore the performance of the eigenspace perturbation approach on various flow-related QoIs: mean velocity, local wall shear stress and pressure, Reynolds shear stress, and turbulent production rate. The in-house DNS database of \cite{zhang2021turbulent} for the SD7003 configuration is used to support our exploration.  

\section{Methodology}
\subsection{\label{sec:level2}Governing equations}
The flow was assumed to be two-dimensional and incompressible. The RANS formulation of the continuity and momentum equations is as follows:

\begin{equation} \label{p_Continuity}
   \frac{\partial \left\langle U_{i} \right\rangle}{\partial x_{i}}=0,
\end{equation}
\begin{equation} \label{p_Momentum}
   \frac{ D \left\langle U_{j}\right\rangle}{\mathrm{Dt}}=-\frac{1}{\rho} \frac{\partial \left\langle P \right\rangle}{\partial x_{j}}+\nu \frac{\partial^{2} {\left\langle U_{j} \right\rangle}}{\partial x_{i} \partial x_{i}}-\frac{\partial \left\langle u_{i} u_{j}\right\rangle}{\partial x_{i}}
\end{equation}

\noindent where $\left\langle \ \right\rangle$ represents time-averaging. $\rho$ is the density, $\left\langle P \right\rangle$ is the time-averaged pressure, and $\nu$ is the kinematic viscosity. The $\left\langle U_{i}\right\rangle$ are the time-averaged velocity components. Reynolds stress terms in Eqs. \ref{p_Continuity} - \ref{p_Momentum}, i.e., $\left\langle u_{i}u_{j}\right\rangle$, are unknowns that need to be approximated using a RANS model. In this study, two-equation linear eddy viscosity models are used, which are based on the Boussinesq turbulent viscosity hypothesis as follows:

\begin{equation}\label{Eq:noMark_uiuj}
    \left\langle{u_{i} u_{j}}\right\rangle=\frac{2}{3} k \delta_{i j}-2 \nu_{\mathrm{t}} \left\langle S_{i j} \right\rangle,
\end{equation}

\noindent where $k$ is the turbulence kinetic energy, $\delta_{i j}$ is the Kronecker delta, $\nu_\mathrm{t}$ is the turbulent viscosity, and $\left\langle S_{i j} \right\rangle$ is the rate of mean strain tensor. In Eq. \ref{Eq:noMark_uiuj}, the deviatoric anisotropic part is 

\begin{equation}\label{Eqn:Bou_Ani_Tensor}
\begin{aligned}
a_{i j} & \equiv\left\langle u_{i} u_{j}\right\rangle-\frac{2}{3} k \delta_{i j} \\
&=-\nu_{\mathrm{t}}\left(\frac{\partial\left\langle U_{i}\right\rangle}{\partial x_{j}}+\frac{\partial\left\langle U_{j}\right\rangle}{\partial x_{i}}\right) \\
&=-2 \nu_{\mathrm{t}} \left\langle S_{i j} \right\rangle.
\end{aligned}
\end{equation}

The normalized anisotropy tensor (used extensively) is defined by 

\begin{equation}\label{Eq:noMark_AnisotropyTensor}
    b_{i j}= \frac{a_{ij}}{2k} = \frac{\big \langle {u_{i} u_{j}} \big \rangle }{2 k}-\frac{\delta_{i j}}{3} = -\frac{\nu_{t} }{k}\big \langle {S_{i j}} \big \rangle. 
\end{equation}

In the results presented hereafter for the flat plate, three different linear eddy viscosity models were considered: the shear-stress transport (SST) $k - \omega$ \cite{menter1993zonal,hellsten1998some,menter2001elements,menter2003ten}, the modified version of SST $k - \omega$ for transitional flow simulations by Langtry and Menter (SST $k - \omega$ LM) \cite{menter2004correlation,menter2006correlation,langtry2009correlation}, and the $k - \varepsilon$ model \cite{el1983k,launder1983numerical}. By considering three different models, we intend to contrast the uncertainty bounds generated by the transition model (SST $k - \omega$ LM) with the two turbulence models (SST $k - \omega$ and $k - \varepsilon$). Results corresponding to these linear eddy viscosity models bereft of any perturbations are referred to as ``baseline'' solutions.

\subsection{Eigenspace perturbation method}
The Reynolds stress tensor $\left\langle u_{i} u_{j}\right\rangle$ is symmetric positive semi-definite \cite{pope2001turbulent}, thus it can be eigen-decomposed as follows:

\begin{equation} \label{Eq:noMarker_Rij}
    \left\langle u_{i} u_{j}\right\rangle=2 k\left(\frac{\delta_{i j}}{3}+v_{i n}  \hat{b}_{n l} v_{j l}\right),
\end{equation}

\noindent in which $k \equiv {u_{i} u_{i}} / 2$, $v$ represents the matrix of orthonormal eigenvectors, $\hat{b}$ represents the diagonal matrix of eigenvalues ($\lambda_{i}$), which are arranged in a non-increasing order such that $\lambda_{1} \geq \lambda_{2} \geq \lambda_{3}$. The amplitude, the shape and the orientation of $\left\langle u_{i}u_{j} \right\rangle$ are explicitly represented by $k$, $\lambda_{i}$, and $v_{i j}$, respectively. Equations \ref{Eq:noMark_AnisotropyTensor} and \ref{Eq:noMarker_Rij} lead to 

\begin{equation}\label{Eq:noMarker_bij}
    b_{i j}=-\frac{\nu_{t} }{k}\big \langle {S_{i j}} \big \rangle = v_{i n} \hat{b}_{n l} v_{j l}.
\end{equation}

Equation \ref{Eq:noMarker_bij} indicates that the Boussinesq turbulent viscosity hypothesis requires that the shape and orientation of $\left\langle u_{i}u_{j} \right\rangle$ to be determined by $(\nu_{t}/k)\big \langle {S_{i j}} \big \rangle$. This assumption implies the $a_{i j}$ tensor is aligned with the $\big \langle {S_{i j}} \big \rangle$ tensor, which is not true in most circumstances in practice, in particular, complex flows, e.g., strongly swirling flows, flow with significant streamline curvature, and flow with separation and reattachment, and thus a source of the model form uncertainty.

The eigenspace perturbation method was first proposed in \cite{emory2011modeling,gorle2012epistemic}. To model errors introduced in the model form uncertainty, perturbation is injected into the eigen-decomposed Reynolds stress defined in Eq. \ref{Eq:noMarker_Rij}. The perturbed Reynolds stress are defined as

\begin{equation}\label{Eqn_Rij_perturbed}
    \left\langle u_{i} u_{j}\right\rangle^{*}=2 k^{*}\left(\frac{1}{3} \delta_{i j}+v_{i n}^{*} \hat{b}_{n l}^{*} v_{j l}^{*}\right),
\end{equation}

\noindent where $k^{*}=k+\Delta k$ is the perturbed turbulence kinetic energy, $\hat{b}_{k l}^{*}$ is the diagonal matrix of perturbed eigenvalues, and $v_{i j}^{*}$ is the perturbed eigenvector matrix. Perturbing $k$ is an important component of the eigenspace perturbation framework. In reality, the coefficient of turbulent viscosity in the Boussinesq turbulent viscosity hypothesis varies between different turbulent flow cases and even between different regions in the same turbulent flow \cite{mishra2019theoretical}; however, the Boussinesq turbulent viscosity hypothesis espouses a universal constant coefficient, which fails to capture the true physics of turbulent flow. According to Mishra and Iaccarino \cite{mishra2019theoretical}, perturbations to turbulence kinetic energy spatially vary the coefficient of turbulent viscosity, and in a sense change the original Boussinesq turbulent viscosity hypothesis to anisotropy viscosity hypothesis. Thus, turbulence kinetic energy perturbation is important to describe the real behavior of turbulent flow. While few studies addressing the perturbations to turbulence kinetic energy have been conducted, it becomes a valuable future research direction. Eigenvector perturbations rotate the eigenvectors of the anisotropy Reynolds stress tensor with respect to the principal axes of the mean rate of strain. Recall that the eigenvectors of the anisotropy Reynolds stress tensor are forced to align along the principal axes of the mean rate of strain due to the limitations of the Boussinesq turbulent viscosity hypothesis \cite{pope2001turbulent}. This again violates the true physics of turbulent flow. Consequently, eigenvector perturbations extend the Boussinesq turbulent viscosity hypoethsis to anisotropy turbulent viscosity hypothesis. Unlike eigenvalue perturbations, which are strictly constrained by realizability. Eigenvector perturbations are more difficult to be physically constrained in a local sense. In the current study, eigenvector perturbations are omitted for brevity. For these reasons, the present study restricts the contribution to eigenvalue perturbations $\hat{b}_{i j}^{*}$. Due to the realizability constraint of the semi-definiteness of $\left\langle u_{i}u_{j} \right\rangle$, there are three extreme states of componentiality of $\left\langle u_{i}u_{j} \right\rangle$: one component limiting state ($1C$), which has one non-zero principal fluctuation, i.e., $\hat{b}_{1c}=\operatorname{diag}[2 / 3,-1 / 3,-1 / 3]$; two component limiting state ($2C$), which has two non-zero principal fluctuations of the same intensity, i.e., $\hat{b}_{2c}=\operatorname{diag}[1 / 6,1 / 6, -1 / 3]$; and three component (isotropic) limiting state (3C), which has three non-zero principal fluctuations of the same intensity, i.e., $\hat{b}_{3c}=\operatorname{diag}[0,0,0]$.

\begin{figure} 
\centerline{\includegraphics[width=3.4in]{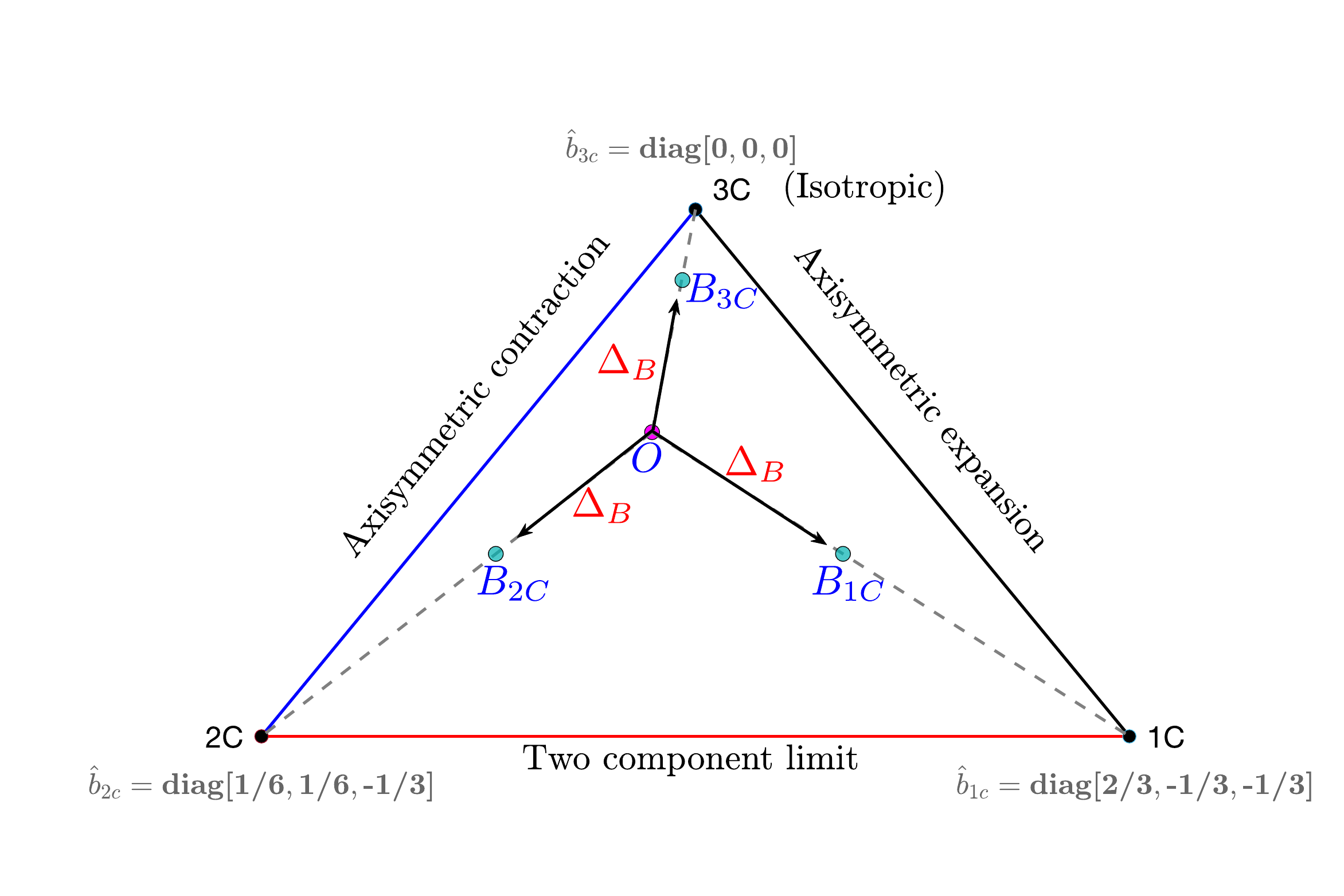}}
\caption{Barycentric map.}
\label{fig:BMap_Sketch.pdf}
\end{figure}

For eigenvalue perturbations, Pecnik and Iaccarino \cite{emory2011modeling} proposed a perturbation approach, which enforces the realizability constraints on $\left\langle u_{i}u_{j} \right\rangle$ via the barycentric map \cite{banerjee2007presentation}, as shown in Fig. \ref{fig:BMap_Sketch.pdf}, because the map contains all realizable sates of $\left\langle u_{i}u_{j} \right\rangle$. In addition, the $\hat{b}_{1c}$, $\hat{b}_{2c}$, and $\hat{b}_{3c}$ limiting states correspond to the three vertices of the barycentric map. Given an arbitrary point $\mathbf{x}$ within the barycentric map, any realizable $\left\langle u_{i}u_{j} \right\rangle$ can be determined by a convex combination of the three vertices $\mathbf{x}_{i c}$ (limiting states) and $\lambda_{l}$ as follows:

\begin{equation}\label{Eq:noMarker_Coordinates_InsideBary}
    \mathbf{x} = \mathbf{x}_{1 \mathrm{c}}\left(\lambda_{1}-\lambda_{2}\right)+\mathbf{x}_{2 \mathrm{c}}\left(2 \lambda_{2}-2 \lambda_{3}\right)+\mathbf{x}_{3 \mathrm{c}}\left(3 \lambda_{3}+1\right).
\end{equation}

In order to define the perturbed eigenvalues $\hat{b}_{i j}^{*}$, first determine the location for the Reynolds stress computed by a linear eddy viscosity model and subsequently inject uncertainty by shifting it to a new location on the barycentric map. In Fig. \ref{fig:BMap_Sketch.pdf}, perturbations toward $1c$, $2c$, and $3c$ vertices of the barycentric map shift point $O$ to $B_{1c/2c/3c}$, respectively, which can be written as 

\begin{equation}\label{Eq:noMarker_xstar}
    \mathbf{x_{B(1c/2c/3c)}^{*}}=\mathbf{x_{O}}+\Delta_{B}\left(\mathbf{x}_{1c/2c/3c}-\mathbf{x_{B(1c/2c/3c)}}\right),
\end{equation}

\noindent where $\Delta_{B}$ is the magnitude of perturbation toward the three vertices. Once the new location is determined, a new set of eigenvalues $\lambda_{i}$ can be computed from Eq. \ref{Eq:noMarker_Coordinates_InsideBary} and to reconstruct $b_{i j}$, which eventually yields $\left\langle u_{i}u_{j} \right\rangle^{*}$.

\subsection{Eigenspace perturbation framework in OpenFOAM}
This study contributes the eigenspace perturbation framework \cite{emory2013modeling} to both the ``simpleFOAM'' (steady) and ``pimpleFOAM'' (transient) solvers in the open source OpenFOAM software \cite{winkelman1980flowfield}. Most previous studies on the model form uncertainty have used the open source OpenFOAM software \cite{winkelman1980flowfield} compounded with the Matlab software to decompose and recompose the Reynolds stress tensor, e.g., see \cite{cremades2019reynolds,hornshoj2021quantifying}. In this study, a new class of the eigenspace perturbation framework written in the $100\%$ OpenFOAM-version of C++ was created and injected in the top level classes in OpenFOAM, which greatly reduces the number of user-defined inputs. This allows the users without much knowledge of the fluid mechanics to use the eigenspace perturbation method in OpenFOAM.

\begin{figure*} 
\centerline{\includegraphics[width=6in]{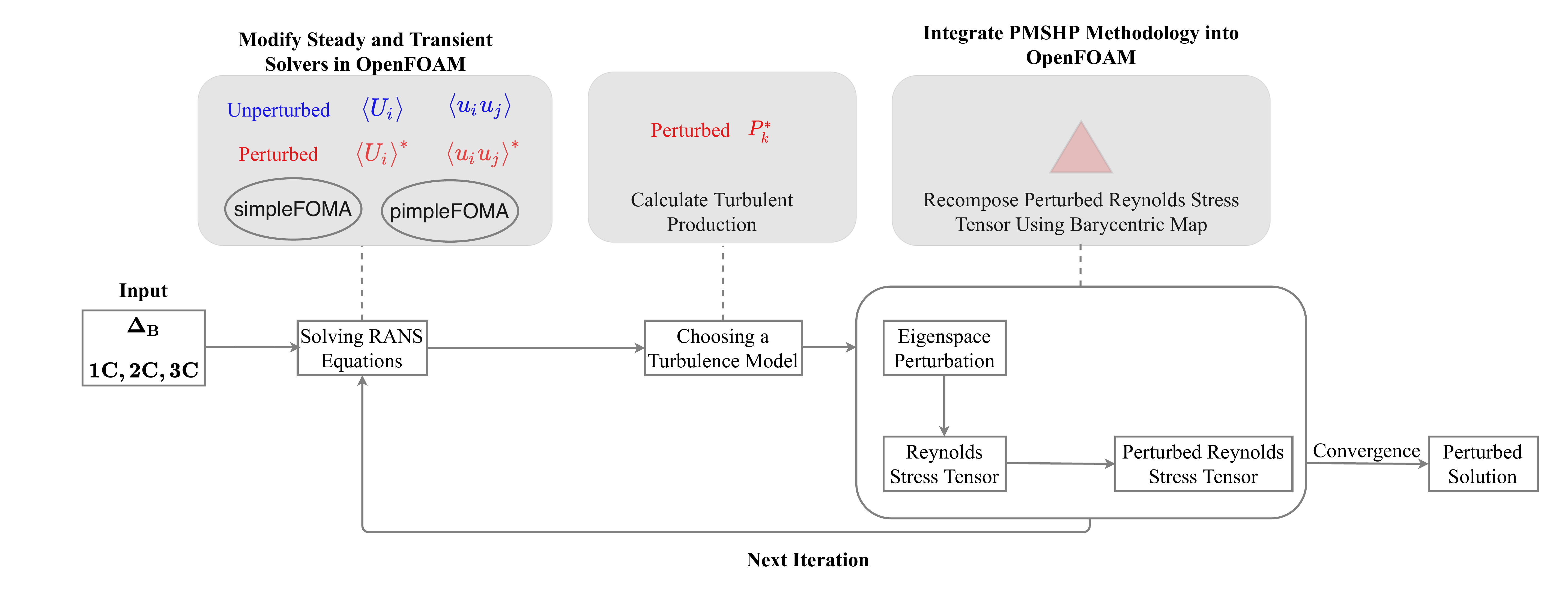}}
\caption{Flow chart showing the implementation of eigenspace perturbation framework within OpenFOAM.}
\label{fig:FlowChart_Method_noMarker.pdf}
\end{figure*}

The magnitude of perturbation $\Delta_{B}$ and which eigenvalue perturbation ($1c$,$2c$,$3c$) to be performed are specified by the user in the input files located under the ``Constant'' directory in OpenFOAM, and C++ code with customized OpenFOAM data type is added to the existing code base that conducts the perturbations during the execution of simulations, as shown in Fig. \ref{fig:FlowChart_Method_noMarker.pdf}. At each control volume (CV), the baseline Reynolds stress tensor is calculated and decomposed into its eigenvalue and eigenvector matrices, which are perturbed using the eigenspace perturbation method as prescribed earlier. The perturbed eigenvalue and eigenvector matrices are then recomposed into a perturbed Reynolds stress tensor for each CV. These perturbed Reynolds stress tensors are then used to compute the perturbed velocity field and the perturbed turbulent production to advance each node to the next (pseudo) time step. At convergence, the Reynolds stress also converges to its perturbed state.

\section{Flow description and numerical method}
In this study, the eigenspace perturbation method was used to test several RANS models in the prediction for the laminar-turbulent transition process that occurs for incompressible transitional flow over two different geometries: ERCOFTAC T3A \cite{roach1990influence} and SD7003 airfoil. Details of these two flow configurations are presented below. 

\subsection{T3A}
\begin{figure} 
\centerline{\includegraphics[width=3.5in]{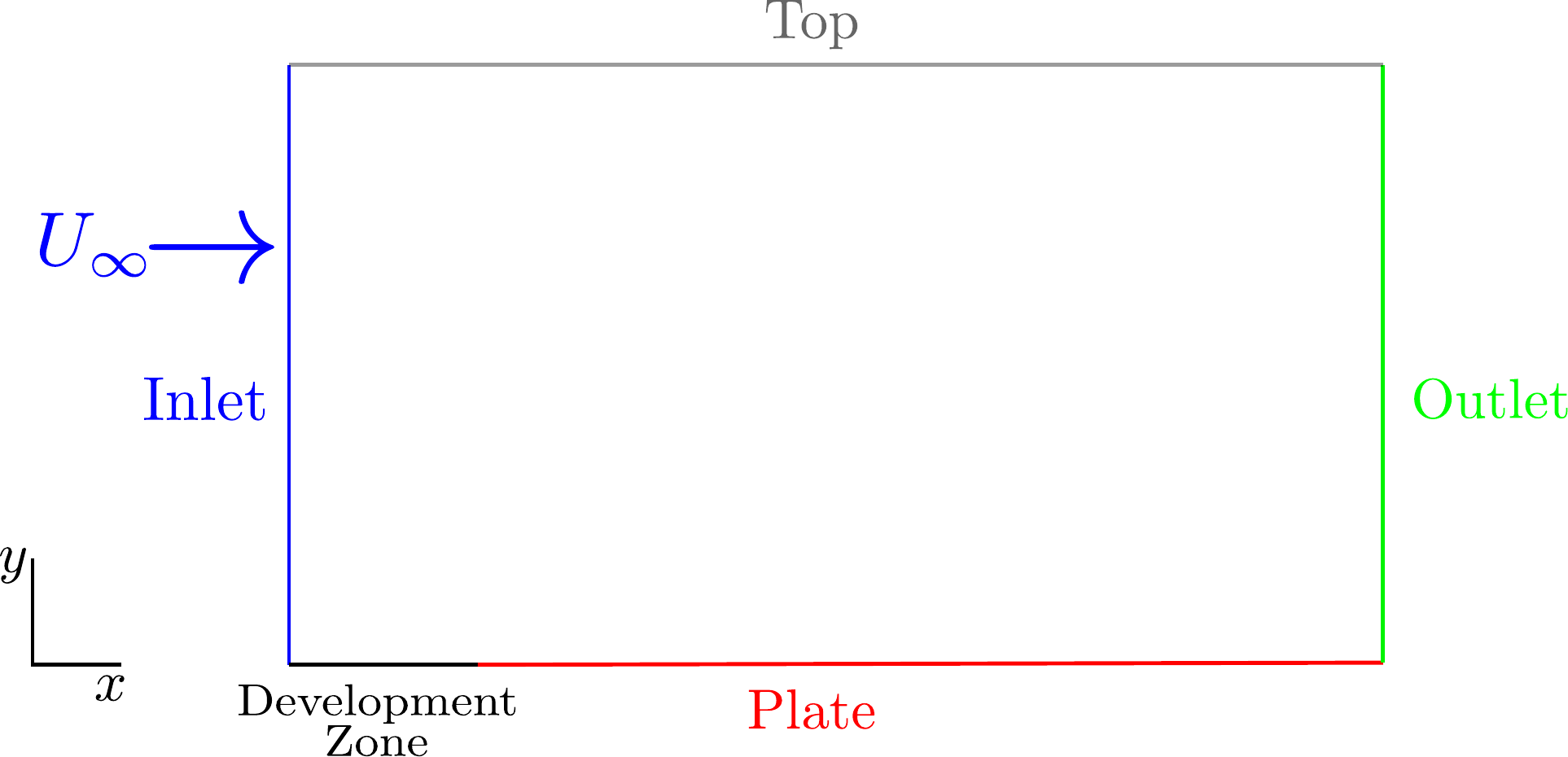}}
\caption{T3A computational domain. {\color{blue} \rule{0.7cm}{0.4mm}} inflow ($U_{\infty}$), {\color{green} \rule{0.7cm}{0.4mm}} outflow, {\color{red} \rule{0.7cm}{0.4mm}}, no-slip wall, {\color{gray} \rule{0.7cm}{0.4mm}}, Top, {\color{black} \rule{0.7cm}{0.4mm}}, and flow-development zone.}
\label{fig:T3A_domain.pdf}
\end{figure}

One flow being considered is the ERCOFTAC T3A case over a flat plate (zero pressure gradient), as shown in Fig. \ref{fig:T3A_domain.pdf}. The solution domain is two-dimensional and uses a Cartesian coordinate system. The computational domain is $3.0 \mathrm{~m}$ (L) long and $1.0 \mathrm{~m}$ (H) high in the streamwise (x) and wall-normal (y) direction, respectively. The simulation results based on the nonuniform grids of $340$ and $60$ control volumes in the streamwise and wall-normal direction, respectively. The first grid node in the wall-normal direction was placed at a value of $y^{+} < 1$ over the entire plate, and the effect of refining the mesh in the $y$-direction on the prediction for the skin friction coefficient was much less than $1\%$. The fluid was assumed to be air, with freestream turbulence intensity of $\operatorname{Tu}=\sqrt{2 k / 3} / U_{\infty} = 3.3\%$, where $U_{\infty}$ represents the freestrem velocity, and kinematic viscosity of $\nu = 1.5 \times 10^{-5} \mathrm{~m}^{2} / \mathrm{s}$. For the smooth flat plate, a no-slip boundary condition was assumed. At the inlet of the domain, the freestream velocity was set equal to $U_{\mathrm{\infty}}=5.4 \mathrm{~m} / \mathrm{s}$. A slip boundary condition was specified for $\left\langle U_{i}\right\rangle$ ($U$ for $x$ direction and $V$ for $y$ direction) at the top of the domain, and a zero-gradient for $k$, $\omega$ and pressure at the outlet and the top of the domain. Upstream of the leading edge has a space that allows the flow to develop before encountering the leading edge, as shown in Fig. \ref{fig:T3A_domain.pdf}. The governing Eqs. \ref{p_Continuity} - \ref{p_Momentum} together with the transport equation for the SST $k-\omega$ LM model were solved using the open source software OpenFOAM \cite{weller1998tensorial}. The transport equations were discretized using finite volume method. The scheme is second order upwind for spatial dicretization of flow quantities, and Gauss linear scheme was used to evaluate the gradients. The pressure-velocity coupling adopted the SIMPLEC (Semi-Implicit Method for Pressure Linked Equations-Consistent) \cite{van1984enhancements} algorithm. The solution fields were iterated until convergence: the residuals leveled out and no discernible change in the solution was observed. 

\subsection{SD7003}
\begin{figure*} 
\centerline{\includegraphics[width=5.0in]{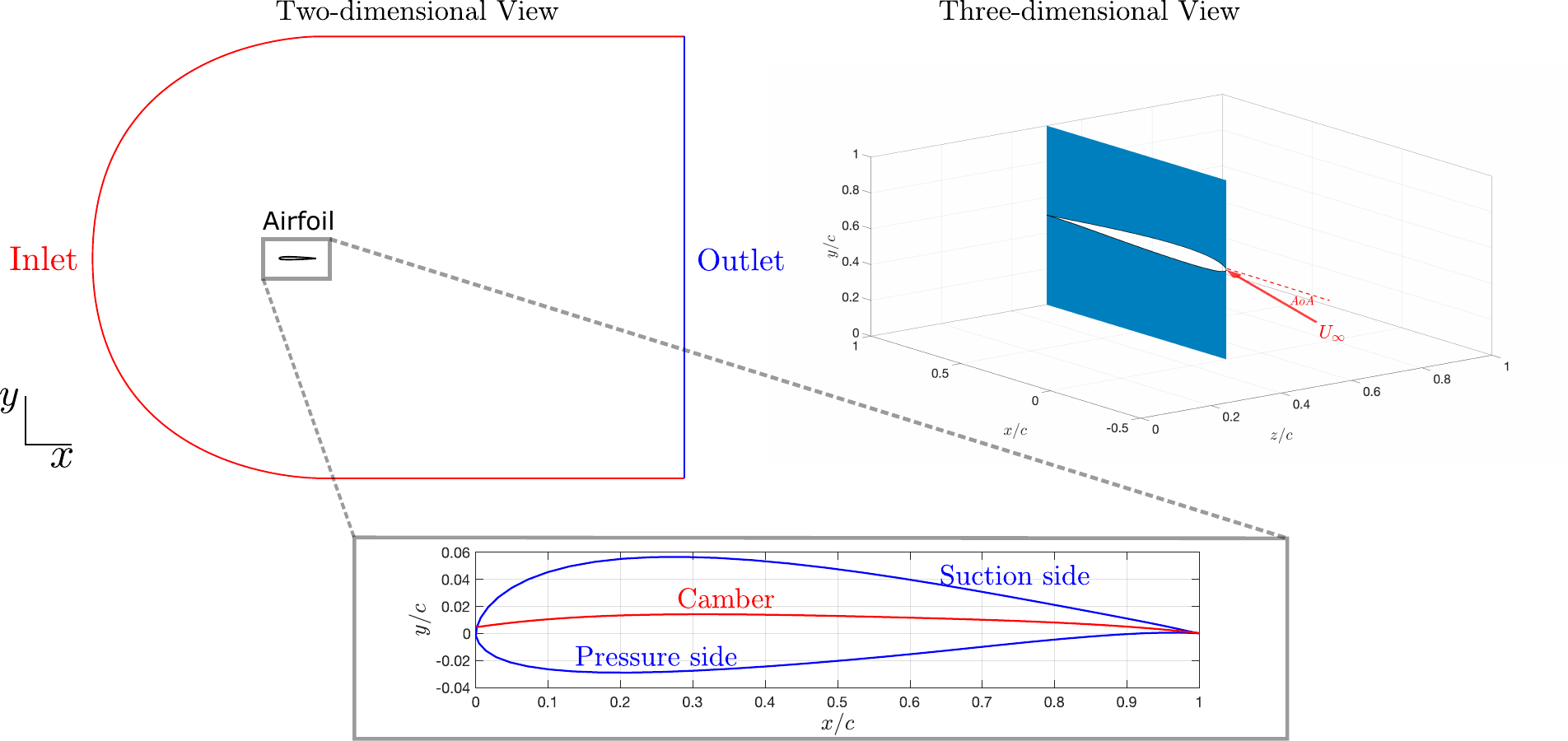}}
  \caption[SD7003 computational domain and boundary conditions: {\color{red} \rule{0.7cm}{0.4mm}} far field, {\color{blue} \rule{0.7cm}{0.4mm}} outflow, {\color{black} \rule{0.7cm}{0.4mm}}, and no-slip walls.]{SD7003 computational domain and boundary conditions: {\color{red} \rule{0.7cm}{0.4mm}} far field, {\color{blue} \rule{0.7cm}{0.4mm}} outflow, {\color{black} \rule{0.7cm}{0.4mm}}, and no-slip walls. Depiction of the suction side, camber, and pressure side of the SD7003 airfoil is displayed in the magnified plot. A three-dimensional version of the computational domain is provided with freestream ($U_{\infty}$) encountering the leading edge at $8^{\circ}$ AoA.}
\label{fig:no_Mk_SD7003domain.pdf}
\end{figure*}

Second flow being considered is around an SD7003 airfoil, as shown in Fig. \ref{fig:no_Mk_SD7003domain.pdf}. At the low Reynolds number based on chord length of $\operatorname{Re}_{c} = 60000$, a laminar separation bubble (LSB) is formed on the suction side of the airfoil. Note that the bubble moves upstream as the angle of attack (AoA) increases \cite{catalano2011rans}. In this study, an $8^{\circ}$ AoA (nearing stall) was considered. Figure \ref{fig:no_Mk_SD7003domain.pdf} schematically shows that the solution domain is a two-dimensional C-topology grid of $389$ (streamwise) $\times$ $280$  (wall-normal) $\times$ $1$ (spanwise) control volumes, which is comparable to the number of control volumes ($768 \times 176$) used in the numerical study of \cite{catalano2011rans}. The magnified view of the two-dimensional SD7003 airfoil labels the camber, suction side and pressure side, as shown in Fig. \ref{fig:no_Mk_SD7003domain.pdf}. The first grid node to the wall was placed at $y^{+} \approx 1.0$ in the turbulent region, in which more than $20$ CVs were placed. The governing Eqs. \ref{p_Continuity} - \ref{p_Momentum} together with the transport equation for the chosen RANS model were solved using the open source software OpenFOAM \cite{weller1998tensorial}. The transport equations were discretized using finite volume method. The scheme was second order upwind for spatial dicretization of flow quantities, and Gauss linear scheme was used to evaluate the gradients. To deal with unsteady flows, PIMPLE algorithm was adopted for pressure-velocity coupling, which is a combination of PISO (Pressure Implicit with Splitting of Operator) \cite{ferziger2002computational} and SIMPLEC \cite{van1984enhancements}. It should be noted that PIMPLE algorithm can deal with large time steps where the maximum Courant (C) number may consistently be above $1$. In this study, the maximum value of C was set consistently equal to $0.6$, and the time step was varied automatically in OpenFOAM to achieve the set maximum. In addition, both residuals and distributions of lift and drag coefficients that vary with respect to time ($T$) were used to track convergence status. The solution fields were iterated until convergence, which required residuals of turbulence kinetic energy and momentum to drop more than four orders of magnitude, and both lift and drag coefficients leveled out with respect to time. This happened at $T \approx 0.3$, which corresponds to a normalized time $T^{*} = T U_{\infty} / c = 6.75$, and similar behavior has been observed by Catalano and Tognaccini \cite{catalano2011rans} in their numerical study for a low-Reynolds number flow over a SD7003 airfoil at $\operatorname{AoA} = 10^{\circ}$. Sampling began at $T = 0.6$ (double the time of convergence) and ended at $T = 1.4$, which required approximately $35000$ iterations for all simulations.   

The fluid was assumed to be air, with freestream turbulence intensity of $\operatorname{Tu} = 0.03\%$ and kinematic viscosity of $\nu = 1.5 \times 10^{-5} \mathrm{~m}^{2} / \mathrm{s}$. Ideally, the value of $\operatorname{Tu}$ should be close to zero. 
From Fig. \ref{fig:no_Mk_SD7003domain.pdf} at the inlet of the domain, the freestream velocity was set equal to $4.5 \ m/s$, which corresponds to $\operatorname{Re}_{c} = 60000$. The chord length was set equal to $c = 0.2 \ m$. At the outlet, a zero-gradient boundary condition was implemented for $\left\langle U_{i}\right\rangle$ ($\left\langle U \right\rangle$ for $x$ direction, $\left\langle V \right\rangle$ for $y$ direction), $k$, $\omega$ and pressure. At the wall, a no-slip boundary condition was used.

\begin{table}
\begin{center}

\caption{Number of computational cells.}

\label{table:Mesh_sensitivity}

\begin{ruledtabular}
\begin{tabular}{c c c c}
{} & Coarse  & Medium &Fine\\
\hline
Number of cells & $93$k &$110$k &$148$k\\

\end{tabular}
\end{ruledtabular}
\end{center}
\end{table}

A grid convergence study of the SST $k-\omega$ LM simulations has been performed to test the influence of the grid resolution on the results. Simulations with three different levels of mesh resolution are summarized in Table \ref{table:Mesh_sensitivity}. The refinement was concentrated in the vicinity of the wall where high spatial gradients are present. Negligible difference in the predictions for the skin friction coefficient $C_{f} = \tau_{w} / {0.5 \rho U_{\infty}^{2}}$, where $\tau_{w}$ is the wall shear stress, and the pressure coefficient $C_{p}=(p-p_{\infty}) / {0.5 \rho U_{\infty}^{2}}$, where $p$ is the static pressure and $p_{\infty}$ is the static pressure in the freestream, were observed among the coarse, medium, and fine meshes. Moreover, only a slight difference in the results based on the coarse mesh and the other two meshes (medium and fine) was observed, i.e., the results based on the medium and fine mesh almost collapsing onto a single curve. In addition, the mean flow ($\left\langle U \right \rangle/U_{\infty}$) and turbulence quantities ($-\left\langle u_{1}u_{2} \right \rangle/U_{\infty}^2$) exhibit a slight difference between the coarse mesh and the other two levels (medium and fine), for which the difference was at almost $1\%$. Therefore, a mesh with $768 \times 176$ (medium) has been considered sufficiently accurate and used for the simulations hereafter. 

\section{Results and discussion}
\subsection{Transitional flow over a smooth flat plate with zero pressure gradient}
In this section, the eigenspace perturbation framework is tested on three different RANS models: SST $k-\omega$ LM \cite{menter2004correlation,menter2006correlation,langtry2009correlation}, SST $k-\omega$ \cite{menter1993zonal,hellsten1998some,menter2001elements,menter2003ten}, and $k - \varepsilon$ \cite{el1983k,launder1983numerical}. The baseline predictions are first presented, followed by eigenvalue perturbations to the anisotropy tensor. 

\subsubsection{Skin friction coefficient}

\begin{figure} 
\centerline{\includegraphics[width=3.5in]{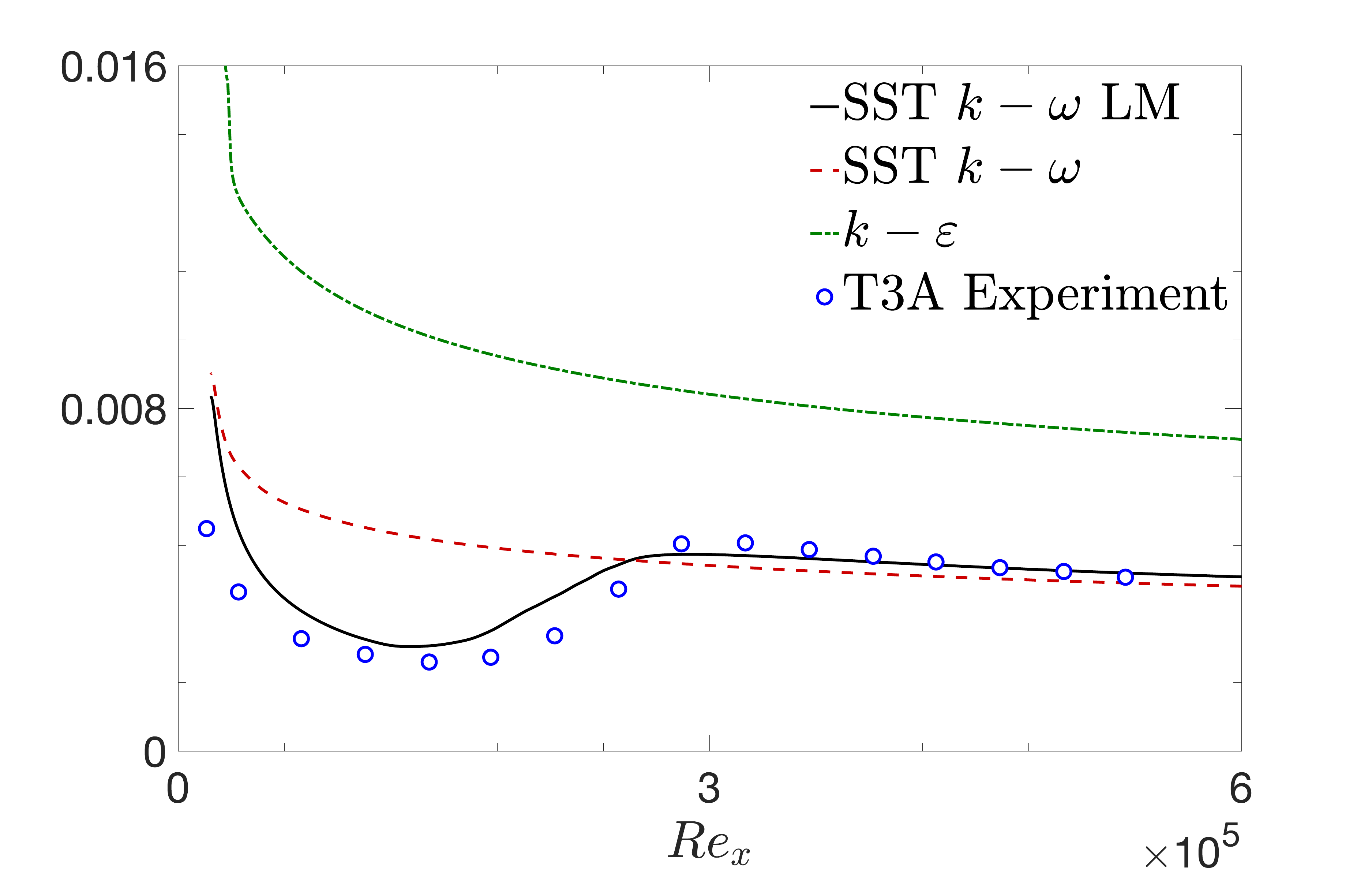}}
\caption{Prediction for skin friction coefficient distribution over a flat plat.}
\label{fig:T3A_cf_baselines.pdf}
\end{figure}

Perhaps the local wall shear stress is the most important parameter for a boundary layer, which in dimensionless form becomes $C_{f}$. In Fig. \ref{fig:T3A_cf_baselines.pdf}, the baseline SST $k-\omega$ LM prediction for $C_{f}$ is plotted with respect to the Reynolds number based on local distance from the leading edge $\operatorname{Re}_{x}$, along with other two popular RANS turbulence models SST \cite{menter1993zonal,hellsten1998some,menter2001elements,menter2003ten} and $k - \varepsilon$ \cite{el1983k,launder1983numerical} for reference. No slip boundary condition was assumed at the solid wall surface for all three RANS models, which are integrated down to the wall. The ERCOFTAC experimental data of \cite{roach1990influence} is included for comparison. Figure \ref{fig:T3A_cf_baselines.pdf} clearly shows a ``trough'' in the experimental data, which marks the laminar-turbulent transition process. The predicted $C_{f}$ profile of SST $k-\omega$ LM \cite{menter2004correlation,menter2006correlation,langtry2009correlation} lies somewhat above the experimental data of \cite{roach1990influence} in the transitional region, while lies slightly below the experimental data as the flow moves further downstream in the fully turbulent region, but overall shows good agreement with the dataset across the entire flat plate. On the other hand, the predicted $C_{f}$ profile of SST $k-\omega$  \cite{menter1993zonal,hellsten1998some,menter2001elements,menter2003ten} shows good agreement with the ERCOFTAC data of \cite{roach1990influence} in the fully turbulent region downstream of the trough, but fails to capture the behavior of laminar-turbulent transition. However, the $k - \varepsilon$ model \cite{el1983k,launder1983numerical} significantly over-predicts the value of $C_{f}$ across the entire flat plate compared to the ERCOFTAC data of \cite{roach1990influence}, as shown in Fig. \ref{fig:T3A_cf_baselines.pdf}.

\subsubsection{Mean flow field}
\begin{figure*} 
\centerline{\includegraphics[width=6.5in]{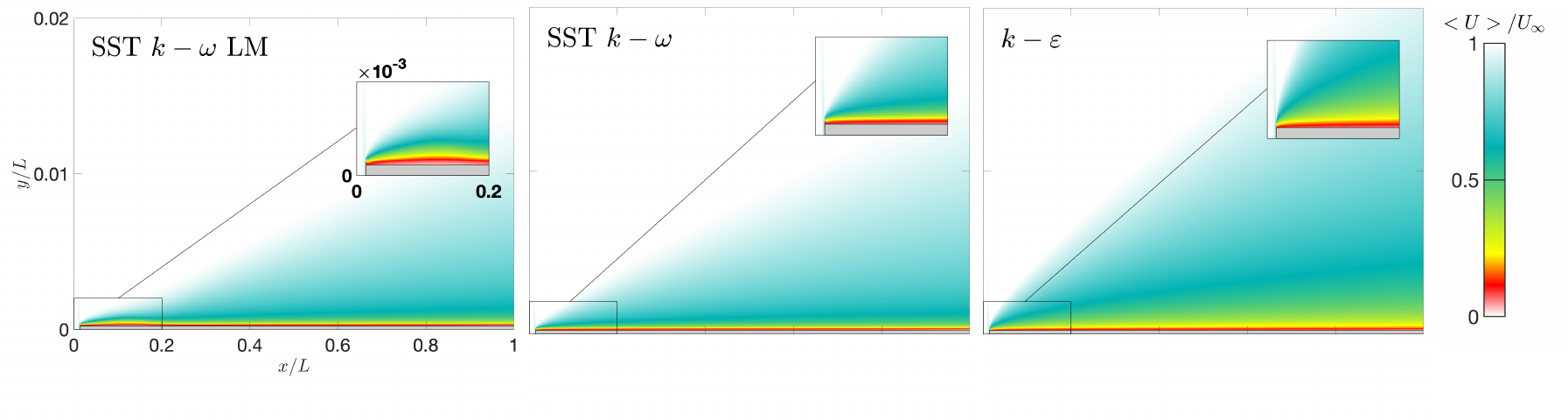}}
\caption[Contours of streamwise mean velocity over a flat plate.]{Contours of streamwise mean velocity over a flat plate. The region close to the leading edge is enlarged for $0 < x/c < 0.2$.}
\label{fig:UQT3A_contour.pdf}
\end{figure*}

The baseline predictions for contours of the mean velocity normalized with the free stream velocity, $\left\langle U \right \rangle/U_{\infty}$ in an $xy$ plane are presented in Fig. \ref{fig:UQT3A_contour.pdf}. Figure \ref{fig:UQT3A_contour.pdf} magnifies the region close to the leading edge for $0 < x/L < 0.2$, which corresponds to $0 < \operatorname{Re}_{x} < 2.16 \times 10^{5}$ (at the trough). In this region, the SST $k-\omega$ LM transition model \cite{menter2004correlation,menter2006correlation,langtry2009correlation} shows a ``bump'' next to the wall, giving a lower magnitude of $\left\langle U \right \rangle/U_{\infty}$ compared to the SST turbulence model \cite{menter1993zonal,hellsten1998some,menter2001elements,menter2003ten}, although a lower value of $C_{f}$ is found at the trough shown in Fig. \ref{fig:T3A_cf_baselines.pdf}. As the bump alters the effective shape of the geometry, we conjecture that an additional ``form-induced drag'' might be generated, which might explain the reduction in the magnitude of $\left\langle U \right \rangle/U_{\infty}$. However, it is clear that no discernible bump around the leading edge is observed for both the SST  \cite{menter1993zonal,hellsten1998some,menter2001elements,menter2003ten} and $k - \varepsilon$ turbulence models \cite{el1983k,launder1983numerical}, which indicates that both models fail to capture the laminar-turbulent transition process. This confirms the behavior shown in Fig. \ref{fig:T3A_cf_baselines.pdf}. In addition, the $k - \varepsilon$ turbulence model overall gives a much smaller value of $\left\langle U \right \rangle/U_{\infty}$ than the other two models, which confirms the significantly increased value of $C_{f}$ along the entire flat plate, as shown in Fig. \ref{fig:T3A_cf_baselines.pdf}.

\begin{figure} 
\centerline{\includegraphics[width=4.0in]{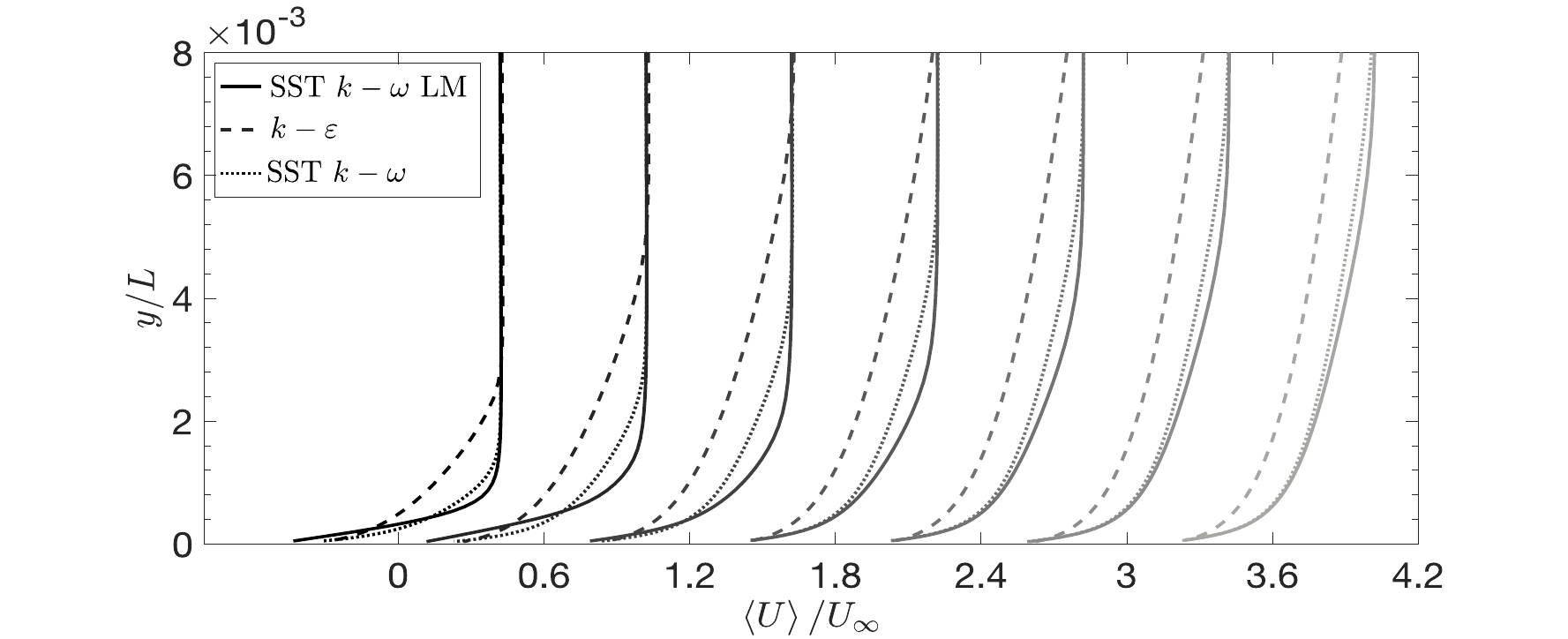}}
\caption[Contours of streamwise mean velocity over a flat plate.]{Profile of normalized mean velocity using three different models. From left to right are profiles at $x = 0.2$, $0.4$, $0.6$, $0.8$, $1.0$, $1.2$, $1.4$ ($x/c = 0.067$, $0.13$, $0.20$, $0.27$, $0.33$, $0.40$, and $0.47$).}
\label{fig:T3A_U_sstlm_ke_sst.pdf}
\end{figure}

The baseline predictions for the $\left\langle U \right \rangle/U_{\infty}$ profiles at different locations using the three different models are plotted in Fig. \ref{fig:T3A_U_sstlm_ke_sst.pdf}. From Fig. \ref{fig:T3A_U_sstlm_ke_sst.pdf}, it is clear that the SST $k-\omega$ LM transition model gives the velocity profiles at $x = 0.2$, $0.4$ and $0.6$ ($x/c = 0.067$, $0.13$ and $0.2$) lagging behind the ones produced by the SST $k-\omega$ turbulence model. This confirms the behavior observed in Fig. \ref{fig:UQT3A_contour.pdf}, indicating that the effect of wall shear stress of the SST $k-\omega$ LM model is enhanced in the vicinity of the wall, i.e., the greatest reduction in momentum by the SST $k-\omega$ LM model. The decrease in momentum propagates higher into the boundary layer as the flow proceeds downstream of the leading edge. In the outer region, all of the velocity profiles recover to the freestream value. In addition, the velocity profile produced by the SST $k-\omega$ LM model begins to lead ahead to the velocity profiles produced by the SST $k-\omega$ and $k - \varepsilon$ models. This reflects that the effect of wall shear stress weakens more quickly for the SST $k-\omega$ LM model than the SST $k-\omega$ and $k - \varepsilon$ models. It is interesting to note that the difference between the profiles produced by the SST $k-\omega$ and SST $k-\omega$ LM models becomes smaller as the flow moves further downstream in the turbulent boundary layer. This confirms the behavior shown in Fig. \ref{fig:T3A_cf_baselines.pdf}, i.e., difference in $C_{f}$ is rather small downstream of the trough. This is due to the fact that in the turbulent region only are the SST $k-\omega$ model formulations triggered in the SST $k-\omega$ LM model \cite{langtry2009correlation}. On the other hand, the velocity profile produced by the $k - \varepsilon$ model overall lags behind the other two models, indicating the effect of wall shear is rather significant throughout the entire boundary layer, which confirms the behavior shown in Fig. \ref{fig:T3A_cf_baselines.pdf}.

\subsubsection{Wall shear stress: uniform $\Delta_{B}$}

\begin{figure} 
\centerline{\includegraphics[width=3.8in]{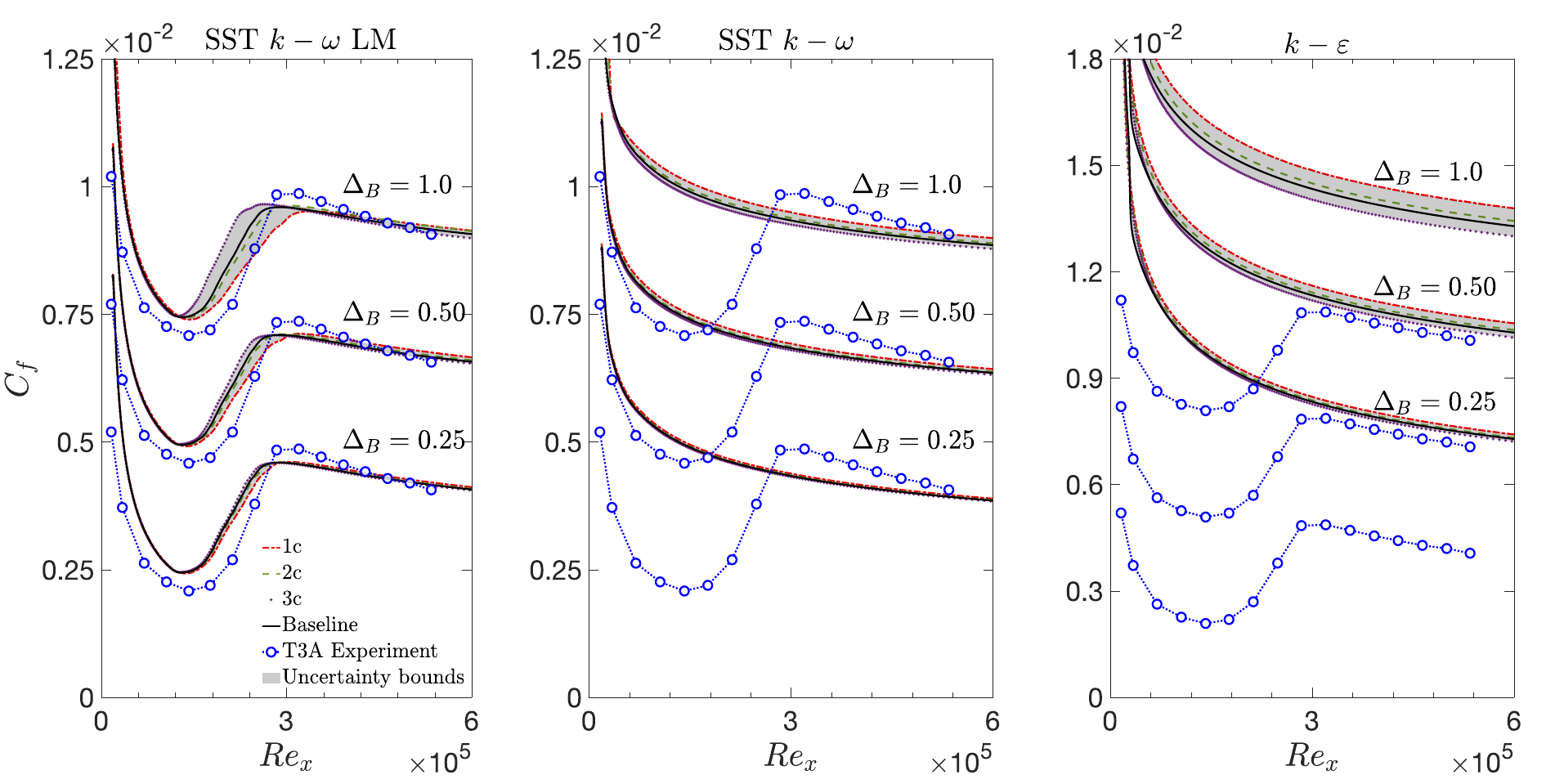}}
\caption[Distribution of skin friction coefficient over a flat plat. Uncertainty bounds (gray envelope) with different magnitudes of $\Delta_{B}$.]{Distribution of skin friction coefficient over a flat plat. Uncertainty bounds (gray envelope) with different magnitudes of $\Delta_{B}$. The baseline prediction (solid line) is provided for reference. Each profile is shifted vertically by a same distance for each model.}
\label{fig:cf_UQT3A_SSTLM_SST_kEpsilon.pdf}
\end{figure}

The eigenspace perturbation framework is tested on three different RANS models: SST $k-\omega$ LM \cite{menter2004correlation,menter2006correlation,langtry2009correlation}, SST $k-\omega$ \cite{menter1993zonal,hellsten1998some,menter2001elements,menter2003ten}, and $k - \varepsilon$ \cite{el1983k,launder1983numerical}.
Figure \ref{fig:cf_UQT3A_SSTLM_SST_kEpsilon.pdf} shows the predictions for $C_{f}$ as a function of $\operatorname{Re}_{x}$. Also included is the ERCOFTAC experimental data of \cite{roach1990influence} for comparison. In Fig. \ref{fig:cf_UQT3A_SSTLM_SST_kEpsilon.pdf}, an ``enveloping behavior'' with respect to the baseline prediction is observed, and this behavior has been observed by other researchers as well, e.g., see \cite{emory2013modeling,mishra2017rans,gorle2019epistemic,iaccarino2017eigenspace}. In Fig. \ref{fig:cf_UQT3A_SSTLM_SST_kEpsilon.pdf}, the simulation's response to different values of $\Delta_{B}$ varies with which model is being used. For SST $k-\omega$ LM, there are two important observations: first, the uncertainty bound is almost concentrated in the transitional region where the $C_{f}$ profile begins to recover toward the fully turbulent profile. Second, in the recovery region the $1c$ and $2c$ perturbations under-predict the baseline prediction ($\downarrow$ $C_{f}$), while the $3c$ perturbation does the opposite ($\uparrow$ $C_{f}$); however, in the fully turbulent region it is the complete opposite of the behavior observed in the recovery region. The $1c$ perturbation shows an tendency to encompass the reference data, which reflects that the increased streamwise stresses contribute to a reduced $C_{f}$ value in the recovery region. The $3c$ perturbation does the opposite, reflecting isotropic stresses tend to increase the value of $C_{f}$ in the recovery region. It is clear that the eigenvalue perturbations are not sufficient to encompass all the reference data. This might be due to the exclusion of the amplitude and eigenvector perturbation of the Reynolds stress tensor, and the parametric uncertainty introduced in the model coefficients. In addition, the response to $\Delta_{B}$ appears to be linear, i.e., the $\Delta_{B} = 1$ envelope is twice the envelope of $\Delta_{B} = 0.50$, which also is twice the envelope of $\Delta_{B} = 0.25$. With increasing $\Delta_{B}$, the $1c$ and $2c$ perturbations from SST $k-\omega$ LM tend to encompass the experimental data, while the $3c$ perturbation deviates from the experimental data. For SST $k-\omega$, the response to $\Delta_{B}$ is relatively small, and the uncertainty bound tends to increase linearly with $\Delta_{B}$, with $1c$ and $2c$ profiles sitting above the baseline prediction while the $3c$ profile sitting slightly below the baseline prediction. In addition, more experimental data in the turbulent region are encompassed when the value of $\Delta_{B}$ is increased. On the other hand, the $k - \varepsilon$ model's response to $\Delta_{B}$ is larger than that for SST $k-\omega$, and again appears to be linear with $\Delta_{B}$; however, it is clear that the $k - \varepsilon$ model overall gives much larger predicted value of $C_{f}$ than the experimental data; consequently, no experimental data are encompassed by the uncertainty bound generated from the $k - \varepsilon$ model. It should be noted that thus far most UQ studies have focused on turbulent flow simulations, from which the $1c$ and $2c$ perturbations always increase the value of $C_{f}$, while the opposite is true for the $3c$ perturbation, e.g., see \cite{mishra2017rans,alonso2017scalable,hornshoj2021quantifying}, which is consistent with the behavior observed with the SST $k-\omega$ and $k - \varepsilon$ turbulence models, as well as the behavior observed in the fully turbulent region with the SST $k-\omega$ LM transition model. 

\subsubsection{Turbulence intensity: $1c$, $2c$, and $3c$ ($\Delta_{B} = 1$)}
\begin{figure} 
\centerline{\includegraphics[width=2.5in]{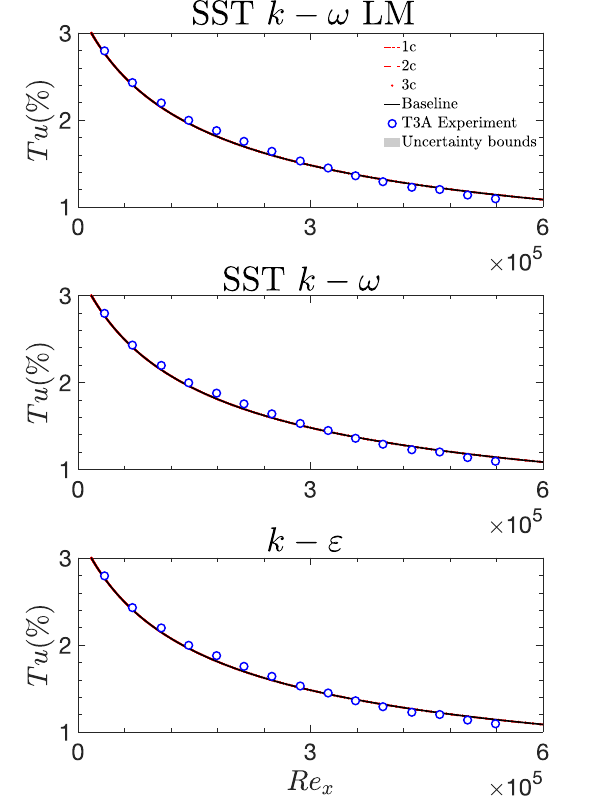}}
\caption[Freestream turbulence intensity decay for T3A. Uncertainty bounds (gray region) for $1c$, $2c$, and $3c$ perturbations (uniform $\Delta_{B} = 1$) are displayed.]{Freestream turbulence intensity decay for T3A. Uncertainty bounds (gray region) for $1c$, $2c$, and $3c$ perturbations (uniform $\Delta_{B} = 1$) are displayed. The baseline prediction (solid line) is provided for reference.}
\label{fig:Tu_deltaB1_all.pdf}
\end{figure}

Figure \ref{fig:Tu_deltaB1_all.pdf} shows decaying of $\operatorname{Tu}$ in freestream as a function of $\operatorname{Re}_{x}$. Also included is the  ERCOFTAC experimental data of \cite{roach1990influence}. Figure \ref{fig:Tu_deltaB1_all.pdf} clearly shows that the baseline predictions for the $\operatorname{Tu}$ profiles using SST $k-\omega$ LM \cite{menter2004correlation,menter2006correlation,langtry2009correlation}, SST $k-\omega$ \cite{menter1993zonal,hellsten1998some,menter2001elements,menter2003ten}, and $k - \varepsilon$ \cite{el1983k,launder1983numerical} are almost indistinguishable from each other, indicating a type of similarity. This indicates that the effect of these different linear eddy viscosity models is restricted to the near-wall region, and the flow becomes insensitive to which model is being used in the freestream region far from the wall. In addition, these three models show good agreement with the experimental data, and show little response to the eigenvalue perturbations ($1c$, $2c$, $3c$), i.e., negligible uncertainty bounds. This indicates a low level of model form uncertainty in the baseline predictions for $\operatorname{Tu}$ decaying in freestream.

\subsubsection{Turbulence kinetic energy: $1c$, $2c$, and $3c$ ($\Delta_{B} = 1$)}

\begin{figure} 
\centerline{\includegraphics[width=3.5in]{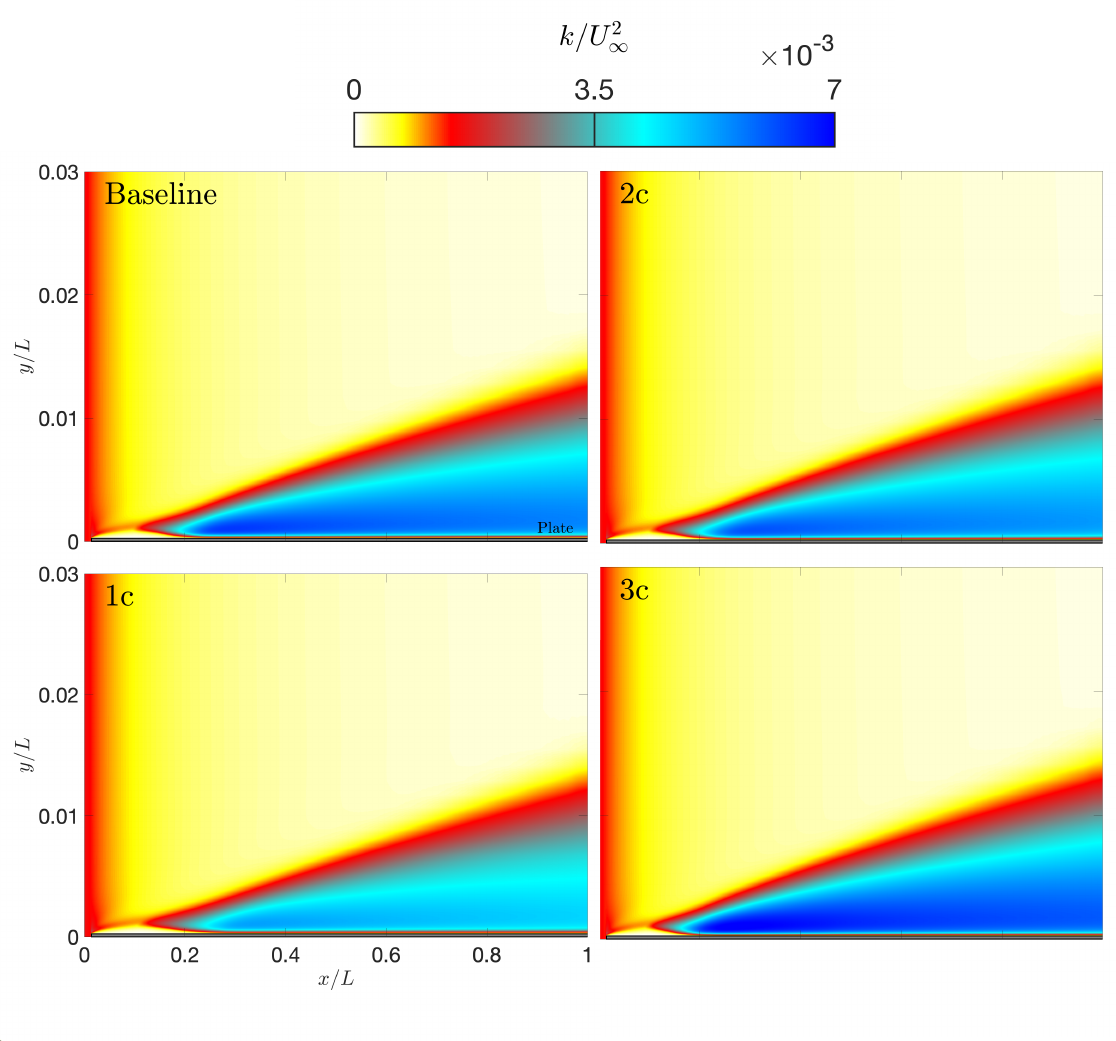}}
\caption{Contours of turbulence kinetic energy for $1c$, $2c$, and $3c$ perturbations using the SST $k-\omega$ LM model over a flat plate. The baseline prediction is provided for reference.}
\label{fig:SSTLM_UQT3A_contour_k.pdf}
\end{figure}

Figure \ref{fig:SSTLM_UQT3A_contour_k.pdf} shows contours of the turbulence kinetic energy normalized with the freestream velocity squared, $k/U_{\infty}^2$ from the baseline, eigenvalue perturbations ($1c$, $2c$, and $3c$) in an $xy$ plane. It is clear that a ``bump'' again appears close to the leading edge, which is similar to the behavior for SST $k-\omega$ LM shown in Fig. \ref{fig:UQT3A_contour.pdf}. The laminar-turbulent transition happens in the bump, where it tends to induce a lower value of $k/U_{\infty}^2$. Compared to the baseline prediction, an overall reduction in the magnitude of $k/U_{\infty}^2$ is observed for the $1c$ and $2c$ perturbations, while the $3c$ perturbation does the opposite.  In addition, the $1c$ perturbation increases the bump length more than the $2c$ perturbation does (bolstering the laminar-turbulent transition). On the other hand, the $3c$ perturbation shortens the bump length (suppressing the laminar-turbulent transition). We note that the $1c$ and $2c$ perturbations overall reduced the velocity magnitude during the lamianr-turbulent transition (in the bump), while the $3c$ perturbation increased the velocity magnitude (perturbed velocity contours are omitted for brevity). Given this, it is interesting to conclude that the size of the transitional region is in a sense inversely related to the mean velocity magnitude under eigenvalue perturbations, e.g., $\uparrow$ transitional region  and $\downarrow$ $\left\langle U \right \rangle$. This suggests that lower local fresstream velocity yields less shear stress, which in turn slows down the progression of transition. 

\subsection{Transitional flow over a SD7003 airfoil}
In this section, the eigenspace perturbation framework is used to introduce uncertainty in the SST $k-\omega$ LM model \cite{menter2004correlation,menter2006correlation,langtry2009correlation}, and the eigenvalue perturbations with $\Delta_{B} = 1$ ($1c$, $2c$, $3c$) are conducted. To begin with, two different values of $\operatorname{Tu} = 0.03\%$ (very low freestream turbulence intensity) and $\operatorname{Tu} = 0.5\%$ (low freestream turbulence intensity) are used to test the sensitivity to the framework. Note that special focus is given to the eigenvalue perturbations for $\operatorname{Tu} = 0.03\%$.

\subsubsection{Crucial transition parameters}

\begin{figure*} 
\centerline{\includegraphics[width=5.0in]{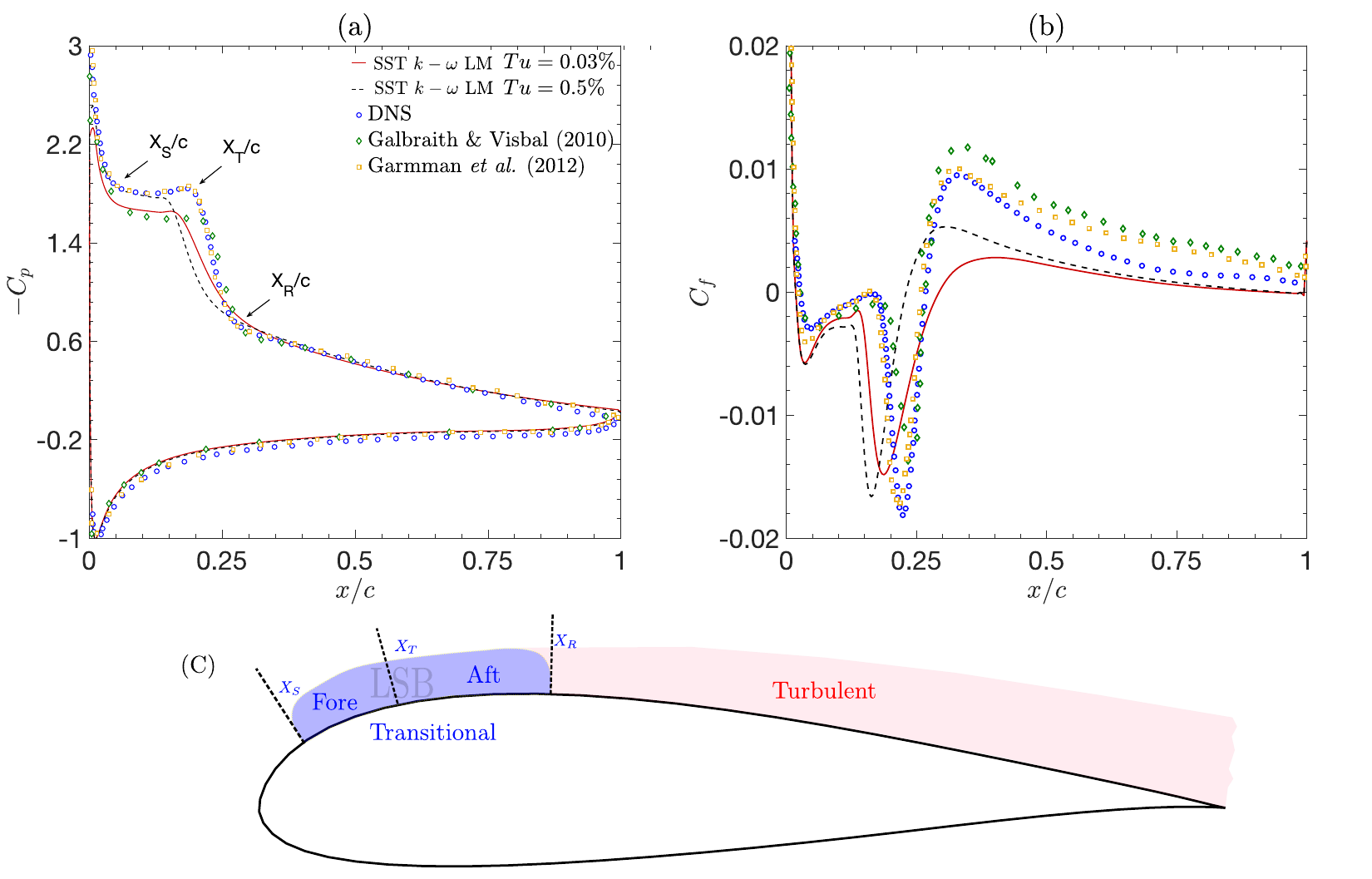}}
\caption{Distribution of (a) pressure coefficient and (b) skin friction coefficient over a SD7003 airfoil at $\operatorname{Re}_{c}=6 \times 10^{4}$ and $\operatorname{AoA} = 8^{\circ}$ for $\operatorname{Tu} = 0.03\%$ and $\operatorname{Tu} = 0.5\%$. (c) Schematic of transitional and turbulent regions over the SD7003 airfoil with important transitional parameters highlighted.}
\label{fig:cfcp_two_Tu0027Tu05.pdf}
\end{figure*}

Figures \ref{fig:cfcp_two_Tu0027Tu05.pdf} (a) and (b) show the baseline predictions for the distributions of $C_{f}$ and $C_{p}$. Also included are the in-house DNS \cite{zhang2021turbulent} and Implicit LES (ILES)/LES data of \cite{galbraith2010implicit} and \cite{garmann2013comparative} for comparison. Figure \ref{fig:cfcp_two_Tu0027Tu05.pdf} (c) schematically defines a transitional region and a turbulent region on the upper surface of the airfoil. Figure \ref{fig:cfcp_two_Tu0027Tu05.pdf} (a) shows a region of nearly constant pressure or a ``flat spot'' \cite{gaster1967structure} that marks the presence of the LSB \cite{tani1964low,o1987laminar}. The separation, transition and reattachment points are the critical transition parameters that can be approximated with either $C_{p}$ or $C_{f}$ plot. According to the technique described by Boutilier and Yarusevych \cite{boutilier2012parametric}, Fig. \ref{fig:cfcp_two_Tu0027Tu05.pdf} (a) shows three ``kinks'' as representatives of the separation, transition and reattachment points, denoted $X_{S}/c$, $X_{T}/c$ and $X_{R}/c$, respectively. From Fig. \ref{fig:cfcp_two_Tu0027Tu05.pdf} (b), finding the zeros of the skin friction coefficient is a different technique that can be used to approximate the $X_{S}/c$, $X_{T}/c$ and $X_{R}/c$ points as well \cite{de2021model}. Based on the reference data, both techniques showed good agreement with each other, and a summary of these transition parameters is presented in Table \ref{T:transi_parameters}. Overall, the baseline predictions for $\operatorname{Tu} = 0.03\%$ show better agreement with the reference data than $\operatorname{Tu} = 0.5\%$. This is due to the fact that the value of $\operatorname{Tu = 0.03\%}$ more closely matches that for the reference data, i.e., $\operatorname{Tu} \approx 0\%$.  In this study, the LSB is treated to be composed of a ``fore'' (from $X_{S}/c$ to $X_{T}/c$) and an ``aft'' (from $X_{T}/c$ to $X_{R}/c$) portion for the sake of investigation simplicity, followed by a fully turbulent region, as shown in Fig. \ref{fig:cfcp_two_Tu0027Tu05.pdf} (c). 

\begin{table}
\begin{center}
\caption{Comparison of transition parameters.}

\label{T:transi_parameters}

\begin{ruledtabular}

\begin{tabular}{c c c c}
Method & $X_{S}/c$ &$X_{T}/c$ &$X_{R}/c$\\
\hline
$\operatorname{Tu} = 0.03\%$/$\operatorname{Tu}=0.5\%$ \cite{langtry2009correlation} & $0.03$/$0.03$& $0.15/0.13$ &$0.29$/$0.23$ \\

In-house DNS \cite{zhang2021turbulent} & $0.02$ &$0.16$&$0.27$\\

LES \cite{garmann2013comparative}& $0.02$ &$0.16$&$0.27$\\

ILES \cite{galbraith2010implicit}& $0.03$ &$0.18$&$0.27$\\

\end{tabular}
\end{ruledtabular}
\end{center}
\end{table}

In Fig. \ref{fig:cfcp_two_Tu0027Tu05.pdf} (a), the predictions for $C_{p}$ with $\operatorname{Tu} = 0.03 \%$ and $\operatorname{Tu} = 0.5 \%$ show good agreement with the ILES data of \cite{galbraith2010implicit}, as well as with the in-house DNS \cite{zhang2021turbulent} and LES data of \cite{garmann2013comparative} at the flat spot (in the fore portion of the LSB), respectively, followed by a discrepancy in the pressure recovery region in the aft portion of the LSB. After the $X_{R}/c$ point, the $C_{p}$ profile for $\operatorname{Tu} = 0.03\%$ and $\operatorname{Tu} = 0.5\%$ shows a collapse onto a single curve, reflecting good agreement with the reference data. 

In Fig. \ref{fig:cfcp_two_Tu0027Tu05.pdf} (b), both sets of results show a negative ``trough'' of $C_{f}$ around $x/c = 0.2$ for $\operatorname{Tu} = 0.03\%$ and around $x/c = 0.15$ for $\operatorname{Tu} = 0.5 \%$, respectively, followed by the trough is the recovery to positive skin-friction values within the aft portion of the LSB. After the $X_{R}/c$ point, a ``crest'' appears, with $C_{f}$ peaking around the values of $0.003$ and $0.005$ for $\operatorname{Tu} = 0.03\%$ and $\operatorname{Tu} = 0.5\%$, respectively. Note that the predicted $C_{f}$ profiles sit significantly below the reference data at the crest, and a similar behavior has been observed by other researchers in their numerical studies, e.g., see \cite{bernardos2019rans,tousi2021active}. There are two interesting observations: first, a shift of the predicted $C_{f}$ profile in the upstream direction is observed for both $\operatorname{Tu} = 0.03\%$ and $\operatorname{Tu} = 0.5\%$, causing a discrepancy in the aft portion of the LSB, and similar behavior has been observed by other researchers as well, see \cite{catalano2011rans,bernardos2019rans,tousi2021active}; second, increasing freestream turbulence intensity results in earlier transition and reattachment, contributing to a reduction in the LSB length, which is consistent with the observation by Mark \textit{et al.} \cite{istvan2018turbulence} in their experimental study. 

\subsubsection{Sensitivity to $\Delta_{B}$}
\begin{figure} 
\centerline{\includegraphics[width=3.5in]{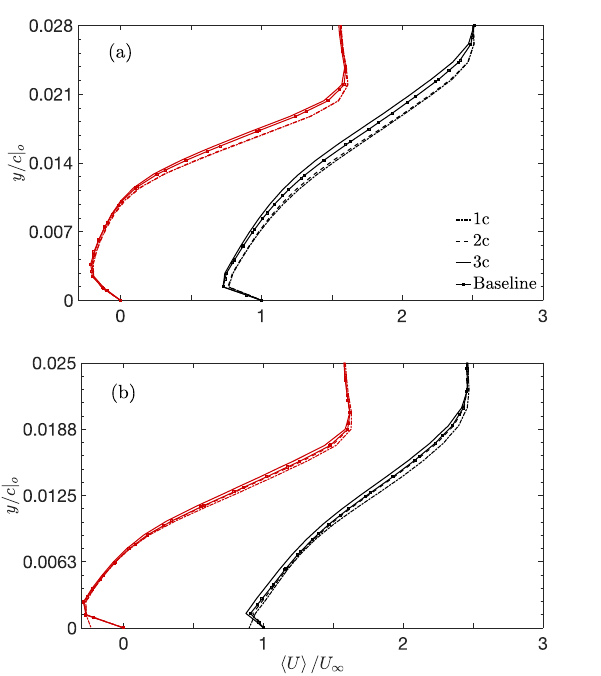}}
\caption[Profile of normalized mean velocity at (a) $\operatorname{Tu} = 0.03\%$ and (b) $\operatorname{Tu} = 0.5\%$ for $y/c|_{o} > 0$.]{Profile of normalized mean velocity at (a) $\operatorname{Tu} = 0.03\%$ and (b) $\operatorname{Tu} = 0.5\%$ for $y/c|_{o} > 0$. For both cases, the baseline (solid line with solid squares) prediction is \textit{enveloped} by the eigenvalue perturbations with $\Delta_{B} = 1.0$. The $3c$ perturbation (solid line) under-predicts the baseline, while both $2c$ (dashed) and $1c$ (dashed-dotted-dashed) perturbations over-predict the baseline. From left to right are the profiles at $x/c = 0.15$ (red) and $x/c = 0.2$ (black).}
\label{fig:Tu0027_Tu05_U_sensitivity_line_x_c015_020.pdf}
\end{figure}

\begin{figure} 
\centerline{\includegraphics[width=3.5in]{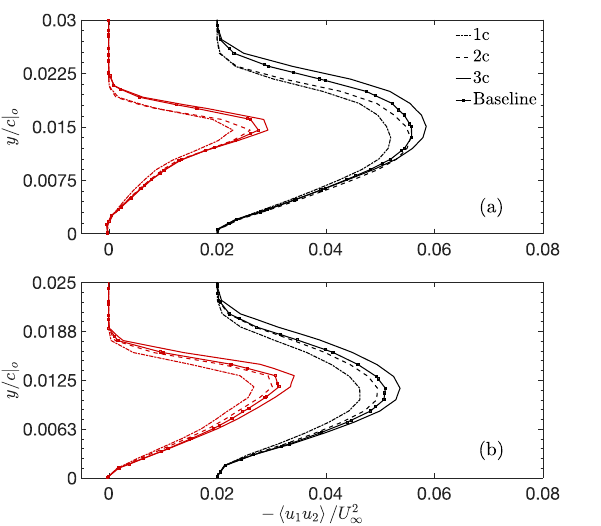}}
\caption[Profile of normalized Reynolds shear stress at (a) $\operatorname{Tu} = 0.03\%$ and (b) $\operatorname{Tu} = 0.5\%$ for $y/c|_{o} > 0$.]{Profile of normalized Reynolds shear stress at (a) $\operatorname{Tu} = 0.03\%$ and (b) $\operatorname{Tu} = 0.5\%$ for $y/c|_{o} > 0$. For both cases, the baseline (solid line with solid squares) prediction is \textit{enveloped} by the eigenvalue perturbations with $\Delta_{B} = 1.0$. The $3c$ perturbation (solid line) under-predicts the baseline, while both $2c$ (dashed) and $1c$ (dashed-dotted-dashed) perturbations over-predict the baseline. From left to right are the profiles at $x/c = 0.15$ (red) and $x/c = 0.2$ (black).}
\label{fig:Tu0027_Tu05_uv_sensitivity_line_x_c015_020.pdf}
\end{figure}

The predicted mean velocity and Reynolds shear stress profiles for eigenvalue perturbations with $\Delta_{B} = 1$ for $\operatorname{Tu} = 0.03\%$ and $\operatorname{Tu} = 0.5\%$ are presented in Figs. \ref{fig:Tu0027_Tu05_U_sensitivity_line_x_c015_020.pdf} (a), (b) and \ref{fig:Tu0027_Tu05_uv_sensitivity_line_x_c015_020.pdf} (a), (b), respectively. Due to the airfoil curved upper surface, both the mean velocity and Reynolds shear stress profiles are shifted down to the origin of $y/c$, denoted $y/c|_{o} = (y-y_{w})/c$ for sake of better contrast, where $y_{w}$ is the vertical location of the upper surface of the airfoil. The focus of this study is on the model form uncertainty in the LSB, especially the aft portion of the LSB where noticeable discrepancies are prevalent (see Fig. \ref{fig:cfcp_two_Tu0027Tu05.pdf}). This is consistent with the analysis of Davide \textit{et al.} \cite{lengani2014pod}, whose focus was also on the aft portion of the LSB where the unstable shear layer was present with shed vortices. Two locations, i.e., $x/c = 0.15$ (around $X_{T}$) and $0.2$ (near $X_{R}$), in the aft portion of the LSB were selected to investigate the effects of the extreme states of componentiality ($1c$, $2c$, $3c$) on both the normalized mean velocity ($\left\langle U \right \rangle/U_{\infty}^2$) profile and normalized Reynolds shear stress ($-\left\langle u_{1}u_{2} \right \rangle/U_{\infty}^2$) profile.
In Figs. \ref{fig:Tu0027_Tu05_U_sensitivity_line_x_c015_020.pdf} (a) and (b), an enveloping behavior with respect to the baseline prediction is observed for both $\operatorname{Tu} = 0.03\%$ and $\operatorname{Tu} = 0.5\%$, in the sense that the $3c$ perturbation reduces the magnitude of mean velocity profile, while $1c$ and $2c$ perturbations do the opposite. In addition, $\operatorname{Tu} = 0.03\%$ yields velocity profile that is less sensitive to the perturbations than $\operatorname{Tu} = 0.5\%$, as shown in Figs. \ref{fig:Tu0027_Tu05_U_sensitivity_line_x_c015_020.pdf} (a) and (b). 

In Figs. \ref{fig:Tu0027_Tu05_uv_sensitivity_line_x_c015_020.pdf} (a) and (b), the enveloping behavior is again observed for $\operatorname{Tu} = 0.03\%$ and $\operatorname{Tu} = 0.5\%$. This time the $3c$ perturbation gives a larger value of $-\left\langle u_{1}u_{2} \right \rangle/U_{\infty}^2$, while the $1c$ and $2c$ perturbations do the opposite. A similar behavior in terms of eigenvalue perturbation to the mean velocity and Reynolds shear stress profiles was observed by Luis \textit{et al.} \cite{cremades2019reynolds} as well.
 In addition, comparable sensitivity level to the eigenvalue perturbation is observed for $\operatorname{Tu} = 0.03\%$ and $\operatorname{Tu} = 0.5\%$.
 
\begin{figure*} 
\centerline{\includegraphics[width=4.5in]{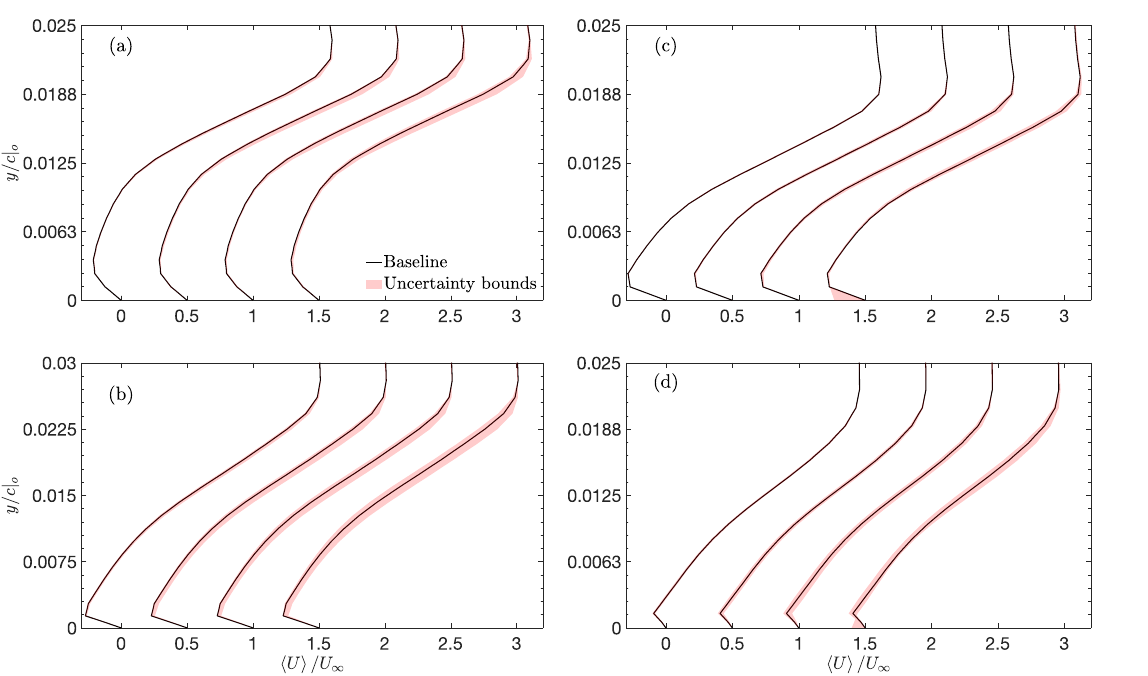}}
\caption[Profile of normalized mean velocity at (a) $x/c = 0.15$ and (b) $x/c = 0.2$ for $\operatorname{Tu} = 0.03\%$; profile of normalized mean velocity at (c) $x/c = 0.15$ and (d) $x/c = 0.2$ for $\operatorname{Tu} = 0.5\%$.]{Profile of normalized mean velocity at (a) $x/c = 0.15$ and (b) $x/c = 0.2$ for $\operatorname{Tu} = 0.03\%$; profile of normalized mean velocity at (c) $x/c = 0.15$ and (d) $x/c = 0.2$ for $\operatorname{Tu} = 0.5\%$. Uncertainty bounds, i.e., red region between $1c$ and $3c$ eigenvalue perturbations, with different magnitudes of $\Delta_{B}$ are displayed. From left to right are $\Delta_{B} = 0.25, 0.5, 0.75,$ and $1.0$ perturbations, respectively. The unperturbed baseline prediction (solid line) is provided for reference.}
\label{fig:UQ_U_x_c015_020_sensitivity_Tu0027_Tu05.pdf}
\end{figure*}

 \begin{figure} 
\centerline{\includegraphics[width=2.5in]{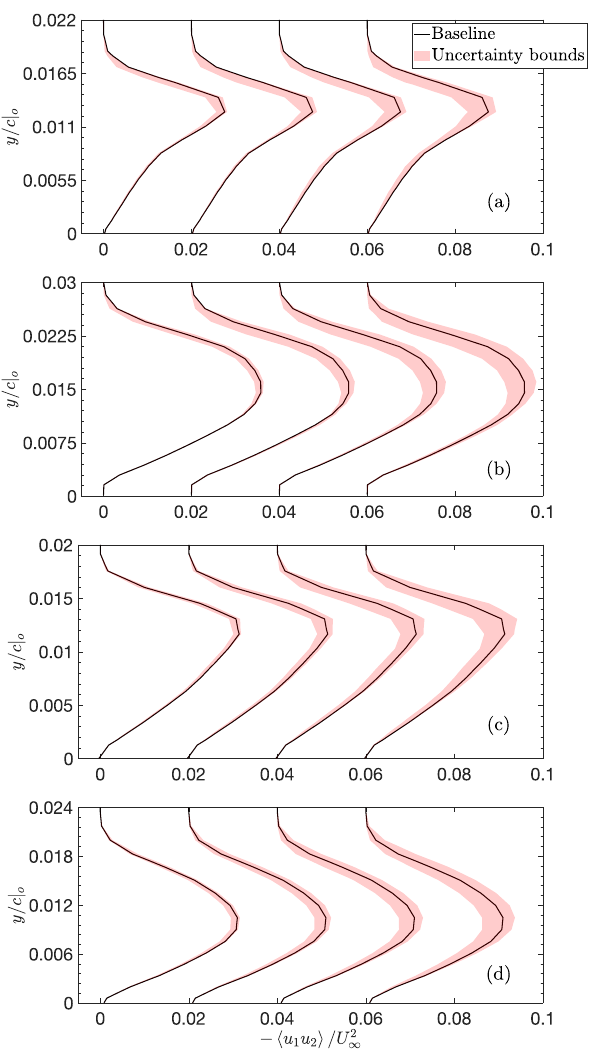}}
\caption[Profile of normalized Reynolds shear stress at (a) $x/c = 0.15$ and (b) $x/c = 0.2$ for $\operatorname{Tu} = 0.03\%$; (c) profile of normalized Reynolds shear stress at (c) $x/c = 0.15$ and (d) $x/c = 0.2$ for $\operatorname{Tu} = 0.5\%$.]{Profile of normalized Reynolds shear stress at (a) $x/c = 0.15$ and (b) $x/c = 0.2$ for $\operatorname{Tu} = 0.03\%$; (c) profile of normalized Reynolds shear stress at (c) $x/c = 0.15$ and (d) $x/c = 0.2$ for $\operatorname{Tu} = 0.5\%$. Uncertainty bounds with different magnitudes of $\Delta_{B}$, i.e., red region between $1c$ and $3c$ profiles, are displayed. From left to right are $\Delta_{B} = 0.25, 0.5, 0.75,$ and $1.0$ perturbations. The unperturbed baseline prediction (solid line) is provided for reference.}
\label{fig:UQ_uv_x_c015_020_sensitivity_Tu0027_Tu05.pdf}
\end{figure}

In Figs. \ref{fig:UQ_U_x_c015_020_sensitivity_Tu0027_Tu05.pdf} (a), (b), (c), (d) and \ref{fig:UQ_uv_x_c015_020_sensitivity_Tu0027_Tu05.pdf} (a), (b), (c), (d), the predicted mean velocity normalized by the freestream velocity, and the Reynolds shear stress normalized by the freestream velocity squared, $\left\langle U \right \rangle/U_{\infty}$ and $-\left\langle u_{1}u_{2}\right\rangle/U_{\infty}^2$ for uncertainty bounds with $\Delta_{B}$ in $0.25$ increments, starting at $0.25$ and increasing up to $1$, for $\operatorname{Tu} = 0.03\%$ and $\operatorname{Tu} = 0.5\%$ and at $x/c = 0.15$ and $x/c = 0.2$, respectively, are presented. It is interesting to note that both mean velocity and Reynolds shear stress show approximately linear responses to these increments in $\Delta_{B}$: a series of linear increases in the size of uncertainty bound with $\Delta_{B}$ is observed. In addition, an enveloping behavior is observed for each value of $\Delta_{B}$. Overall, it shows that the increase in $\Delta_{B}$ leads to more increased $-\left\langle u_{1}u_{2}\right\rangle/U_{\infty}^2$ in magnitude compared to $\left\langle U \right \rangle/U_{\infty}$. From Figs. \ref{fig:UQ_U_x_c015_020_sensitivity_Tu0027_Tu05.pdf} (a), (b), (c) and (d), the sensitivity of these uncertainty bounds to $\Delta_{B}$ for $\operatorname{Tu} = 0.03\%$ is somewhat larger than that for $\operatorname{Tu} = 0.5\%$, which is consistent with the behavior shown in Figs. \ref{fig:Tu0027_Tu05_U_sensitivity_line_x_c015_020.pdf} (a) and (b). Physically, increased value of turbulence intensity suppresses the strength of eigenvalue perturbation to the Reynolds stress tensor, therefore resulting in a smaller size of uncertainty bound. In general, the simulation's response to $\Delta_{B}$ is stronger for $-\left\langle u_{1}u_{2} \right \rangle/U_{\infty}^2$ than that for $\left\langle U \right \rangle/U_{\infty}$. Therefore, it can be concluded that the degree of response to the injection of eigenvalue perturbation depends on which quantities-of-interest (QoIs) are being observed.  

\subsubsection{Visual representation for states of turbulence}
\begin{figure} 
\centerline{\includegraphics[width=3.5in]{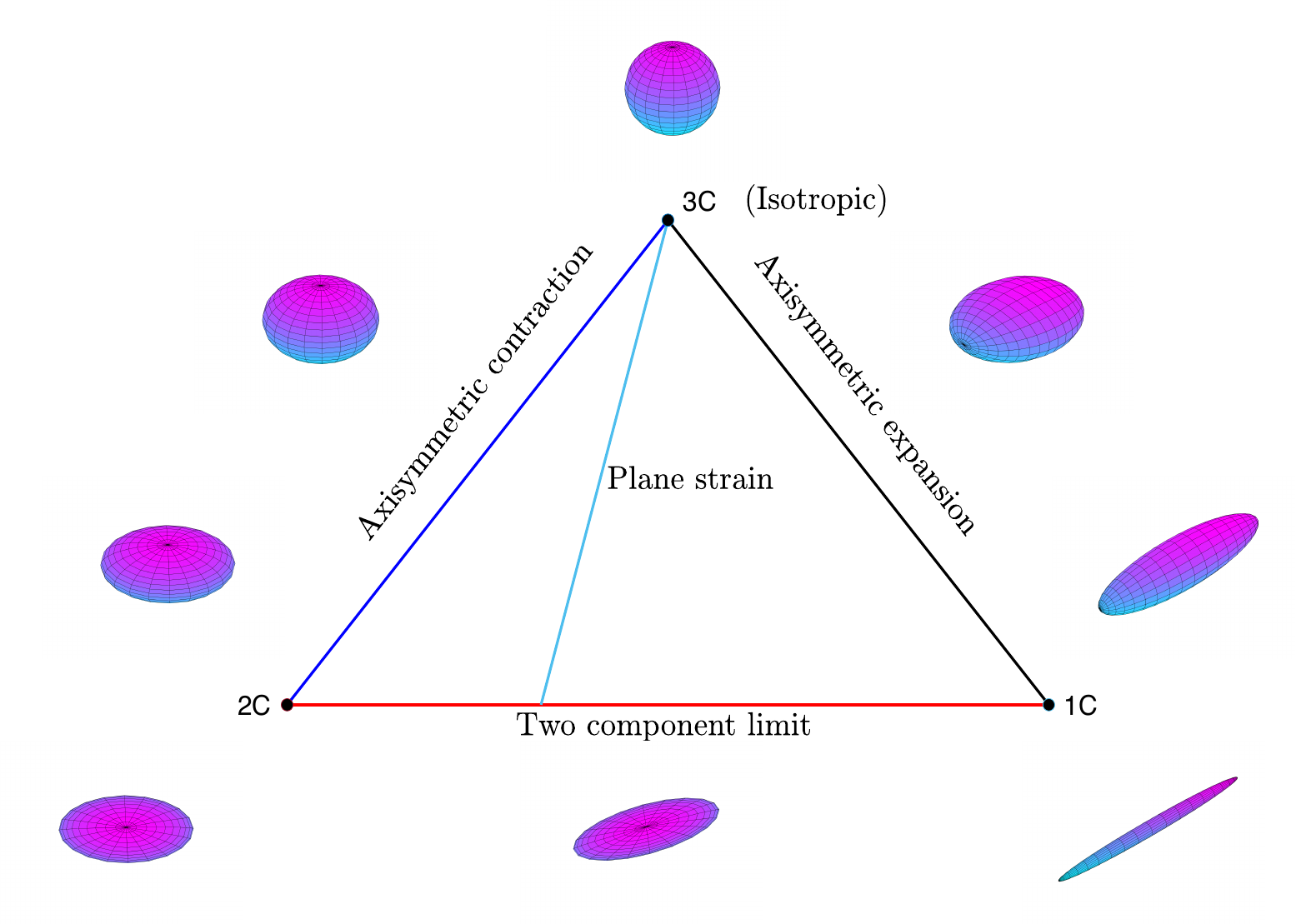}}
\caption[Shapes of limiting anisotropy states in the barycentric map corresponding to varying componentiality of the flow.]{Shapes of limiting anisotropy states in the barycentric map corresponding to varying componentiality of the flow: one-component ($1c$), two-component ($2c$), and three-component ($3c$), which are joint by three limiting boundary states (axisymmetric contraction, axisymmetric expansion, and two-component limit). Note that the axisymmetric contraction and expansion states describe turbulent fluctuations in the shape of oblate and prolate spheroids, respectively. 
}
\label{fig:BaryM_tensorShapes.pdf}
\end{figure}

\begin{figure*} 
\centerline{\includegraphics[width=6.0in]{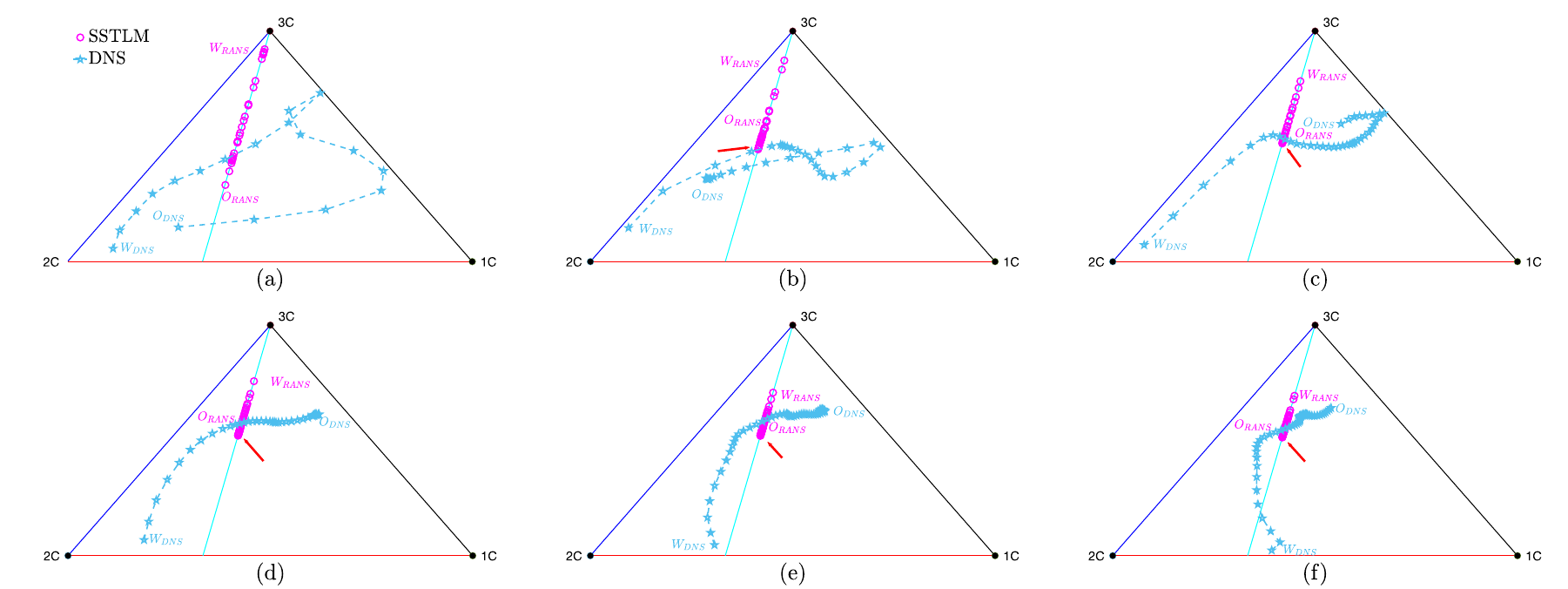}}
\caption[Anisotropy states based on eigenvalues of anisotropy tensor for Reynolds stress on the barycentric map for $\operatorname{Tu} = 0.03\%$. (a) - (c): $x/c = 0.15, 0.2$ and $0.3$. (d) - (e): $x/c = 0.4, 0.5$ and $0.6$.]{Anisotropy states based on eigenvalues of anisotropy tensor for Reynolds stress on the barycentric map \cite{banerjee2007presentation} for $\operatorname{Tu} = 0.03\%$. (a) - (c): $x/c = 0.15, 0.2$ and $0.3$. (d) - (e): $x/c = 0.4, 0.5$ and $0.6$. Anisotropy trajectory follows the path from $W_{\color{magenta}{RANS}/\color{cyan}{DNS}}$ (at the wall) to $O_{\color{magenta}{RANS}/\color{cyan}{DNS}}$ (OBL). Red arrows are added to indicate the point at which the SST $k-\omega$ LM trajectories reverse back in the opposite direction toward $W_{\color{magenta}{RANS}}$ along the plane strain line.}
\label{fig:BMap_six_All.pdf}
\end{figure*}

\begin{figure*} 
\centerline{\includegraphics[width=6.0in]{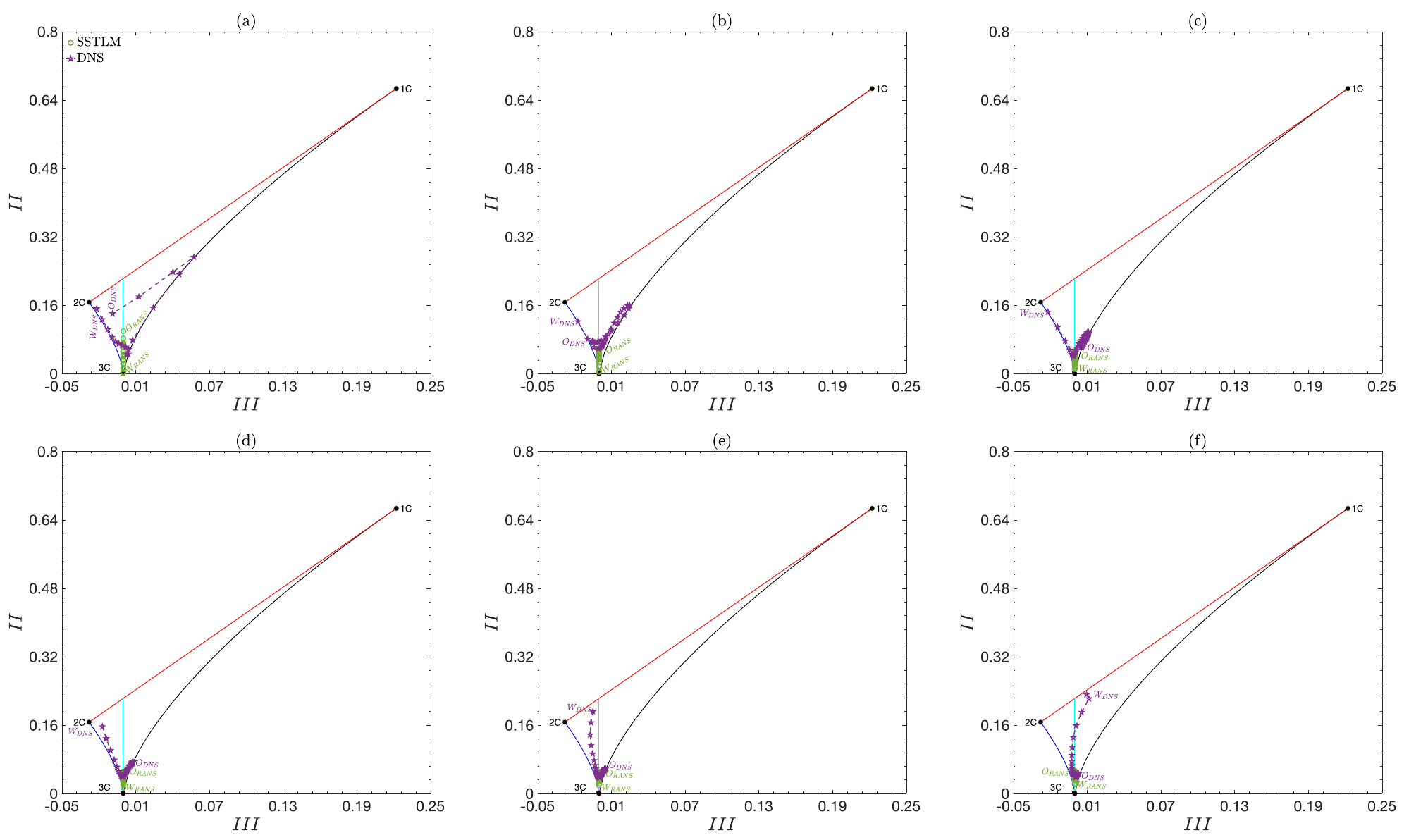}}
\caption[Anisotropy states characterized by invariants ($II$,$III$) of anisotropy tensor for Reynolds stress on the Lumley's invariant map for $\operatorname{Tu} = 0.03\%$. (a) - (c): $x/c = 0.15, 0.2$ and $0.3$. (d) - (e): $x/c = 0.4, 0.5$ and $0.6$.]{Anisotropy states characterized by invariants ($II$,$III$) of anisotropy tensor for Reynolds stress on the Lumley's invariant map \cite{lumley1979computational} for $\operatorname{Tu} = 0.03\%$. (a) - (c): $x/c = 0.15, 0.2$ and $0.3$. (d) - (e): $x/c = 0.4, 0.5$ and $0.6$. Anisotropy trajectory follows the path from $W_{\color{green}{RANS}/\color{violet}{DNS}}$ (at the wall) to $O_{\color{green}{RANS}/\color{violet}{DNS}}$ (OBL).}
\label{fig:AMap_six_All.pdf}
\end{figure*}

Essentially, what eigenvalue perturbations do is to shift a baseline Reynolds stress anisotropy state to a new location on the barycentric map. The Reynolds stress anisotropy itself is an abstract concept, as defined in Eq. \ref{Eq:noMark_AnisotropyTensor}. This concept can be better understood if one considers barycentric map \cite{banerjee2007presentation}. Figures \ref{fig:BMap_six_All.pdf} (a) - (f) show the baseline predictions for the anisotropy states using the SST $k-\omega$ LM model at several locations selected along the suction side of the airfoil. Also included are the in-house DNS data \cite{zhang2021turbulent} for comparison. In addition, Lumley's invariant map \cite{lumley1979computational,pope2001turbulent,durbin2011statistical} as an equivalent of the barycentric map is provided for reference, as shown in Fig. \ref{fig:AMap_six_All.pdf}. In addition, Fig. \ref{fig:BaryM_tensorShapes.pdf} shows the visual representations of turbulent fluctuations for the limiting states on the barycentric map and all those colorful objects are representations of the Reynolds stress ellipsoid: edges of the triangle and the three vertices of componentiality ( $1c$, $2c$, $3c$). Among them, $1c$ (one-component) describes a flow where turbulent fluctuations only exist along one direction, referred to as ``rod-like'' turbulence \cite{emory2014visualizing}; $2c$ (two-component) describes a flow where turbulent fluctuations exist along two directions, referred to as ``pancake-like'' turbulence \cite{emory2014visualizing}; and $3c$ (isotropic) represents turbulence with equal fluctuations along three directions, referred to as ``spherical like'' turbulence \cite{emory2014visualizing}. All interior states on the barycentric map are smooth transitioning of turbulence between these limiting states.  

Figures \ref{fig:BMap_six_All.pdf} (a) - (f) show Boussinesq anisotropy trajectories in a barycentric map from the wall surface ($W_{RANS/DNS}$) to the outer edge of the boundary layer (OBL) ($O_{RANS/DNS}$) at several locations along the suction side of the airfoil. In Figs. \ref{fig:BMap_six_All.pdf} (a) - (f), red arrow is provided to highlight the turning point where Boussinesq anisotropy state is reversing back toward $W_{RANS}$, i.e., the $3c$ vertex. This indicates that the baseline RANS (SST $k-\omega$ LM) predictions for anisotropy state tend to become more isotropic with increasing distance away from the wall. In Figs. \ref{fig:BMap_six_All.pdf} (a) - (f), the Boussinesq anisotropy states are clustered around the plane strain line. The behavior is qualitatively similar to that observed by Edeling \textit{et al.} \cite{edeling2018data} and Simon D \textit{et al.} \cite{hornshoj2021quantifying}. Note that a simulation will follow the plane strain line if at least one eigenvalue of $b_{ij}$ is zero \cite{banerjee2007presentation}, and the Boussinesq anisotropy tensor, see Eq. \ref{Eq:noMark_AnisotropyTensor}, will give a zero eigenvalue when one of $S_{\alpha \alpha}$ equals zero. Therefore, all two-dimensional simulations will yield anisotropy trajectories along the plane strain when the Bousinessq turbulent viscosity hypothesis is adopted, i.e., $\hat{b}_{i j}=\operatorname{diag}\left(\lambda_{1}, 0,-\lambda_{1}\right)$. This behavior was also observed by Edeling \textit{et al.} \cite{edeling2018data} and Simon \textit{et al.} \cite{hornshoj2021quantifying}. Overall, the SST $k-\omega$ LM anisotropy states are more isotropic. 

More interestingly, DNS yields realistic anisotropy states. For the aft portion of the LSB, Fig. \ref{fig:BMap_six_All.pdf} (a) shows that DNS produces turbulence in axisymmetric contraction state at the wall for $x/c = 0.015$, shifting to an axiymmetric expansion state and then reversing back toward the axisymmetric contraction state at the OBL. Refer to Fig. \ref{fig:BaryM_tensorShapes.pdf}, this indicates that turbulence exhibits an oblate spheroid at the wall, then gradually transitions to a prolate spheroid, and eventually exhibits an oblate spheroid at the OBL. A similar behavior is observed at $x/c = 0.2$ as well, as shown in Fig. \ref{fig:BMap_six_All.pdf} (b). This indicates that the laminar-turbulent transition process tends to suppress the streamwise stresses in the regions both close to the wall and in the outer region of the boundary layer, while fosters the streamwise stresses in between. Within the attached turbulent boundary layer from $x/c = 0.3$ to $0.6$, there is a tendency for turbulence at the wall to shift from the axisymmetric contraction state to the two-component state (wall-normal Reynolds stresses damped out at the wall); while away from the wall turbulence gradually shifts toward the axisymmetric expansion, reflecting increased streamwise stresses. Figures \ref{fig:AMap_six_All.pdf} (a) - (f) present anisotropy trajectories in the Lumley's invariant map, which contain the same information as the barycentric map, except based on a non-linear domain (non-linearity in variables $II$ and $III$). Overall, same trend is observed using the Lumley's invariant map; however clusters of anisotropy states appear around the $3c$ vertex.

\begin{figure} 
\centerline{\includegraphics[width=3.5in]{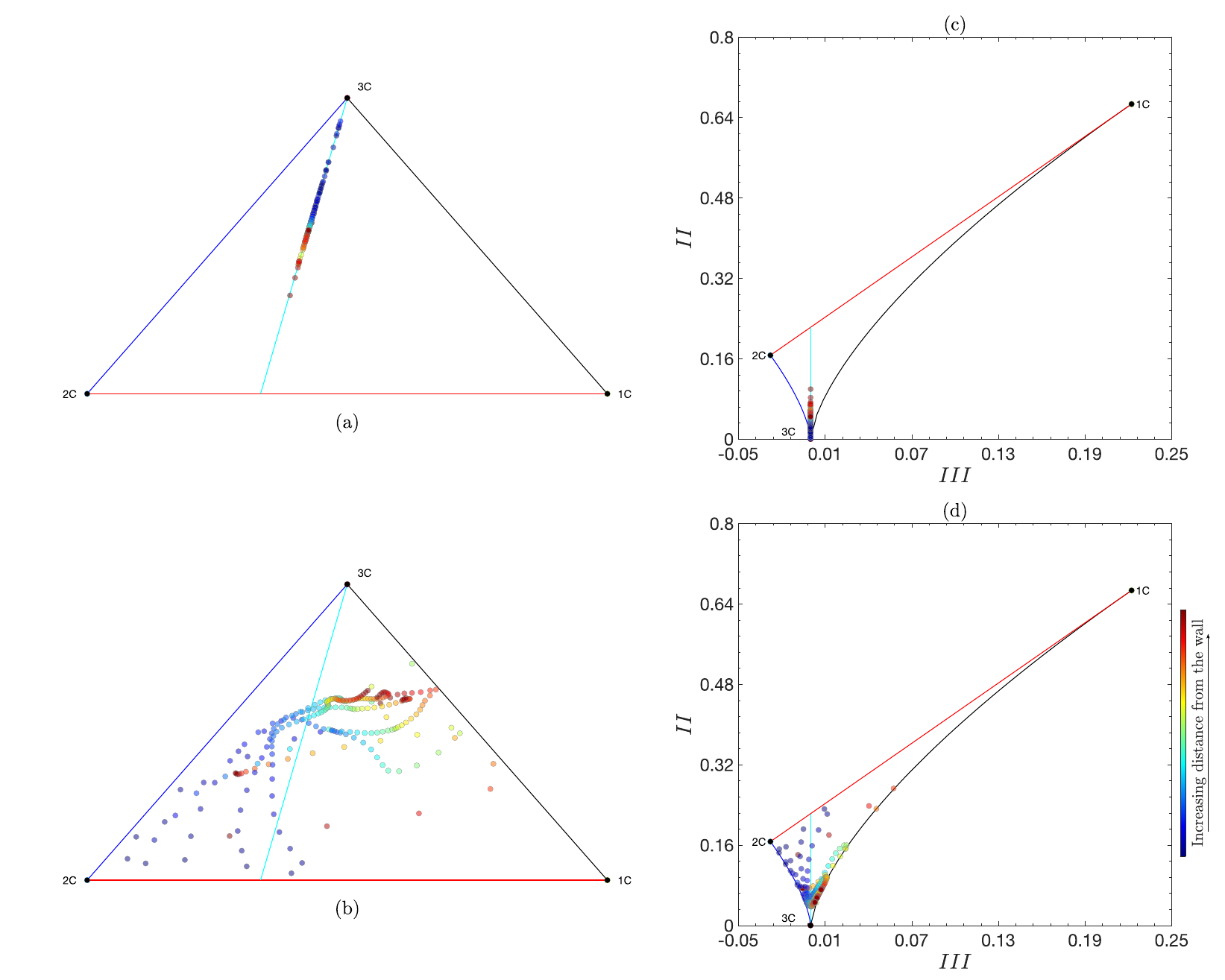}}
\caption{Anisotropy states for $x/c = 0.15, 0.2, 0.3, 0.4, 0.5$ and $0.6$ on both the barycentric and Lumley's invariant maps. (a) and (c): SST $k-\omega$ LM simulation ($\operatorname{Tu} = 0.03\%$); (b) and (d): in-house DNS. Note that anisotropy states are painted based on the distance from the wall surface.}
\label{fig:AIM_BMap_RANSDNS_all.pdf}
\end{figure}

Because anisotropy states are defined based on $\lambda_{i}$, its trajectories do not contain information regarding the physical domain \cite{emory2014visualizing}. This shortcoming is addressed by painting the points based on physical coordinates, as shown in Figs. \ref{fig:AIM_BMap_RANSDNS_all.pdf} (a) - (d). Figures \ref{fig:AIM_BMap_RANSDNS_all.pdf} (a) - (d) present the anisotropy states for all locations on both the barycentric map and the Lumley's invariant map, respectively. Overall, the Boussinesq anisotropy states are more isotropic, clustering around the plane strain line, as shown in Figs. \ref{fig:AIM_BMap_RANSDNS_all.pdf} (a) and (c). On the other hand, the DNS anisotropy states scatter on both maps, reflecting a tendency of turbulence shifting from more axisymmetric contraction/two-component (oblate spheroid or pancake-like) to axisymmetric expansion (prolate spheroid) when the distance from the wall is increased, as shown in Figs. \ref{fig:AIM_BMap_RANSDNS_all.pdf} (b) and (d).   

\subsubsection{Instantaneous velocity field}
\begin{figure} 
\centerline{\includegraphics[width=3.5in]{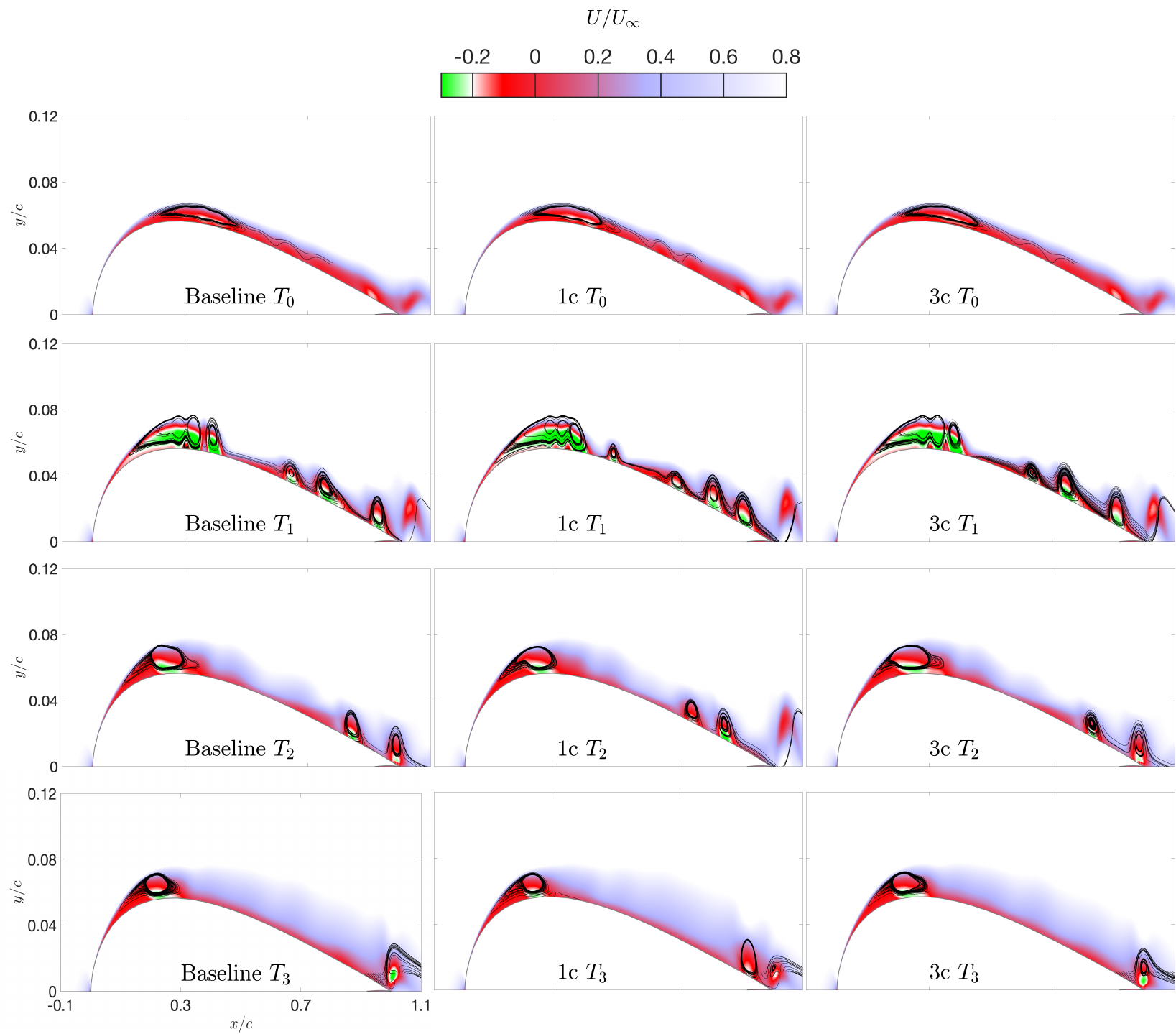}}
\caption[Contours of initial transient behavior of instantaneous mean streamwise velocity for baseline and eigenvalue perturbations ($1c$ and $3c$) at different times: $T_{0} = 0.1$, $T_{1} = 0.13$, $T_{2} = 0.15$, $T_{3} = 0.17$.]{Contours of initial transient behavior of instantaneous mean streamwise velocity for baseline and eigenvalue perturbations ($1c$ and $3c$) at different times: $T_{0} = 0.1$, $T_{1} = 0.13$, $T_{2} = 0.15$, $T_{3} = 0.17$. Streamlines are superimposed on the plot to capture the vortex shedding process. $\operatorname{Tu} = 0.03\%$.}
\label{fig:Instant_U.pdf}
\end{figure}

The unsteady visualization of ``vortex shedding'' \cite{watmuff1999evolution} from the baseline and the eigenvalue perturbations of $1c$ and $3c$ are presented in Fig. \ref{fig:Instant_U.pdf}. Energy losses are accompanied with a vortex shedding phenomenon that involves high unsteadiness within the flow \cite{marxen2003combined,mcauliffe2007transition,alam2000direct,wissink2003dns,simoni2012transition}. Streamlines are provided to capture the shed vortices on the suction side of the airfoil. Figure \ref{fig:Instant_U.pdf} shows the snapshots of the flow at four different times: $T_{0}$, $T_{1}$, $T_{2}$, and $T_{3}$. The shed vortices convect downstream, and eventually breakdown to smaller structures \cite{lengani2014pod,kurelek2016coherent}. All simulations show that the LSB first originates at $T_{0}$ for $0.3 < x/c < 0.6$, then gradually moves nearer to the leading edge at $T_{3}$. In Fig. \ref{fig:Instant_U.pdf}, vortex paring begins at $T_{1}$, and the coalesced vortices become a single, larger vortex at $T_{2}$, which significantly increases the boundary layer thickness, and similar behavior has been observed by other researchers \cite{lin1996low,lengani2014pod} as well. This large-scale vortex or coherent structure \cite{lengani2014pod} remains near the leading edge, followed by vortex shedding breaking down to turbulence further downstream. Note that this large-scale vortex structure indicates the evolution of a LSB at an early stage. At $T_{3}$, the turbulence downstream is composed of stochastic small-scale structures, which are smeared out by the unsteady SST $k-\omega$ LM model. Compared to the baseline prediction, the $1c$ perturbation tends to suppress the size of the coherent structure near the leading edge, while the $3c$ perturbation tends to bolster the size of it, as can be seen at $T_{3}$.   

\subsubsection{Turbulent production}
\begin{figure} 
\centerline{\includegraphics[width=3.5in]{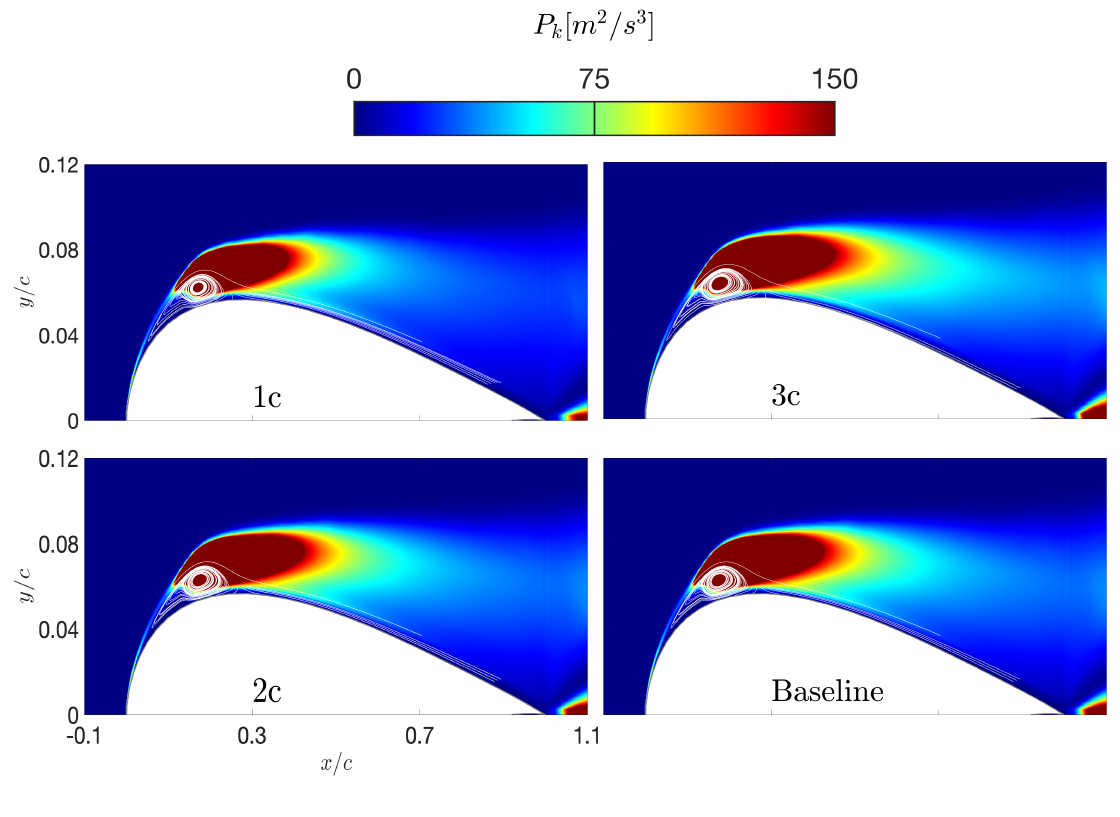}}
\caption[Contours of turbulent production for $1c$, $2c$ and $3c$ perturbations at $\operatorname{Tu} = 0.03\%$.]{Contours of turbulent production for $1c$, $2c$ and $3c$ perturbations at $\operatorname{Tu} = 0.03\%$. Baseline prediction is provided for reference. Streamlines show the size of the LSB on the suction side of the airfoil.}
\label{fig:PkMean_1c2c3cBase.pdf}
\end{figure}

Figure \ref{fig:PkMean_1c2c3cBase.pdf} shows contours of turbulent production from the baseline and the $1c$, $2c$, and $3c$ perturbations. Also included are the time-averaged streamlines to highlight the location for reverse flow incurred within the LSB. In Fig. \ref{fig:PkMean_1c2c3cBase.pdf}, values of $P_{k}$ appear to be very small, i.e., close to zero, in the vicinity of the wall near the leading edge $x/c < 0.15$, and in the outer region of the flow as well. This suggests a low level of turbulence is produced in these regions, where mean-flow kinetic energy is more prevalent. From Fig. \ref{fig:PkMean_1c2c3cBase.pdf}, for all contour plots it clearly shows a peak of $P_{k}$ found in the dark red region where the LSB is present, and the level of $P_{k}$ gradually deteriorates downstream. In addition, the $1c$ perturbation produces a smaller value of $P_{k}$ than the baseline prediction. On the other hand, the $3c$ perturbation produces a larger value of $P_{k}$ than the baseline prediction. This behavior is qualitatively similar to the very recent study of Clara \textit{et al.} \cite{de2021model}, which focused on enhancing the turbulent production in the LSB. On the other hand, $P_{k}$ produced by the $2c$ perturbation has a comparable magnitude to that for the baseline prediciton, and hence in between the $1c$ and $3c$ perturbations. 

\subsubsection{Skin friction coefficient}
\begin{figure} 
\centerline{\includegraphics[width=3.5in]{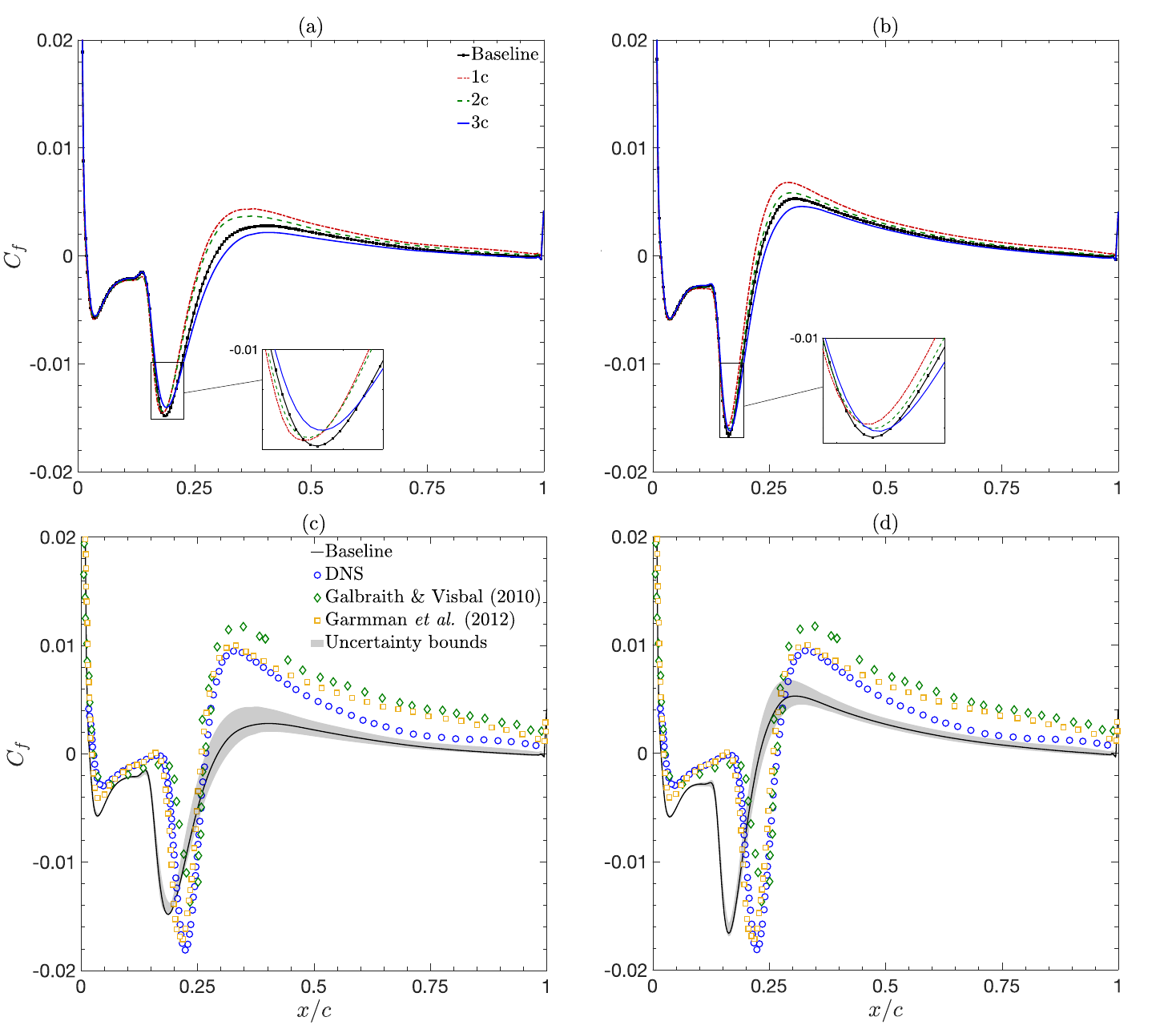}}
\caption[Distribution of skin friction coefficient for $1c$, $2c$ and $3c$ perturbations with enlarged region at trough and uncertainty bounds for eigenvalue perturbations with uniform value of $\Delta_{B} = 1$: (a) and (c) for $\operatorname{Tu} = 0.03\%$, (b) and (d) for $\operatorname{Tu} = 0.5\%$.]{Distribution of skin friction coefficient for $1c$, $2c$ and $3c$ perturbations with enlarged region at trough and uncertainty bounds for eigenvalue perturbations with uniform value of $\Delta_{B} = 1$: (a) and (c) for $\operatorname{Tu} = 0.03\%$, (b) and (d) for $\operatorname{Tu} = 0.5\%$. Baseline prediction is provided for reference. $\circ$ in-house DNS data \cite{zhang2021turbulent}.}
\label{fig:cf_four_line_Tu0027Tu05.pdf}
\end{figure}

The baseline prediction along with the eigenvalue perturbations ($1c$, $2c$, $3c$) for $C_{f}$ are shown in Figs. \ref{fig:cf_four_line_Tu0027Tu05.pdf} (a), (c) ($\operatorname{Tu} = 0.03\%$), (b), (d) ($\operatorname{Tu} = 0.5\%$). Also included are the in-house DNS \cite{zhang2021turbulent} and ILES/LES data by \cite{galbraith2010implicit} and \cite{garmann2013comparative} for comparison.  

Figures \ref{fig:cf_four_line_Tu0027Tu05.pdf} (a) and (b) enlarge the region for the trough to distinguish the clusters of $C_{f}$ profiles, where a negative peak is present around $x/c = 0.2$. In this region, a trough of $C_{f}$ appears, reflecting the significantly increased $C_{f}$ in magnitude within the LSB. In this region, enveloping behavior with respect to the baseline prediction occurs except around the negative peak in the trough, where the $1c$ perturbation sits slightly below the baseline prediction, while the $3c$ perturbation sits somewhat above the baseline prediction. A similar behavior has been observed by Mishra and Iaccarino \cite{mishra2017rans} and Iaccarino \textit{et al.} \cite{iaccarino2017eigenspace} in their UQ study for the canonical case of a turbulent flow over a backward-facing step. Downstream of the peak, the value of $C_{f}$ begins to sharply recover and approaches a value of $C_{f} = 0$ near $X_{R}$. In the recovery region, the $1c$ and $2c$ perturbations sit above the baseline prediction while the $3c$ perturbation does the opposite, indicating a reduction of $C_{f}$ for the $1c$ and $2c$ perturbations (enhancing $\left\langle U \right \rangle$), while an increase of $C_{f}$ for the $3c$ perturbation (suppressing $\left\langle U \right \rangle$) in the aft portion of the LSB. This is consistent with the behavior of the perturbed $C_{f}$ for T3A with the SST $k-\omega$ LM model, as shown in Fig. \ref{fig:cf_UQT3A_SSTLM_SST_kEpsilon.pdf}. As a consequence, shortening of the LSB length is observed under the $1c$ and $2c$ perturbations, which leads to an upstream shift of $X_{R}$ and shows a tendency to approach closer to the reference data. On the other hand, the $3c$ perturbation shifts the $X_{R}$ point further downstream, resulting in a larger length LSB. A similar behavior has been observed by other researchers in their numerical studies for the canonical case of a turbulent flow over a backward-facing step, e.g., see \cite{mishra2017rans,iaccarino2017eigenspace,gorle2019epistemic}.

Further downstream of $X_{R}$ is the attached turbulent boundary layer, where the $1c$ and $2c$ perturbations sit consistently above the baseline prediction, while the $3c$ perturbation does the opposite but in a less intense manner. In addition, these three eigenvalue perturbations gradually approach the baseline prediction as the flow moves further downstream of $X_{R}$. 
This indicates that the model form uncertainty becomes smaller as the flow proceeds further downstream, hence an increase in trustworthiness in the baseline prediction. In Figs. \ref{fig:cf_four_line_Tu0027Tu05.pdf} (c) and (d), the uncertainty bounds ($1c$, $2c$, $3c$) depicted by the gray region are constructed with respect to the baseline prediction. In the fore portion of the LSB, it is clear that negligibly small responses to $1c$, $2c$ and $3c$ perturbations are observed for the region $x/c < 0.15$, collapsing onto the baseline prediction and indicating relatively high trustworthiness, i.e., showing relatively good agreement with the in-house DNS \cite{zhang2021turbulent} and ILES/LES data of \cite{galbraith2010implicit, garmann2013comparative}. However, the size of the uncertainty bounds begins to increase in the aft portion of the LSB (further downstream of $X_{T}$). Figure \ref{fig:cf_four_line_Tu0027Tu05.pdf} (c) clearly shows that the uncertainty bounds encompass $X_{R}$ based on the in-house DNS and ILES/LES data of \cite{galbraith2010implicit, garmann2013comparative} for $\operatorname{Tu} = 0.03\%$. On the other hand, an overall shift in the upstream direction is observed for $\operatorname{Tu} = 0.5\%$, failing to encompass $X_{R}$.  Downstream of $X_{R}$, a noticeable discrepancy is observed for both $\operatorname{Tu} = 0.03\%$ and $\operatorname{Tu} = 0.5\%$, and uncertainty bounds are insufficient to encompass the reference data. In addition, the size of the uncertainty bounds for $\operatorname{Tu} = 0.5\%$ is overall smaller than that for $\operatorname{Tu} = 0.03\%$, as shown in \ref{fig:cf_four_line_Tu0027Tu05.pdf} (d). The important conclusion is that the simulation's response to the injection of eigenvalue perturbation varies with the magnitude of initial turbulent condition being used.

\begin{figure} 
\centerline{\includegraphics[width=3.5in]{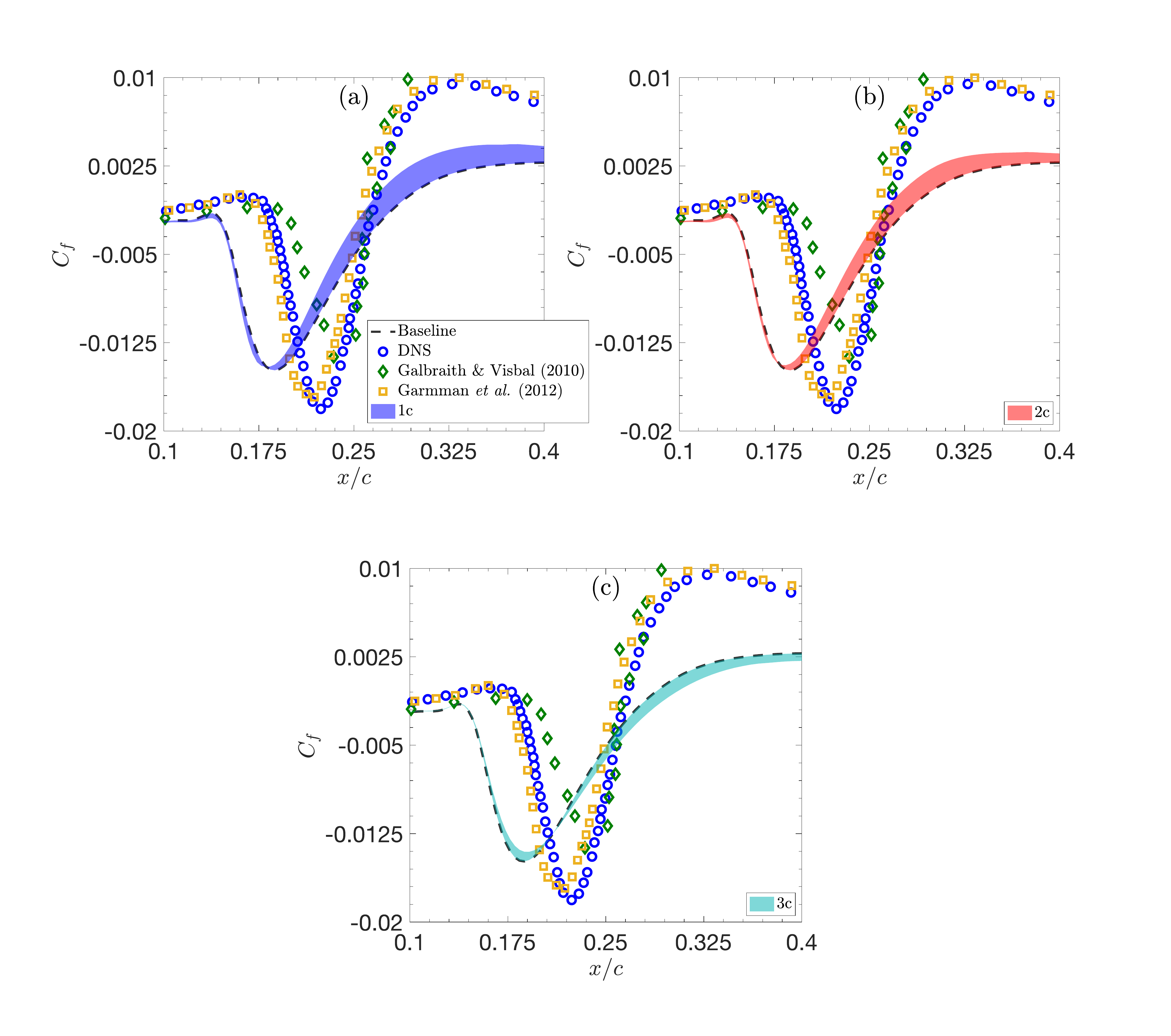}}
\caption[Enlarged version of Fig. \ref{fig:cf_four_line_Tu0027Tu05.pdf} (c) ($\operatorname{Tu} = 0.03\%$) at trough. Displayed are the uncertainty bounds for (a) $1c$, (b) $2c$ and (c) $3c$ perturbations.]{Enlarged version of Fig. \ref{fig:cf_four_line_Tu0027Tu05.pdf} (c) ($\operatorname{Tu} = 0.03\%$) at trough. Displayed are the uncertainty bounds for (a) $1c$, (b) $2c$ and (c) $3c$ perturbations. Baseline prediction is provided for reference. $\circ$ in-house DNS data \cite{zhang2021turbulent}.}
\label{fig:subTu0027_cf.pdf}
\end{figure}

Figures \ref{fig:subTu0027_cf.pdf} (a) - (c) enlarge the region for the trough to highlight the individual effects of these eigenvalue perturbations ($1c$, $2c$, $3c$) for $\operatorname{Tu} = 0.03\%$. In the fore portion of the LSB ($0.02 < x/c < 0.15$), all of the eigenvalue perturbations and the baseline prediction collapse onto a single curve, indicating a low sensitivity to the injection of eigenvalue perturbation and hence low level of model form uncertainty. On the other hand, the uncertainty bounds for $1c$ and $2c$ sit above the baseline prediction, reflecting a decrease in magnitude, with $2c$ less strength than $1c$, in the aft portion of the LSB ($0.15 < x/c < 0.26$). This contributes to a reduction of the discrepancy at the crest (the aft portion of the LSB), as shown in Figs. \ref{fig:subTu0027_cf.pdf} (a) and (b). In contrast to $1c$ and $2c$ perturbations, $3c$ perturbation sits somewhat below the baseline prediction, giving a lower value of $C_{f}$, while in a weaker strength (smaller size of the uncertainty bound) compared to the $1c$ and $2c$ perturbations. Because $3c$ perturbation retains the isotropic nature of the turbulent viscosity model, it has limited effect on the perturbed results. This explains the smaller size of uncertainty bound generated from the $3c$ perturbation compared to the $1c$ and $2c$ perturbations. Such inefficacy of $3c$ perturbation has been observed by Emory \textit{et al.} \cite{emory2013modeling} as well. As the result, this increases the discrepancy by deviating away from the reference data. 

\subsubsection{Pressure coefficient}
\begin{figure} 
\centerline{\includegraphics[width=3.5in]{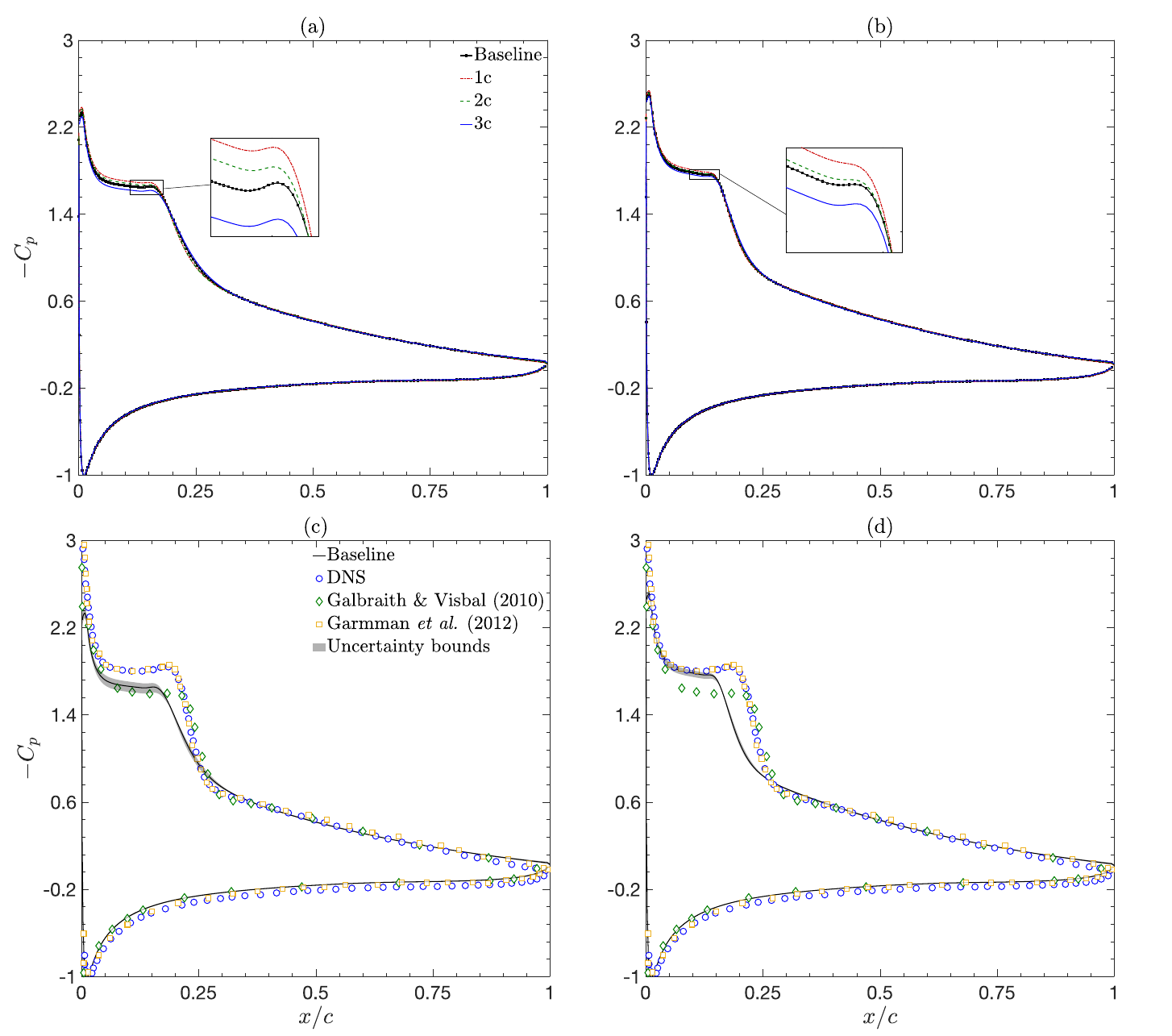}}
\caption[Distribution of pressure coefficient for $1c$, $2c$ and $3c$ perturbations with enlarged region at flat spot and uncertainty bounds for eigenvalue perturbations with uniform value of $\Delta_{B} = 1$: (a) and (c) for $\operatorname{Tu} = 0.03\%$, (b) and (d) for $\operatorname{Tu} = 0.5\%$.]{Distribution of pressure coefficient for $1c$, $2c$ and $3c$ perturbations with enlarged region at flat spot and uncertainty bounds for eigenvalue perturbations with uniform value of $\Delta_{B} = 1$: (a) and (c) for $\operatorname{Tu} = 0.03\%$, (b) and (d) for $\operatorname{Tu} = 0.5\%$. Baseline prediction is provided for reference. $\circ$ in-house DNS data \cite{zhang2021turbulent}.}
\label{fig:cp_four_Tu0027Tu05.pdf}
\end{figure}

Figures \ref{fig:cp_four_Tu0027Tu05.pdf} (a), (c) ($\operatorname{Tu} = 0.03\%$), (b), (d) ($\operatorname{Tu} = 0.5\%$) present the distribution of pressure coefficient $C_{p}$ over the SD7003 airfoil: baseline prediction and eigenvalue perturbations ($1c$, $2c$, $3c$). The in-house DNS \cite{zhang2021turbulent} and ILES/LES data of  \cite{galbraith2010implicit} and \cite{garmann2013comparative} are included for comparison. In Figs. \ref{fig:cp_four_Tu0027Tu05.pdf} (a) and (b), the region at the flat spot where the laminar-turbulent transition process occurs is enlarged. In this region, the $1c$ and $2c$ perturbations lie somewhat above the baseline prediction, with $2c$ less strength than $1c$. On the other hand, the $3c$ perturbation shows comparable strength with the $1c$ perturbation, lying below the baseline prediction, as shown in Figs. \ref{fig:cp_four_Tu0027Tu05.pdf} (a) and (b). Except for the flat spot, all of the profiles show a collapse, indicating a type of similarity. 

\begin{figure} 
\centerline{\includegraphics[width=3.5in]{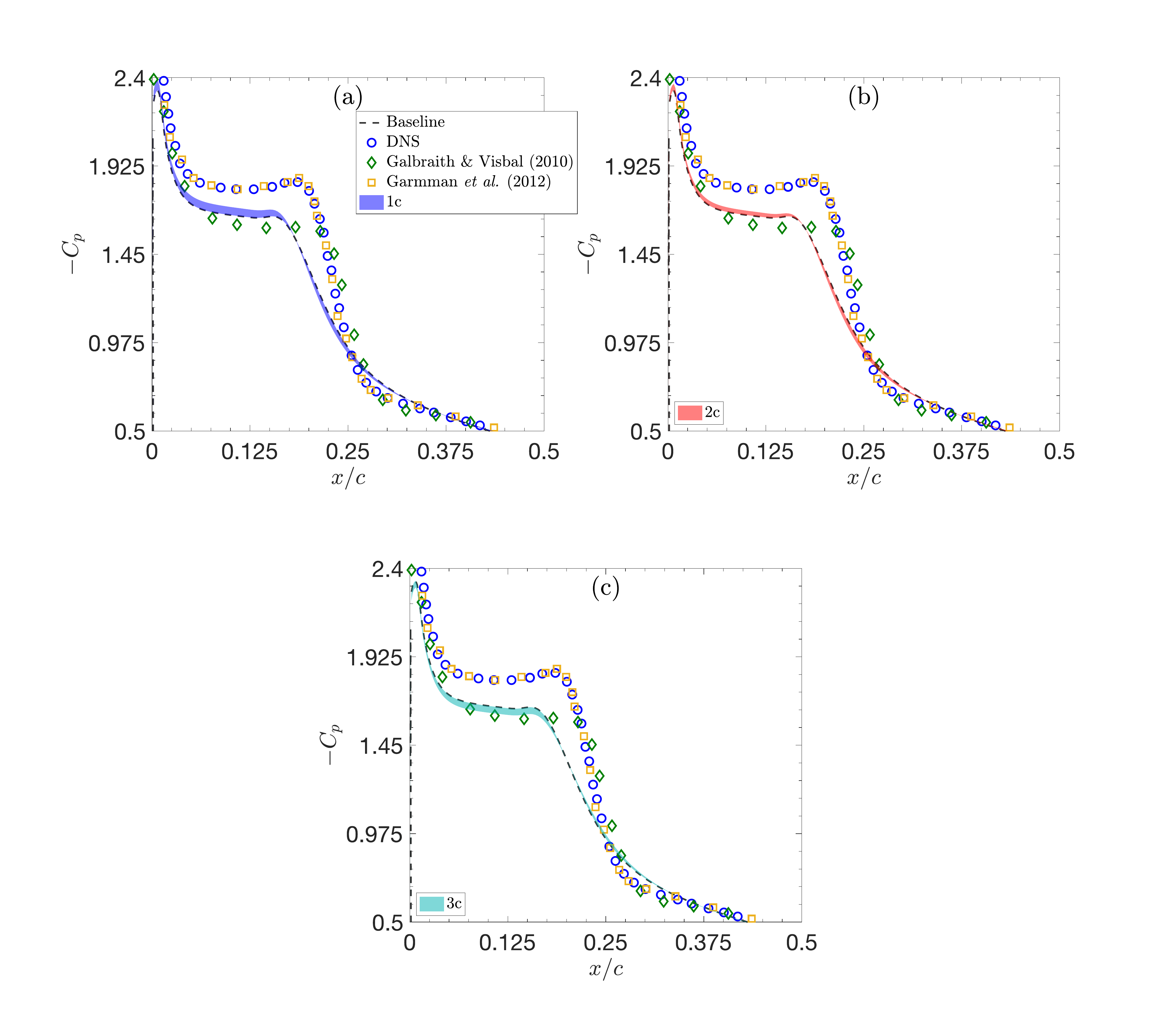}}
\caption[Enlarged version of Fig. \ref{fig:cp_four_Tu0027Tu05.pdf} (c) ($\operatorname{Tu} = 0.03\%$) at flat spot. Displayed are the uncertainty bounds for (a) $1c$, (b) $2c$ and (c) $3c$ perturbations.]{Enlarged version of Fig. \ref{fig:cp_four_Tu0027Tu05.pdf} (c) ($\operatorname{Tu} = 0.03\%$) at flat spot. Displayed are the uncertainty bounds for (a) $1c$, (b) $2c$ and (c) $3c$ perturbations. Baseline prediction is provided for reference. $\circ$ in-house DNS data \cite{zhang2021turbulent}.}
\label{fig:subTu0027_cp.pdf}
\end{figure}

In Figs. \ref{fig:cp_four_Tu0027Tu05.pdf} (c) and (d), the uncertainty bounds ($1c$, $2c$, $3c$) depicted by the gray region are constructed with respect to the baseline prediction, and no discernible uncertainty bounds is observed except at the flat spot, which is within the fore portion of the LSB. In addition, an enveloping behavior with respect to the baseline prediction is observed at the flat spot. This indicates that the model form uncertainty is most prevalent at the flat spot, indicating relatively low trustworthiness in the prediction for $C_{p}$. It is interesting to note that the uncertainty bounds for $\operatorname{Tu} = 0.03\%$ tend to encompass the ILES data of \cite{galbraith2010implicit} at the flat spot, while the uncertainty bounds for $\operatorname{Tu} = 0.5\%$ tend to encompass the in-house DNS \cite{zhang2021turbulent} and LES data of \cite{garmann2013comparative}. These trends do not happen in the prediction for $C_{f}$, as shown in Fig. \ref{fig:cf_four_line_Tu0027Tu05.pdf}. In the aft portion of the LSB, a small discrepancy that marks a lower value of $C_{p}$ is observed except in the region around $X_{R}$, where it gives a slightly larger value of $C_{p}$. On the other hand, no discernible uncertainty bounds is observed for the rest regions, reflecting a low level of the model form uncertainty and hence relatively high trustworthiness. Overall, the size of the uncertainty bounds for $\operatorname{Tu} = 0.03\%$ is larger than that for $\operatorname{Tu} = 0.5\%$. This again confirms the conclusion drawn previously: the degree of response to the injection of eigenvalue perturbation varies with the magnitude of initial turbulent condition being used. There is an important observation: $C_{p}$ is clearly much less sensitive to the eigenvalue perturbations than $C_{f}$, as shown in Figs. \ref{fig:cf_four_line_Tu0027Tu05.pdf} (c) and (d). The perturbed $C_{p}$ profiles only deviates from the baseline prediction at the flat spot. This is due to the fact that the wall pressure is determined by the freestream, which is only modified meticulously by the eigenvalue perturbations \cite{emory2013modeling}.

Figures \ref{fig:subTu0027_cp.pdf} (a) - (c) enlarge the region for the flat spot to highlight the individual effects of these eigenvalue perturbations ($1c$, $2c$, $3c$) for $\operatorname{Tu} = 0.03\%$. Figures \ref{fig:subTu0027_cp.pdf} (a) and (b) show that the uncertainty bound for $1c$ over-predicts the baseline prediction more than that for $2c$ at the flat spot, while $3c$ does the opposite in a comparable strength to the $1c$ perturbation. As a result, a tendency for $1c$ and $2c$ to approach toward the in-house DNS \cite{zhang2021turbulent} and LES data of \cite{garmann2013comparative}, and $3c$ to approach closer to the ILES data of \cite{garmann2013comparative}, is observed at the flat spot, as shown in Figs. \ref{fig:subTu0027_cp.pdf} (a) - (c).

\subsubsection{Mean velocity field}
\begin{figure} 
\centerline{\includegraphics[width=3.5in]{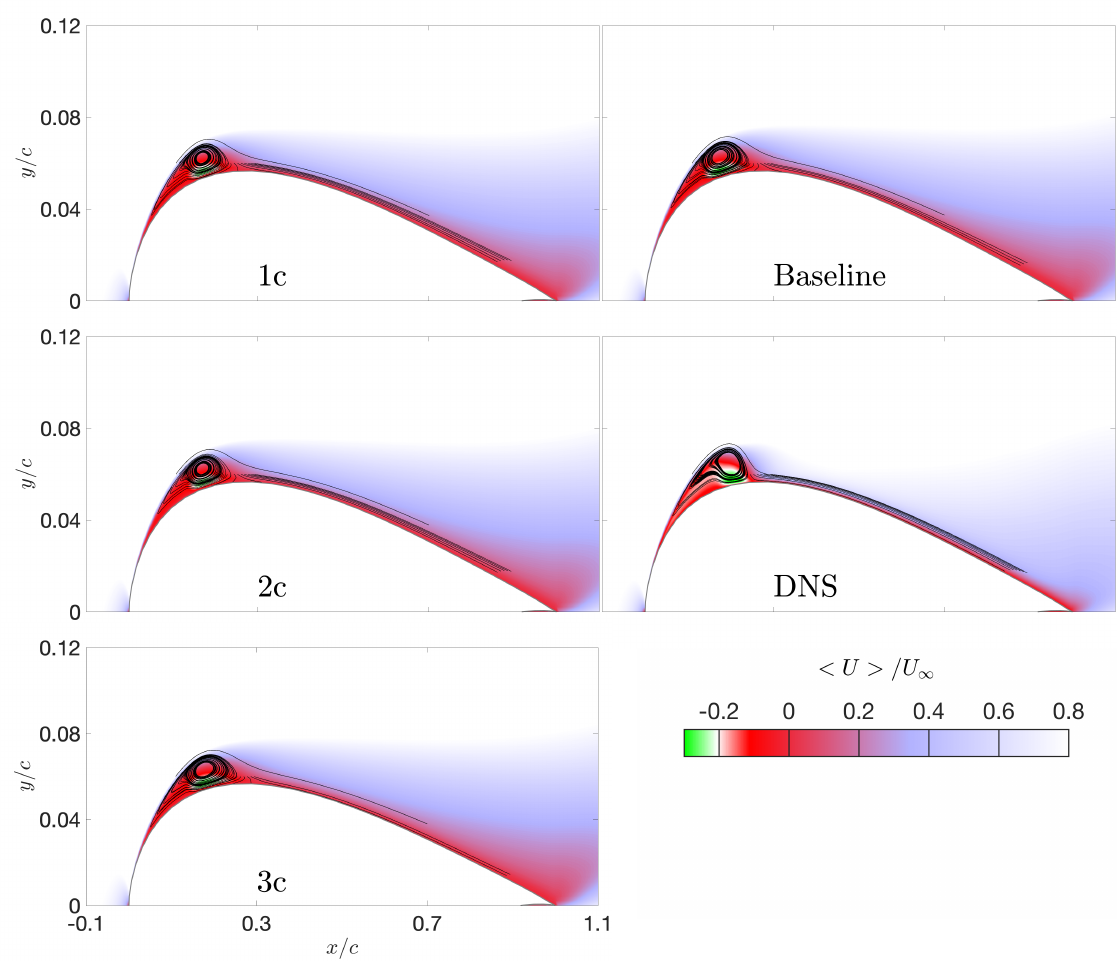}}
\caption[Contours of $\left\langle U \right \rangle/U_{\infty}$ for $1c$, $2c$ and $3c$ perturbations at $\operatorname{Tu} = 0.03\%$.]{Contours of $\left\langle U \right \rangle/U_{\infty}$ for $1c$, $2c$ and $3c$ perturbations at $\operatorname{Tu} = 0.03\%$. Baseline prediction is provided for reference, and in-house DNS data \cite{zhang2021turbulent} are included for comparison. Streamlines show the size of the LSB on the suction side of the airfoil.}
\label{fig:streamlines_U_subplot_focus.pdf}
\end{figure}

Contours of the mean velocity normalized with the freestream velocity, $\left\langle U \right \rangle/U_{\infty}$ from the baseline, eigenvalue perturbations ($1c$, $2c$, $3c$), and in-house DNS of \cite{zhang2021turbulent} in an $xy$ plane are shown in Fig. \ref{fig:streamlines_U_subplot_focus.pdf}. Moreover, included mean streamlines for the depiction of reverse flow ($\left\langle U \right \rangle/U_{\infty} < 0$) clearly show a large recirculating region located upstream in the region near the leading edge. The large recirculating region contains large-temporal-scale events (coherent structures), which are at low-frequency fluctuations due to very-large scale of unsteadiness of the recirculating region itself \cite{kiya1985structure}. Consequently, these large-scale coherent structures survive after time-averaging, namely LSB. The contours confirm the behavior observed in the prediction for $C_{f}$. First, the $1c$ and $2c$ perturbations reduce the magnitude of $C_{f}$ in the aft portion of the LSB compared to the baseline prediction, which leads to a lower value of $P_{k}$ shown in Fig. \ref{fig:PkMean_1c2c3cBase.pdf}, indicating subdued turbulence. Second, the $3c$ perturbation does the opposite, i.e., increased magnitude of $C_{f}$ bolstering $P_{k}$ and hence an increase in turbulence kinetic energy in the aft portion of the LSB. Recall that turbulence is produced through the straining mechanism, ( $-\left\langle u_{i}u_{j}\right\rangle \frac{\partial\left\langle U_{i}\right\rangle}{\partial x_{j}}$), i.e., a transfer of kinetic energy from mean flow to turbulence \cite{pope2001turbulent,durbin2011statistical}. Therefore, the $1c$ and $2c$ perturbations that enhance the streamwise stresses working against the mean flow gradient in a sense of reversal of energy from turbulence to the mean field have redistributed the flow energy in the LSB, while the $3c$ perturbation enhances the mechanism of transferring kinetic energy from the mean flow to turbulence. Correspondingly, the $1c$ and $2c$ perturbations increase the magnitude of $\left\langle U \right \rangle/U_{\infty}$ as a compromise to the decrease of the turbulence kinetic energy contained in the LSB, while the $3c$ perturbation reveals an reduction in the  magnitude of $\left\langle U \right \rangle/U_{\infty}$ to bolster turbulence. As a result, it can be observed that the size of the region of reverse flow becomes smaller for $1c$ and $2c$ compared to the baseline prediction, characterized by a shallower region of streamlines, while $3c$ gives a larger size of the region of reverse flow, characterized  by a broader region of streamlines. As the flow approaches downstream of $X_{R}$, streamlines for $1c$ and $2c$ get closer than that for the baseline prediction, indicating a larger magnitude of $\left\langle U \right \rangle/U_{\infty}$, while streamlines for $3c$ get thinner, which reflects a smaller magnitude of $\left\langle U \right \rangle/U_{\infty}$, as shown in Fig. \ref{fig:streamlines_U_subplot_focus.pdf}.  In addition, the $2c$ perturbation yields a magnitude of $\left\langle U \right \rangle/U_{\infty}$ in between the $1c$ and $3c$ perturbations. In comparison to the baseline prediction, the in-house DNS data overall show a larger magnitude of $\left\langle U \right \rangle/U_{\infty}$ in the attached turbulent boundary layer. This is characterized by the densely piled-up streamlines next to the wall, showing similar behavior to the $1c$ and $2c$ perturbations. Moreover, a larger region of reverse flow is observed for the in-house DNS data than that for the baseline prediction, and hence a larger recirculating vortex formed, depicted by the mean streamlines. As the height of the LSB produced by in-house DNS is somewhat increased compared to the baseline prediction, which modifies the effective shape of the SD7003 airfoil \cite{gaster1967structure}. This might partly explain the discrepancies that appear in the baseline predictions for $C_{f}$ and $C_{p}$.

\begin{figure} 
\centerline{\includegraphics[width=3.8in]{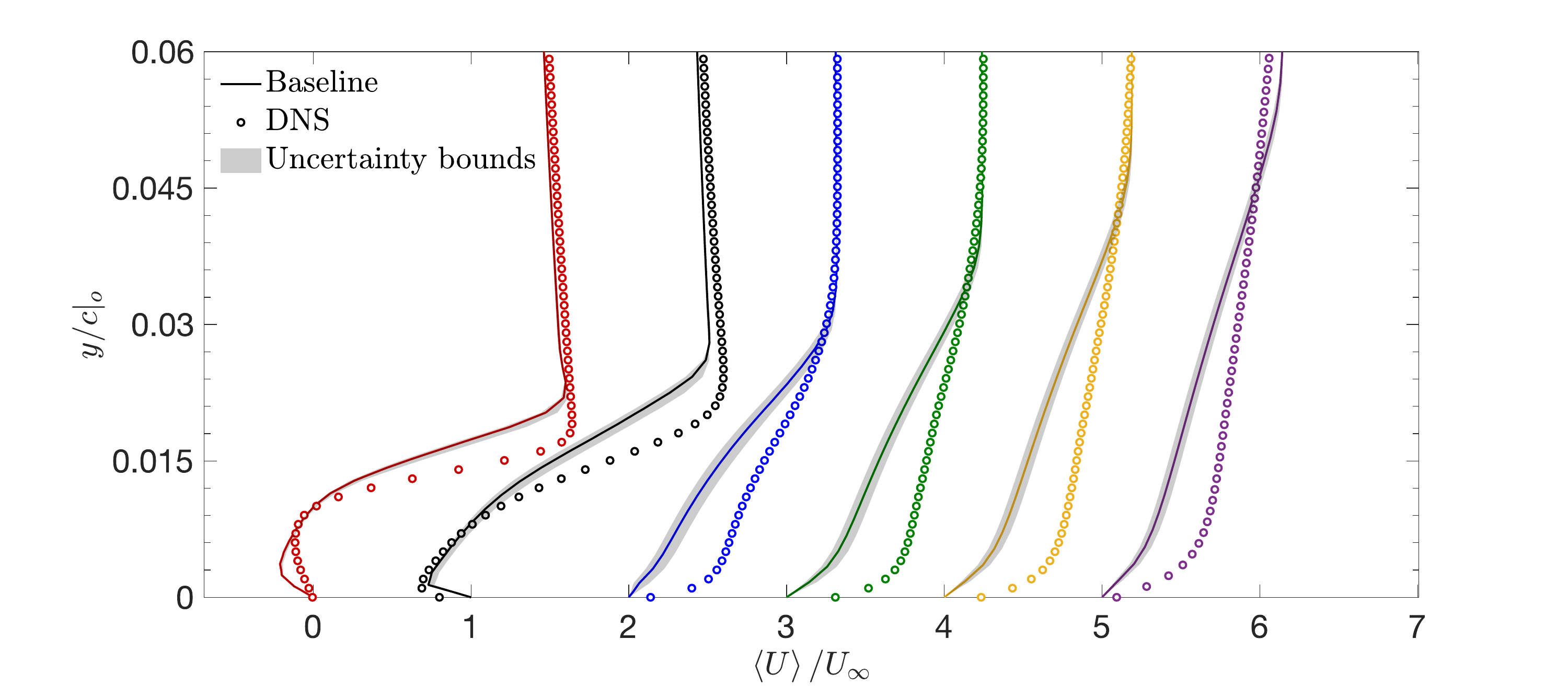}}
\caption[Streamwise mean velocity profiles in the aft portion of the LSB ($x/c = 0.15$ and $0.2$) and in the attached turbulent boundary layer ($x/c = 0.3, 0.4, 0.5$ and $0.6$) at $\operatorname{Tu} = 0.03\%$.]{Streamwise mean velocity profiles in the aft portion of the LSB ($x/c = 0.15$ and $0.2$) and in the attached turbulent boundary layer ($x/c = 0.3, 0.4, 0.5$ and $0.6$) at $\operatorname{Tu} = 0.03\%$. From left to right are $x/c = 0.15, 0.2, 0.3, 0.4, 0.5$ and $0.6$, respectively. Displayed are uncertainty bounds for eigenvalue perturbations with uniform value of $\Delta_{B} = 1$ ($1c$, $2c$ and $3c$). Baseline prediction is provided for reference. $\circ$ in-house DNS data \cite{zhang2021turbulent}.}
\label{fig:U_all_new.pdf}
\end{figure}

The predicted mean $\left\langle U \right \rangle/U_{\infty}$ profiles across the entire boundary layer on the suction side are plotted in Fig. \ref{fig:U_all_new.pdf}. The uncertainty bounds generated from the eigenvalue perturbations ($1c$, $2c$, $3c$) are depicted by gray regions, as shown in Fig. \ref{fig:U_all_new.pdf}. Also included is the in-house DNS data of \cite{zhang2021turbulent}. The boundary layer has separated in the aft portion of the LSB at $x/c =0.15$ and $0.2$, where the reverse flow close to the wall is visible. At $x/c = 0.15$, the predicted $\left\langle U \right \rangle/U_{\infty}$ profile shows good agreement with the in-house DNS data for $0.006 < y/c < 0.012$ and $y/c > 0.021$, except in the region next to the wall $y/c < 0.006$ and far from the wall $0.012 < y/c < 0.021$. The response to the injection of eigenvalue perturbation for $x/c = 0.15$ is negligibly small, and no encompassing of the in-house DNS data is observed. As the flow moves a bit further downstream, the uncertainty bounds tend to encompass the in-house DNS data in the region of reverse flow next to the wall $y/c < 0.012$, while a growing discrepancy is found gradually away from the wall $0.012 < y/c < 0.027$, and no discernible uncertainty bounds is observed above the OBL $y/c > 0.027$. Further downstream of $X_{R}$ is the attached TBL (from $x/c = 0.3$ to $x/c = 0.6$), the recovery to turbulent profile occurs, showing consistent under-predictions for the $\left\langle U \right \rangle/U_{\infty}$ profiles. A similar behavior was observed by Luis \textit{et al.} \cite{bernardos2019rans} in their numerical study for a transitional flow over a NACA 0012 airfoil using the SST $k-\omega$ LM model. This is due to the inaccurate prediction for the period vortical structures that shed downstream. The uncertainty bound reduces in size as the flow proceeds from $x/c = 0.3$ to $x/c = 0.6$. The positive values of $C_{f}$ after $X_{R}$ might partly contribute this behavior. Moreover, it should be noted that the SST $k-\omega$ LM model is inherently incapable of capturing the presence of rotational strains due to the adoption of the Boussinesq turbulent viscosity hypothesis. However, rotation strains make crucial contributions to the evolution of turbulence. For example, on the cambered SD7003 airfoil the boundary layer growth rate is decreased on the convex surface and increased on the concave surface. However, the eigenspace perturbation method strictly adheres to the extended form of the Boussinesq turbulent viscosity hypothesis, therefore is unable to account for the limitations of rotation and curvature effects \cite{mishra2019theoretical}. This might partly explain the insufficient uncertainty bounds to encompass the discrepancies observed in Figs.  \ref{fig:subTu0027_cf.pdf} and \ref{fig:U_all_new.pdf}.

\begin{figure} 
\centerline{\includegraphics[width=3.5in]{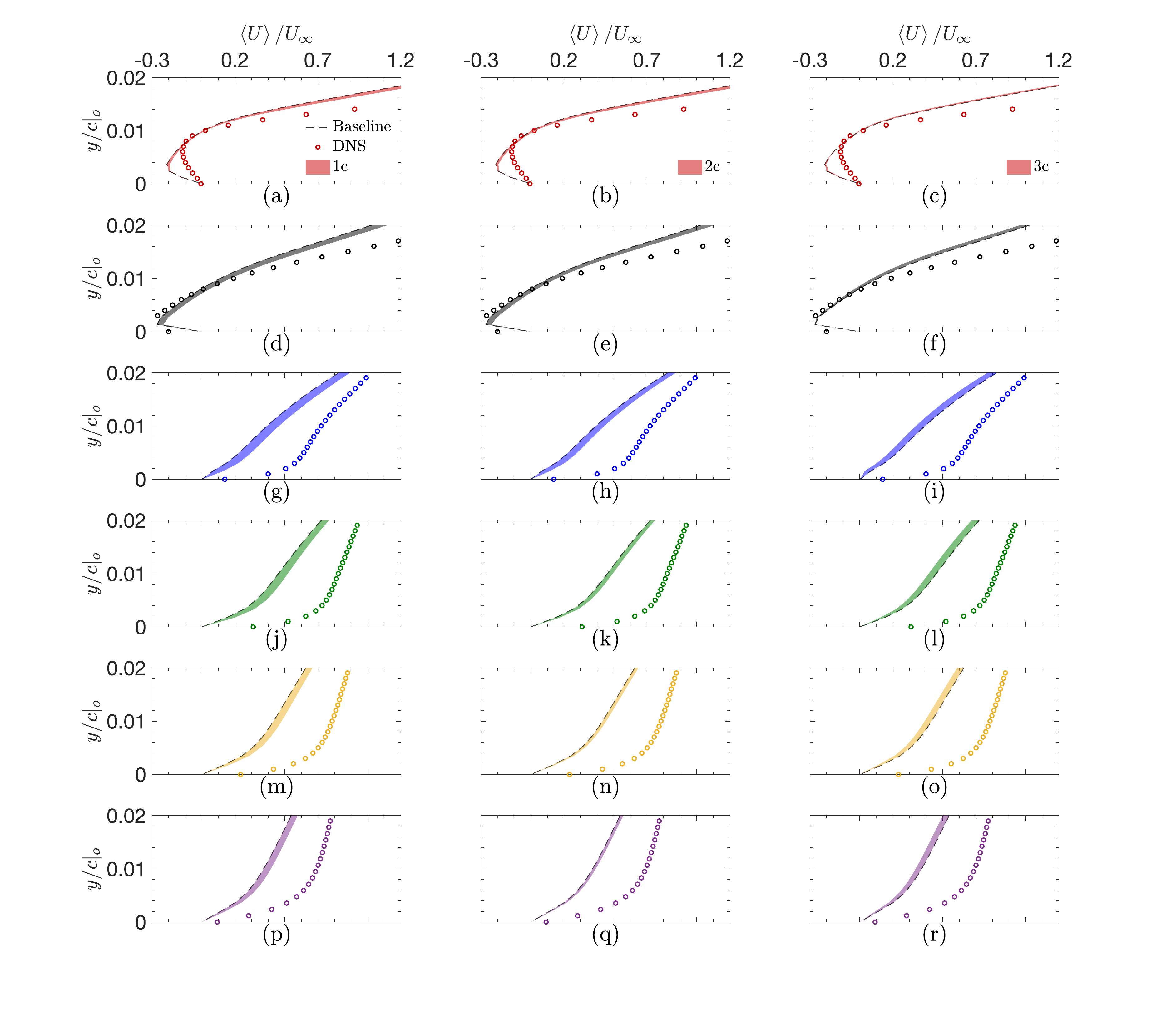}}
\caption[Enlarged version of Fig. \ref{fig:U_all_new.pdf} for region next to the wall $0 < y/c|_{o} < 0.02$.]{Enlarged version of Fig. \ref{fig:U_all_new.pdf} for region next to the wall $0 < y/c|_{o} < 0.02$. From top to bottom are (a), (b), (c) for $x/c = 0.15$, (d), (e), (f) for $x/c = 0.2$, (g), (h), (i) for $x/c = 0.3$, (j), (k), (l) for $x/c = 0.4$, (m), (n), (o) for $x/c = 0.5$ and (p), (q), (r) for $x/c = 0.6$, respectively. From left to right are $1c$, $2c$ and $3c$ perturbation results, displaying a clear relationship between which perturbation and the size of its uncertainty bound. Baseline prediction is provided for reference. $\circ$ in-house DNS data \cite{zhang2021turbulent}.}
\label{fig:sub_U_all.pdf}
\end{figure}

Figures \ref{fig:sub_U_all.pdf} (a) - (r) expand the near-wall region to highlight the individual effects of eigenvalue perturbations ($1c$, $2c$, $3c$) on $\left\langle U \right \rangle/U_{\infty}$. At $x/c = 0.15$, Figs. \ref{fig:sub_U_all.pdf} (a) and (b) show that the $1c$ and $2c$ perturbations tend to over-predict the baseline prediction, although very slightly. On the other hand, the $3c$ perturbation becomes almost indistinguishable from the baseline prediction, as shown in Fig. \ref{fig:sub_U_all.pdf} (c). 
At $x/c = 0.2$, the uncertainty bounds generated from the $1c$ and $2c$ perturbations are noticeably increased in size, sitting below the baseline prediction. On the other hand, the simulation is less sensitive to the $3c$ perturbation, yielding a relatively small uncertainty bound that sits above the baseline prediction, as shown in Figs. \ref{fig:sub_U_all.pdf} (d) - (f). From $x/c = 0.3$ to $x/c = 0.6$, the flow begins to recover to turbulent profile, the size of the uncertainty bounds generated from the $1c$, $2c$, and $3c$ perturbations increases to maximum at $x/c = 0.3$, followed by a steady decrease further downstream. Overall, it is clear that the $1c$ and $2c$ uncertainty bounds over-predict the baseline prediction, showing a tendency to approach closer to the in-house DNS data, while the $3c$ uncertainty bound does the opposite. A similar behavior was also observed by Luis \textit{et al.} \cite{cremades2019reynolds} in their numerical study for a turbulent flow over a backward-facing step. Moreover, it should be noted that the $1c$ and $2c$ perturbations react more favorably with the positive $C_{f}$ values in the region downstream of $X_{R}$ than the $3c$ perturbation, characterized by a rather noticeable decrease in the size of the uncertainty bounds.

\subsubsection{Reynolds shear stress}
\begin{figure} 
\centerline{\includegraphics[width=3.5in]{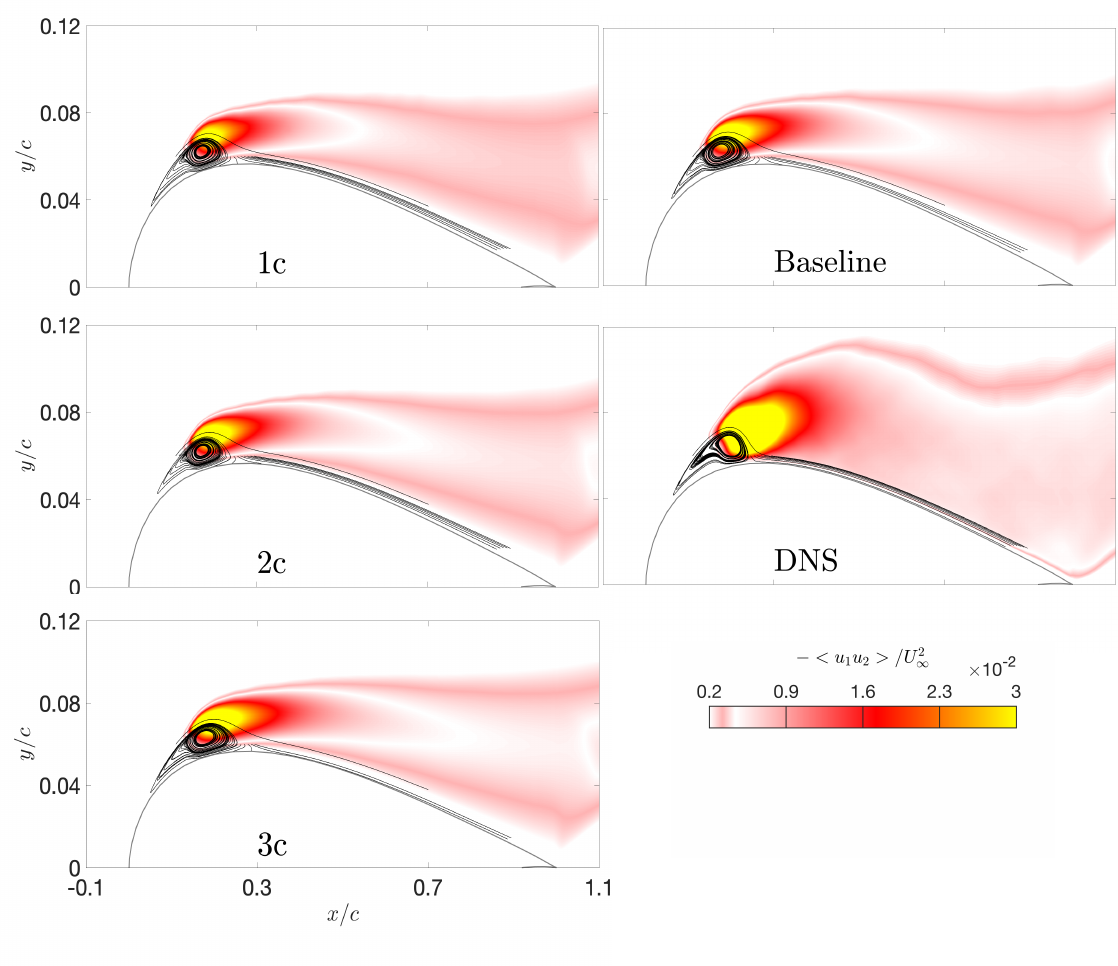}}
\caption[Contours of $-\left\langle u_{1}u_{2} \right \rangle/U_{\infty}^2$ for $1c$, $2c$ and $3c$ perturbations at $\operatorname{Tu} = 0.03\%$.]{Contours of $-\left\langle u_{1}u_{2} \right \rangle/U_{\infty}^2$ for $1c$, $2c$ and $3c$ perturbations at $\operatorname{Tu} = 0.03\%$. Baseline prediction is provided for reference, and in-house DNS data \cite{zhang2021turbulent} are included for comparison. Streamlines show the size of the LSB on the suction side of the airfoil.}
\label{fig:RANS_contour_streamline_normuv_All_subplot_focus.pdf}
\end{figure}

Contours of the Reynolds shear stress normalized with the freestream velocity squared, $-\left\langle u_{1}u_{2} \right \rangle/U_{\infty}^2$ from the baseline, $1c$, $2c$ and $3c$ perturbations, and in-house DNS \cite{zhang2021turbulent} in an $xy$ plane are shown in Fig. \ref{fig:RANS_contour_streamline_normuv_All_subplot_focus.pdf}. Streamlines are included to depict the region of reverse flow, where the LSB is present. From Fig. \ref{fig:RANS_contour_streamline_normuv_All_subplot_focus.pdf}, it is clear that the magnitude of $-\left\langle u_{1}u_{2} \right \rangle/U_{\infty}^2$ is almost zero in the region near the leading edge as well as in the outer region of the flow. A similar behavior was observed by Zhang and Rival \cite{zhang2020direct} in their experimental study. Recall that the Reynolds shear stress is dedicated to the contribution of the turbulent production \cite{pope2001turbulent,durbin2011statistical}, a low level of turbulence should be produced in these regions. 
This confirms the behavior observed in the prediction for $P_{k}$ shown in Fig. \ref{fig:PkMean_1c2c3cBase.pdf}. In addition, Fig. \ref{fig:RANS_contour_streamline_normuv_All_subplot_focus.pdf} shows that $-\left\langle u_{1}u_{2} \right \rangle/U_{\infty}^2$ contours are everywhere positive, which is consistent with the positive magnitude of turbulent production observed in Fig. \ref{fig:PkMean_1c2c3cBase.pdf} (the correlation $\left\langle u_{1}u_{2} \right \rangle$ is almost always negative when the mean gradient is positive, and vice versa). Likewise, the magnitude of $-\left\langle u_{1}u_{2} \right \rangle/U_{\infty}^2$ reduces as the flow approaches further downstream from the LBS where a peak value of $-\left\langle u_{1}u_{2} \right \rangle/U_{\infty}^2$ is found , i.e., bright yellow region for $x/c \sim
0.25$. It should be noted that the highest Reynolds shear stress happens around $X_{R}$, where a high level of momentum transfer due to anisotropy Reynolds stresses is present. A similar behavior was observed in the experimental measurements of Zhang and Rival \cite{zhang2020direct} as well. From Fig. \ref{fig:RANS_contour_streamline_normuv_All_subplot_focus.pdf}, the $1c$ and $2c$ perturbations under-predict the baseline prediction, with $2c$ yielding a relatively larger magnitude of $-\left\langle u_{1}u_{2} \right \rangle/U_{\infty}^2$ than that for $1c$, while the $3c$ perturbation over-predicts the baseline prediction. In addition, it clearly shows that the baseline prediction for the $-\left\langle u_{1}u_{2} \right \rangle/U_{\infty}^2$ contour is prominently reduced in magnitude compared to the in-house DNS data across the entire suction side. Therefore, the $3c$ perturbation that increases the magnitude of $-\left\langle u_{1}u_{2} \right \rangle/U_{\infty}^2$ shows a tendency of approaching closer to the in-house DNS data, which is consistent with the behavior of the prediction for $P_{k}$ shown in Fig. \ref{fig:PkMean_1c2c3cBase.pdf}. One may notice that the $3c$ perturbation with the largest turbulent production does not have the reattachment point occur earlier than the $1c$ perturbation with the least turbulent production. According to Davide \textit{et al.} \cite{lengani2014pod}, the overall turbulence kinetic energy can be decomposed into the large-scale coherent (Kelvin-Helmholtz induced) and stochastic (turbulence-induced) contributions. The $1c$ perturbation suppresses the size of the LSB, during which part of the coherent energy is reversed into the mean flow. On the other hand, the $3c$ perturbation fosters the large-scale coherent vortex-shedding structures by extracting more energy from the mean flow, which results in large coherent energy stored in these large-scale structures. This might partly explain the anomalous behavior. The similar behavior has been observed by Gianluca \textit{et al.}  \cite{iaccarino2017eigenspace} and Luis \textit{et al.} \cite{cremades2019reynolds} in their numerical study for a turbulent flow over a backward-facing step.

\begin{figure} 
\centerline{\includegraphics[width=3.5in]{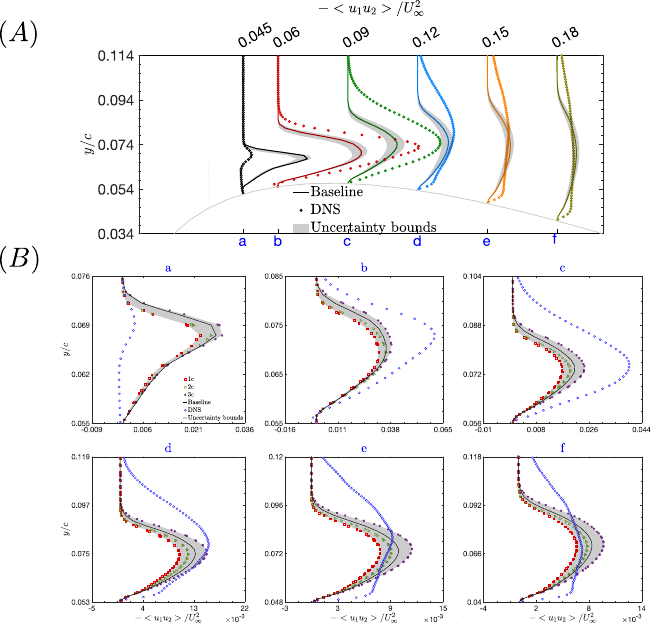}}
\caption[$(A)$ Profile of normalized Reynolds shear stress along the suction side of the SD7003 airfoil (geometry depicted by gray line). $(B)$ Enlarged version of $(A)$ at each location.]{$(A)$ Profile of normalized Reynolds shear stress along the suction side of the SD7003 airfoil (geometry depicted by gray line): from left to right are $\textcolor{blue}{a}$, $\textcolor{blue}{b}$, $\textcolor{blue}{c}$, $\textcolor{blue}{d}$, $\textcolor{blue}{e}$, $\textcolor{blue}{f}$ for $x/c = 0.15$, $0.2$, $0.3$, $0.4$, $0.5$ and $0.6$, respectively. Displayed are uncertainty bounds for eigenvalue perturbations with uniform value of $\Delta_{B} = 1$ at $\operatorname{Tu} = 0.03\%$. Profile of baseline prediction is provided for reference. $(B)$ Enlarged version of $(A)$ at each location. Profiles of baseline prediction and $1c$, $2c$, and $3c$ perturbations are provided for reference. $\circ$ in-house DNS data \cite{zhang2021turbulent}.}
\label{fig:uv_all_six_single_sub.pdf}
\end{figure}

The predicted $-\left\langle u_{1}u_{2} \right \rangle/U_{\infty}^2$ profiles on the suction side are shown in Figs. \ref{fig:uv_all_six_single_sub.pdf} ($A$) and ($B$). For better sense of virtual reality, the actual locations, denoted $a$ - $f$, on the suction side of the SD7003 airfoil are used, as shown in Fig. \ref{fig:uv_all_six_single_sub.pdf} ($A$). Moreover, Fig. \ref{fig:uv_all_six_single_sub.pdf} ($B$) focuses on the effects of eigenvalue perturbations ($1c$, $2c$, $3c$) for each location. In Figs. \ref{fig:uv_all_six_single_sub.pdf} ($A$) and ($B$), uncertainty bounds are depicted by gray regions. Figure \ref{fig:uv_all_six_single_sub.pdf} ($A$) clearly shows an enveloping behavior with respect to the baseline prediction at each location. An increase in the size of uncertainty bounds is observed as the flow moves from $a$ to $b$, reaching a maximum at $c$, which is consistent with the behavior shown in Fig. \ref{fig:U_all_new.pdf}. This indicates that the most model form uncertainty is found in the region downstream of the LSB near $X_{R}$. While the generated uncertainty bounds are not sufficient to encompass the reference data in the aft portion of the LSB, the reason is partly due to the fact that the eigenspace perturbation method is unable to account for the limitations of rotation and curvature effects, as stated in the previous section. In addition, the excluded amplitude perturbation and eigenvector perturbation of the Reynolds stress tensor and the parametric uncertainty introduced in model coefficients might also lead to the insufficiency in the generated uncertainty bounds. Further downstream from $X_{R}$, the size of the uncertainty bounds gradually reduces, which shows a tendency of encompassing the in-house DNS data, reflecting increased trustworthiness within the attached TBL. 

Figure \ref{fig:uv_all_six_single_sub.pdf} ($B$) scrutinizes closely the effects of the $1c$, $2c$ and $3c$ perturbations at each location. From $a$-$f$, it is clear that the $1c$ and $2c$ perturbations under-predict the baseline prediction, while the $3c$ perturbation does the opposite. This confirms well with the behavior observed in the prediction for $P_{k}$, as shown in Fig. \ref{fig:PkMean_1c2c3cBase.pdf}. This implies that Reynolds shear stress plays a key role in contributing to the turbulent production. At $a$ (in the aft portion of the LSB), the $1c$ uncertainty bound shows a tendency to approach closer to the in-house DNS data, although a noticeable discrepancy is observed. At $b$ (in the aft portion of the LSB) and $c$ (downstream of the LSB near $X_{R}$), the $3c$ perturbation increases the uncertainty bound toward the in-house DNS data, while both the $1c$ and $2c$ perturbations increase the uncertainty bounds in an opposite manner, namely, deviating from the in-house DNS data. As the flow proceeds further downstream within the attached turbulent boundary layer, the uncertainty bounds generated from the $1c$, $2c$ and $3c$ perturbations begin to encompass the in-house DNS data, although discrepancies in the near-wall region as well as in the outer region of the flow are observed. Overall, the model form uncertainty is relatively small within the attached turbulent boundary layer, which indicates more trustworthy results than that for the LSB. 

\section{Conclusions}
The goal of this study was to advance our understanding of a physics-based approach to quantify model form uncertainty in transition RANS simulations of incompressible flows over a flat plate (T3A transition with $\operatorname{Tu} = 3.3\%$ and $U_{\infty} = 5.4 m/s$) and an airfoil (SD7003 airfoil with $\operatorname{Re_{c}} = 60,000$). Eigenvalue perturbations to the three extreme states ($1c$, $2c$, and $3c$) of the barycentric map has been investigated using the eigenspace perturbation framework by Emory \textit{et al.} \cite{emory2013modeling}, which has been implemented in a user-friendly fashion into the open-source OpenFOAM software. The eigenspace perturbation method directly injects perturbations to the Reynolds stress tensor, resulting in a perturbed velocity field by solving the momentum equations. The perturbations were conducted through a decomposition of the Reynolds stress tensor, i.e., perturbing eigenvalues of the anisotropy tensor.

\subsection{T3A}
The SST $k-\omega$ LM model successfully captured the laminar-turbulent transition over a flat plate, and showed overall good agreement between the prediction for the skin friction coefficient and the experimental data of \cite{roach1990influence}. On the other hand, the SST $k-\omega$ and $k - \varepsilon$ models clearly failed to capture the transition process, characterized by a trough in the prediction for the skin friction coefficient given by SST $k-\omega$ LM. Most model form uncertainty was concentrated at the trough, in which eigenvalue perturbations exhibited an opposite way compared to that for the turbulent region downstream of the trough. The size of uncertainty bound tended to increase linearly with the magnitude of $\Delta_{B}$ for all three models. An important conclusion was drawn: the degree of response to the eigenvalue perturbations depends on which turbulence model is being used and which QoIs are bieng observed. From the contours of $k/U_{\infty}^2$, a bump at the leading edge of the flat plate marks laminar-turbulent transition, which corresponds to the location of the trough in skin friction coefficient plot.

\subsection{SD7003 airfoil}
The predictions for the important transition parameters ($X_{S}$, $X_{T}$, and $X_{R}$) for $\mathrm{Tu} = 0.03\%$ were overall in good agreement with the reference data of \cite{galbraith2010implicit,garmann2013comparative,zhang2021turbulent} than that for $\mathrm{Tu} = 0.5\%$. Overall, the predictions for the skin friction coefficient for $\mathrm{Tu} = 0.03\%$ in the fore portion of the LSB better matched the reference data than that for $\mathrm{Tu} = 0.5\%$, while a relatively large discrepancy was found in the aft portion. This is consistent with the predictions by \cite{catalano2011rans,bernardos2019rans,tousi2021active}. When the eigenvalue perturbations to the mean velocity profile and the Reynolds shear stress profile were plotted in the aft portion, an enveloping behavior was observed. It was interesting to note that a series of linear increments in $\Delta_{B}$ led to linear responses in the increasing size of uncertainty bound for both the mean velocity profile and the Reynolds shear stress profile. 

The anisotropy states of the Reynolds stress tensor were also analyzed using the Lumley's invariant map and the barycentric map. The Boussinesq anisotropy states based on the SST $k-\omega$ LM model were clustered around the plane strain line due to the two-dimensional flow condition. On the other hand, the anisotropy states for in-house DNS in the aft portion of the LSB showed an oblate spheroid at the wall, and gradually transitioned to a prolate spheroid with increasing distance from the wall. This revealed that the laminar-turbulent transition tended to damp the streamwise stresses in the near-wall region. Downstream of $X_{R}$ in the turbulent boundary layer, anisotropy states shifted from the axisymmetric contraction state to the two-component state. Away from the wall anisotropy states gradually shifed toward the axisymmetric expansion state, indicating increasing streamwsie stresses. 

The contours of the instantaneous velocity field on the suction side of the SD7003 airfoil were plotted at four different times, i.e., $T_{0}$ - $T_{3}$. The LSB structure first originated at $T_{0}$, and vortex paring began at $T_{1}$, then coalesced vortices to become a single and larger vortex at $T_{2}$, followed by vortex shedding breaking down to smaller stochastic small-scale structures at $T_{3}$ when the LSB moved nearer to the leading edge. The $1c$ perturbation showed a tendency to suppress the evolution of the LSB, while the $3c$ perturbation fostered the formation of it instead. 

Turbulent production $P_{k}$ over the SD7003 airfoil peaked in the LSB, and $1c$ perturbation gave a lower production compared to the baseline prediction, while the opposite was true for the $3c$ perturbation. Besides, the $2c$ perturbation gave comparable magnitude of $P_{k}$ to the baseline prediction. 

When the uncertainty bounds for $C_{f}$ over the airfoil for $\operatorname{Tu} = 0.03\%$ and $\operatorname{Tu} = 0.5\%$ were plotted, the $1c$ and $2c$ perturbations yielded a smaller magnitude of $C_{f}$ (enhancing $\left\langle U \right\rangle$) (larger negative value of $C_{f}$) than the $3c$ perturbation (reducing $\left\langle U \right\rangle$) in the trough (aft portion of the LSB). This behavior is qualitatively similar to the UQ analysis for $C_{f}$ for a turbulent flow over a backward-facing step \cite{iaccarino2017eigenspace}. In addition, the $1c$ and $2c$ perturbations shifted the $X_{R}$ in the upstream direction, suppressing the LSB length, which showed a tendency to approach closer to the reference data for $\operatorname{Tu} = 0.03\%$. While, the $3c$ perturbation did the opposite. This behavior has been observed by other researchers as well, e.g., \cite{mishra2017rans,iaccarino2017eigenspace,gorle2019epistemic}. In general, the SST $k-\omega$ LM model was less sensitive to the eigenvalue perturbations for $\operatorname{Tu} = 0.5\%$ than $\operatorname{Tu} = 0.03\%$. $X_{R}$ for $\operatorname{Tu} = 0.03\%$ was well captured within the uncertainty bounds.  

The uncertainty bounds for $C_{p}$ over the airfoil for $\operatorname{Tu} = 0.03\%$ and $\operatorname{Tu} = 0.5\%$ were also analyzed. The model form uncertainty was most prevalent at the flat spot (fore portion of the LSB) as well as in the region around $X_{R}$ (aft portion of the LSB), with the $1c$ and $2c$ perturbations giving a larger value of $C_{p}$ than the baseline prediction, while the $3c$ perturbation did the opposite, hence an enveloping behavior was observed. On the other hand, no discernible uncertainty bounds were observed for the rest regions, indicating a low level of the model form uncertainty. Overall, the size of the uncertainty bounds for $\operatorname{Tu} = 0.03\%$ was larger than that for $\operatorname{Tu} = 0.5\%$.

The contours of the dimensionless mean velocity $\left\langle U \right\rangle/U_{\infty}$ over the airfoil for $\operatorname{Tu} = 0.03\%$ were plotted in an $xy$ plane, a large recirculating region was found in the region of reverse flow ($\left\langle U \right\rangle/U_{\infty} < 0$). The size of reverse flow became smaller under the $1c$ and $2c$ perturbations, while the $3c$ perturbation bolstered the region of reverse flow. The dimensionless mean velocity $\left\langle U \right\rangle/U_{\infty}$ profiles across the suction side of the airfoil were also plotted in shifted coordinates $y/c|_{o}$, the lower sections of the $\left\langle U \right\rangle/U_{\infty}$ profile in the aft portion of the LSB showed relatively good agreement with the in-house DNS data \cite{zhang2021turbulent}. The size of uncertainty bounds was negligible at $X_{T}$, then reached its maximum near $X_{R}$. Downstream of $X_{R}$, $\left\langle U \right\rangle/U_{\infty}$ increased in magnitude under the $1c$ and $2c$ perturbations compared to the baseline prediction, while the $3c$ perturbation did the opposite. This behavior is qualitatively similar to that observed by  Luis \textit{et al.} \cite{cremades2019reynolds}. 

The contours of the dimensionless Reynolds shear stress $-\left\langle u_{1}u_{2} \right\rangle/U_{\infty}^{2}$ over the airfoil for $\operatorname{Tu} = 0.03\%$ were plotted in an $xy$ plane, a peak appeared around the LSB. The $1c$ perturbation under-predicted the baseline prediction, while the $3c$ perturbation increased the magnitude of $-\left\langle u_{1}u_{2} \right\rangle/U_{\infty}^{2}$, showing a tendency to approach closer to the in-house DNS data. The predicted $-\left\langle u_{1}u_{2} \right\rangle/U_{\infty}^{2}$ profiles were also plotted based on the actual locations on the suction side of the airfoil, and an enveloping behavior with respect to the baseline prediction was observed. The size of uncertainty bounds increased to its maximum near $X_{R}$, followed by a smooth reduction as the flow proceeded downstream of $X_{R}$. Overall, the $1c$ and $2c$ perturbations under-predicted the baseline prediction, while the $3c$ perturbation did the opposite. Downstream of $X_{R}$ in the attached turbulent boundary layer, uncertainty bounds generated from the $1c$, $2c$, and $3c$ perturbations began to encompass near-wall sections of the in-house DNS profiles.     

Overall, the eigenspace perturbation framework was effective in constructing uncertainty bounds for a variety of QoIs. Future work will focus on perturbation to the amplitude of Reynolds stress and the eigenvectors of the Reynolds stresses to complete the full range of the model form uncertainty in the SST $k-\omega$ LM model. Also a wider range of RANS-based transition models will be tested using the eigenspace perturbation framework.

\begin{acknowledgments}
The support of the Natural Sciences and Engineering Research Council (NSERC) of Canada for the research program of Professor Xiaohua Wu and Professor David E. Rival is gratefully acknowledged. The author Dr. Minghan Chu thanks Dr. Aashwin Ananda Mishra for providing valuable feedback and helpful discussions.  
\end{acknowledgments}

\section*{Data Availability Statement}
The data that support the findings of this study are available from the corresponding author upon reasonable request. 

\section*{References}
\bibliography{aipsamp}

\end{document}